\documentclass[12pt]{book}

\usepackage[round,authoryear]{natbib}

\usepackage{indentfirst}
\usepackage[T1]{fontenc}
\usepackage{phdthesis}
\usepackage{graphicx}
\usepackage[utf8]{inputenc}
\usepackage{makeidx}
\usepackage{enumerate}
\usepackage{color}
\usepackage{dcolumn}
\usepackage{rotating}
\usepackage{url}
\usepackage{slashed}
\usepackage{txfonts}
\usepackage{lettrine}
\usepackage{amsfonts}
\usepackage{longtable}
\usepackage{bm}

\usepackage{lmodern}

\DeclareMathAlphabet{\mathitbf}{OML}{cmm}{b}{it}

\definecolor{kolor_cyt}{RGB}{49,79,150}
\def\citec#1{\textsc{\textcolor{kolor_cyt}{\cite{#1}}}}
\def\citepc#1{\textsc{\textcolor{kolor_cyt}{\citep{#1}}}}
\def\citealtc#1{\textsc{\textcolor{kolor_cyt}{\citealt{#1}}}}

\def\beq{\begin{equation}}
\def\beqa{\begin{eqnarray}}
\def\beqaa{\begin{eqnarray*}}
\def\eeq{\end{equation}}
\def\eeqa{\end{eqnarray}}
\def\eeqaa{\end{eqnarray*}}

\newcommand{\de}{\mathrm{d}} 
\newcommand{\dep}{\partial}
\newcommand{\del}{\delta}
\newcommand{\ind}{\indent}
\newcommand{\noi}{\noindent}
\newcommand{\Mpc}{\;\mathrm{Mpc}}
\newcommand{\Mpch}{\;\mathrm{Mpc} / h}
\newcommand{\hMpc}{\;h / \mathrm{Mpc}}
\newcommand{\kms}{\;\mathrm{km} / \mathrm{s}}
\newcommand{\kmsMpc}{\;\mathrm{km} / \mathrm{s} / \mathrm{Mpc}}
\newcommand{\bmd}{\mathitbf{d}}
\newcommand{\bmg}{\mathitbf{g}}
\newcommand{\bmk}{\mathitbf{k}}
\newcommand{\bmr}{\mathitbf{r}}
\newcommand{\bmu}{\mathitbf{u}}
\newcommand{\bmv}{\mathitbf{v}}
\newcommand{\bmx}{\mathitbf{x}}
\newcommand{\bmy}{\mathitbf{y}}
\newcommand{\mrc}{\mathrm{c}}
\newcommand{\mre}{\mathrm{e}}
\newcommand{\mri}{\mathrm{i}}
\newcommand{\mrm}{\mathrm{m}}
\newcommand{\mrg}{\mathrm{g}}

\newcommand{\lan}{\langle}
\newcommand{\ran}{\rangle}
\newcommand{\Omm}{\Omega_\mathrm{m}} 
\newcommand{\Omb}{\Omega_\mathrm{b}}
\newcommand{\vLG}{\bmv_\mathrm{_{LG}}}
\newcommand{\gLG}{\bmg_\mathrm{_{LG}}}
\newcommand{\degr}{^{\circ}}
\newcommand{\sig}{\sigma}
\newcommand{\te}{\vartheta}
\newcommand{\calR}{\mathcal{R}}
\newcommand{\calL}{\mathcal{L}}
\newcommand{\eps}{{\epsilon}}
\newcommand{\rmmin}{\mathrm{min}}
\newcommand{\rmmax}{\mathrm{max}}
\newcommand{\Kmin}{K_\rmmin}
\newcommand{\Kmax}{K_\rmmax}
\newcommand{\bmle}{\beta_{\rm MLE}}
\newcommand{\mn}{_{\mu\nu}}
\newcommand{\LCDM}{$\Lambda$CDM}
\newcommand{\intlim}{\int\limits}
\newcommand{\tsig}{\tilde{\sigma}}
\newcommand{\tro}{\tilde{\varrho}}

\newcommand{\CCBCC}{\textsc{\textcolor{kolor_cyt}{C08}}}

\definecolor{mygreen}{RGB}{0,150,0}
\definecolor{myblue}{RGB}{0,191,255}

\newcommand{\aap}{Astronomy \& Astrophysics}

\newcommand{\apjl}{Astrophysical Journal Letters}
\newcommand{\apjs}{Astrophysical Journal Supplement Series}

\makeindex

\begin{document}

\title{\begin{Huge}{\textsf{\textbf{Motion of the Local Group\\as a cosmological probe}}}\end{Huge}}

\author{\vspace{1cm}\\
\begin{LARGE}{Maciej Bilicki}\end{LARGE}\\
\vspace{2cm}\\
\textit{Doctoral thesis written under the supervision of}\\
\textit{prof.\ Micha{\l} Chodorowski}\\
\vspace{3cm}\\
Nicolaus Copernicus Astronomical Center\\
Polish Academy of Sciences}

\date{Warsaw, December 2011}

\maketitle

\newpage
\thispagestyle{empty}

\vspace*{9cm}

\begin{flushright}
\begin{scriptsize}
\noi Illustration to the right:\\distribution of galaxies and stars\\in the Two Micron All Sky Survey.\\
Figure courtesy of Thomas Jarrett.
\end{scriptsize}
\end{flushright}

\newpage
\pagenumbering{roman}
\thispagestyle{empty}

\vspace*{5cm}

\begin{center}
\includegraphics[width=0.9\textwidth]{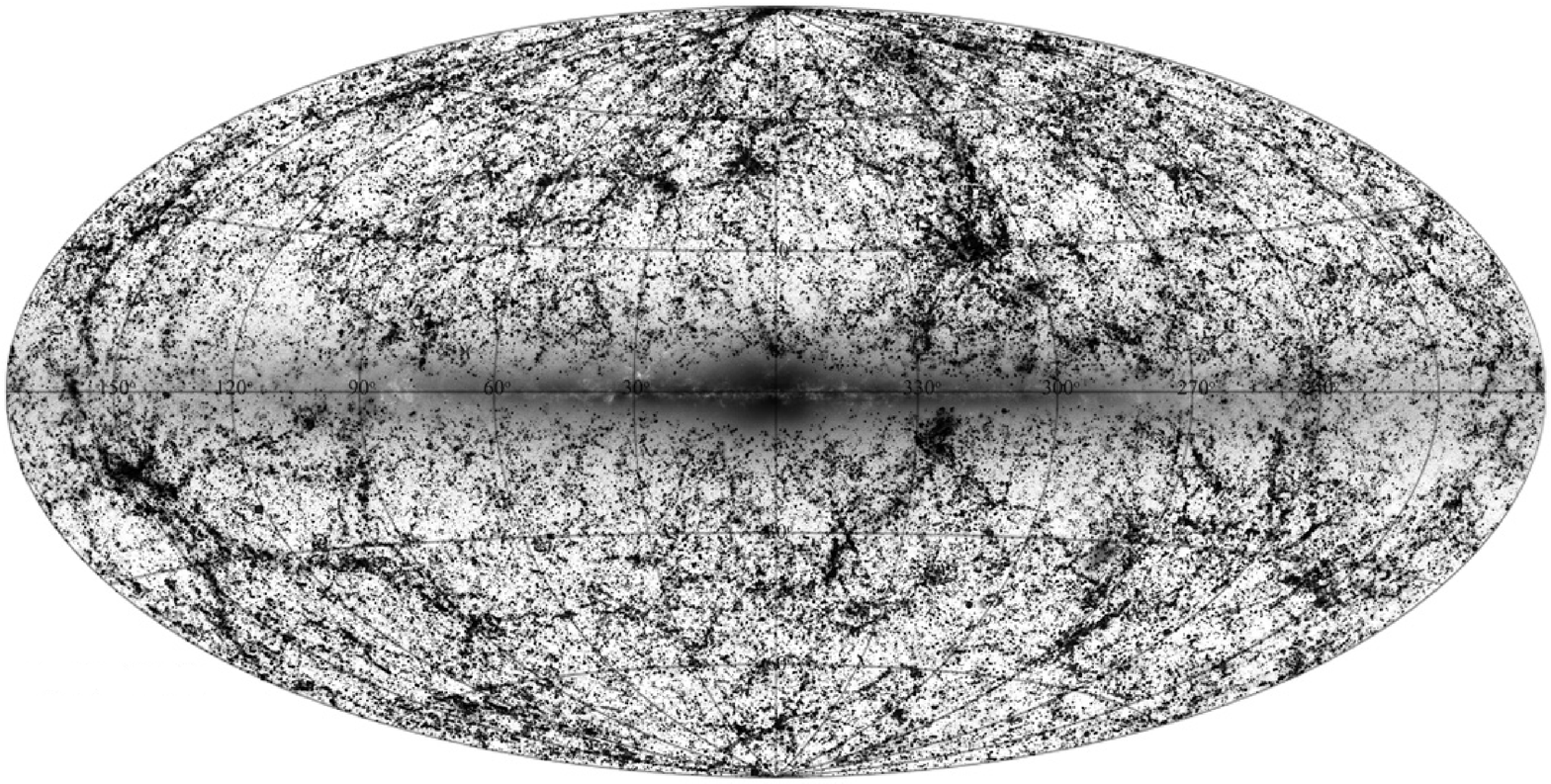} 

\end{center}
\vspace{5cm}

\hspace{1cm}\textit{``What makes you think you can discover anything? Who are you?''}

\hspace{1cm}\textit{``Nobody. Nobody at all. But the secrets of the Universe don't mind.}

\hspace{1cm}\textit{They reveal themselves to nobodies who care.''}

\hspace{5cm}\begin{small}{-- The Outer Limits: The Galaxy Being}\end{small}

\newpage
\thispagestyle{empty}
\mbox{}

\newpage

\begin{flushright}
\begin{Large}
\textbf{\textsf{Podzi\c{e}kowania}}
\end{Large}
\end{flushright}
\begin{small}
Pragn\c{e} podzi\c{e}kowa\'{c} mojemu Promotorowi, prof.\ Micha{\l}owi Chodorowskiemu, za ide\c{e} tego projektu oraz sta{\l}\c{a} pomoc i wsparcie podczas moich studi\'{o}w doktoranckich. Bez Niego praca ta nigdy by nie powsta{\l}a.

\noi Dzi\c{e}kuj\c{e} r\'{o}wnie\.{z} mojej \.{Z}onie, Agnieszce, za cierpliwo\'{s}\'{c}, naszemu Synowi, Miko{\l}ajowi, za rado\'{s}\'{c} i u\'{s}miech, a tak\.{z}e Rodzicom, Babciom i Siostrze.

\noi Podzi\c{e}kowania nale\.{z}\c{a} si\c{e} r\'{o}wnie\.{z} moim Kolegom ze studium doktoranckiego w Centrum Astronomicznym im.\ Miko{\l}aja Kopernika, a w szczeg\'{o}lno\'{s}ci Magdzie i Przemkowi za mi{\l}\c{a} atmosfer\c{e} w pracowni oraz Wojtkowi za dostarczenie szablonu pracy doktorskiej i pomoc w korzystaniu z niego. 

\noi Komfort pracy i mo\.{z}liwo\'{s}\'{c} wyjazd\'{o}w zagranicznych gwarantowa{\l}y mi dwa projekty grantowe Ministerstwa Nauki i Szkolnictwa Wy\.{z}szego: nr nr N N203 025333 i N N203 509838. Wyrazy wdzi\c{e}czno\'{s}ci dla ich Kierownik\'{o}w: prof.\ Ewy {\L}okas oraz prof.\ Micha{\l}a Cho\-do\-row\-skie\-go. 
\end{small}

\vspace{0.3cm}
\begin{flushright}
\begin{Large}
\textbf{\textsf{Acknowledgements}}
\end{Large}
\end{flushright}
\begin{small}
I am indebted to my advisor, prof.\  Micha{\l} Chodorowski, for the idea behind this project, as well as his constant assistance and support along the way. This thesis would not have been possible without him.

\noi Great thanks to my wife Agnieszka for patience, our son Miko{\l}aj for the joy and smiles, as well as to my Parents, Grandmothers and Sister.

\noi Many people helped me a lot during this project. Special acknowledgements go to Tom Jarrett for his invaluable assistance in revealing the secrets of 2MASS galaxies and to Gary Mamon for long discussions and explanations, as well as to Adi Nusser for useful comments. I am also grateful to Tom, Gary, Adi, Guilhem Lavaux and Roya Mohayaee for their hospitality during my research visits. Constructive suggestions were provided by Rien van de Weygaert, Marc Davis and Pirin Erdo{\u g}du. R.\ Brent Tully gave the original idea of the study concerning the Local Void.

\noi Last but not least, I want to thank my fellow PhD students from the Copernicus Astronomical Center, especially my officemates Magda Otulakowska-Hypka and Przemek Jacewicz for great atmosphere in the office, as well as Wojtek Hellwing for the thesis template and his help with its usage.
\end{small}

\newpage

\begin{small}
\noi I have made use of:\\ 
-- data products from the Two Micron All Sky Survey (2MASS), which is a joint project of the University of Massachusetts and the Infrared Processing and Analysis Center/California Institute of Technology, funded by the National Aeronautics and Space Administration and the National Science Foundation;\\
-- the NASA/IPAC Extragalactic Database (NED), which is operated by the Jet Propulsion Laboratory, California Institute of Technology, under contract with the National Aeronautics and Space Administration;\\
-- the NASA/IPAC Infrared Science Archive, which is operated by the Jet Propulsion Laboratory, California Institute of Technology, under contract with the National Aeronautics and Space Administration.\\
I also  acknowledge the use of the TOPCAT software \citep{TOPCAT}.\\

\noi The research presented here was partially supported by the Polish Ministry of Science and Higher Education under grants nos.\ N N203 025333 and N N203 509838. I thank their PIs, respectively prof.\ Ewa {\L}okas and prof. Micha{\l} Chodorowski. Part of this work was carried out within the framework of the European Associated Laboratory ``Astrophysics Poland-France''. Acknowledgements to prof.\ Pawe{\l} Haensel.
\end{small}

\newpage
\begin{flushright}
\begin{LARGE}
\textbf{\textsf{Summary}}
\end{LARGE}
\end{flushright}

In this thesis, we use the motion of the Local Group of galaxies (LG) through the Universe to measure the cosmological parameter of non-relativistic matter density, $\Omm$. For that purpose, we compare the peculiar velocity of the LG with its gravitational acceleration. The former is known from the dipole of the cosmic microwave background radiation and the latter is estimated here from the clustering dipole of galaxies in the Two Micron All Sky Survey (2MASS) Extended Source Catalog. We start by presenting the general framework of perturbation theory of gravitational instability in the expanding Universe and how it applies to the  peculiar motion of the LG. Next, we study a particular effect for the dipole measurement, related to the fact that a nearby Local Void is partially hidden behind our Galaxy. We then describe in detail how we handled the 2MASS extragalactic data for the purpose of our analysis. Finally, we present two methods to estimate the density $\Omm$, combined with the linear biasing $b$ into the parameter $\beta=\Omm^{0.55}/b$, from the comparison of the LG velocity and acceleration. The first approach is to study the growth of the 2MASS clustering dipole with increased depth of the sample and compare it with theoretical expectations. The second is to apply the maximum-likelihood method in order to improve the precision of the measurement. With both these methods we find $\beta\simeq0.4$ and $\Omm\simeq0.2$, which is consistent with various independent estimates. We also  briefly mention some future prospects in the field.

\vspace{2cm}

\begin{small}
\noi\textit{This Thesis presents and expands the results included in the following publications:}\\
\noi\cite{BiCh10}, \textit{MNRAS}, \textbf{406}, 1358 \textit{-- Chapter \ref{Ch:Nonlin},}\\
\noi\cite{BCJM11}, \textit{ApJ}, \textbf{741}, 31 \textit{-- Chapters \ref{Ch:Data} and \ref{Ch:Growth}.}

\vspace{0.2cm}

\noi\textit{Additionally, in Chapters \ref{Ch:Growth} and \ref{Ch:MLE} we apply the findings of} \cite{CCBCC}, \textit{MNRAS}, \textbf{389}, 717.

\noi\textit{Material unpublished at the time of Thesis submission is contained in Chapter \ref{Ch:MLE}.}
\end{small}

\textsf{\tableofcontents\label{TOC}}


\chapter{\textsf{\textbf{Introduction}}}
\label{Ch:Intro}
\pagenumbering{arabic}
\setcounter{page}{1}
\lettrine[lraise=0.5,lines=2,findent=1pt,nindent=0em]{O}bservational cosmology has been developing rapidly in the recent years. Progress in detector technology, usage of still bigger telescopes in more and more astronomically favorable places (including Antarctica and space), {supported by numerous theoretical investigations and computer simulations,} resulted in the emergence of a model of the Universe, which is considered to be \textit{standard}: the Lambda--Cold Dark Matter cosmological model (\LCDM). It is based on the \textit{cosmological principle}: the Universe is homogeneous and isotropic when smoothed over large scales. This assumption is supported by various observations, including these of the cosmic microwave background (CMB) and the large-scale distribution of galaxies.\\ 
\ind Among the great successes of the \LCDM\ model, one should mention its ability to perfectly reproduce the observed properties of the CMB. The angular power spectrum of the temperature distribution of this radiation is now accurately measured up to the multipole moment of $\ell\simeq3000$, thanks to several years of observations of the WMAP satellite \citepc{Jarosik11}, as well as new ground-based facilities, such as the Atacama Cosmology Telescope \citepc{ACT11} or the South Pole Telescope \citepc{SPt11}. These measurements, as well as those of the CMB temperature-polarization power spectrum (currently known up to $\ell\simeq700$, \citealtc{Jarosik11}), are perfectly fitted by a simple spatially flat model with non-zero cosmological constant $\Lambda$. Another evidence strongly supporting the validity of this model on the largest scales are numerous measurements of \textit{baryon acoustic oscillations} (BAO), i.e.\ patterns of baryonic matter clustering at certain length scales due to acoustic waves which propagated in the early universe. BAO give us a `standard ruler', whose size measured at various cosmological epochs is consistent with predictions of \LCDM\ (e.g.\ \citealtc{Percival10,Blake11,Beutler11}). Last but not least, observations of `standard candles' such as supernovae Ia, which actually gave first firm arguments for a non-zero value of $\Lambda$ \citepc{Riess98,Perl99}, additionally confirm that this model is a very successful general description of our Universe (e.g.\ \citealtc{Kessler09}).\\
\ind On smaller scales {we cannot of course treat the Universe as perfectly homogeneous, because galaxies not only do exist, but also} are not arranged randomly and form structures such as clusters, superclusters, filaments or walls, separated by huge voids. For that reason, there are claims that the observed inhomogeneities cannot be neglected when analyzing the big picture of the Cosmos, {for instance by accounting for the so-called \textit{backreaction} (see e.g.\ \citealtc{Buch11} and references therein). However, in general these effects are believed to be weak and owing to the considerations presented in the previous paragraph} we will not address them here. On the contrary, from now on we will assume that the cosmological background is \textit{statistically} homogeneous and isotropic over large scales. What is more, {it is known that} gravity is the major force responsible for large-scale structure evolution {-- the other three fundamental interactions are negligible on cosmic scales}. This means altogether that the Universe as a whole can be described in the framework of  \textit{Friedman-Lema\^{i}tre models}, obtained by solving the \textit{Einstein equations} under these assumptions, {and any perturbations with respect to this idealized picture will be considered within this model as a background.}
\section{Friedman-Lema\^{i}tre models}
The cosmological principle expressed mathematically says that the metric of the Universe (the `background') is given by the \textit{Robertson-Walker form}:
\beq
\label{eq:RW metric}
\de s^2=c^2 \de t^2-a^2(t)\left[\frac{\de x^2}{1-k x^2}+x^2\left(\de \theta^2+\sin^2\theta\, \de \phi^2\right)\right]\;
\eeq
 \citepc{Rob29,Walk33}, where $a(t)$ is the scale factor, $t$ is the cosmic time, $k$ is the curvature constant and $(x, \theta, \phi)$ are spatial (spherical polar) coordinates.

Another assumption underlying the relativistic cosmology is that the energy-momentum tensor of the matter field, $T\mn$, takes the form proper for the ideal fluid:
\begin{equation}\label{eq:e-m tensor}
T\mn=(\rho+p)u_\mu u_\nu+p g\mn\;,
\end{equation}
where $u_\mu$ is the four-velocity of matter, $\rho$ stands for energy density, $p$ is pressure and $g\mn$ the metric tensor.

Writing down the general form of the Einstein equations \citepc{OTW},
\begin{equation}\label{eq:Einstein}
R\mn-\tfrac{1}{2}R\, g\mn+\Lambda\, g\mn=8\pi G\, T\mn
\end{equation}
{(with $R\mn$ the Ricci tensor and $R$ the Ricci scalar)} and plugging Eqs.\ (\ref{eq:RW metric}) and (\ref{eq:e-m tensor}) into it, we get the following two independent \textit{Friedman-Lema\^{i}tre equations}:
\beqa
H^2 \,\, \equiv \,\, \left(\frac{\dot{a}}{a}\right)^2 & = & \frac{8\pi G}{3}\rho-\frac{k}{a^2}+\frac{\Lambda}{3} \label{eq:Friedman dot} \\
\frac{\ddot{a}}{a} & = & -\frac{4\pi G}{3}(\rho+3 p)+\frac{\Lambda}{3} \label{eq:Friedman doubledot}
\eeqa
\citepc{Frie22,Lem27}, where $H$ (\textit{Hubble parameter}), $a$, $\rho$ and $p$ are functions of the cosmic time.

{The major components of the energy-density budget of the Universe can be treated as perfect fluids with the equation of state (EOS), $p(\rho)$, given by a simple relation
\begin{equation}
p=w \rho\;.
\end{equation}
In particular, for the purpose of cosmological analyses the pressure of non-relativistic matter (cold dark matter and baryons) can be neglected and we can describe it as \textit{dust} with $w=0$. Further on, for radiation and neutrinos we have  $w={1}\slash{3}$ and finally, for the cosmological constant the EOS parameter is $w=-1$.}

{Except for the very early universe, interactions between different species of the cosmic fluid do not change their relative number density. Therefore, we can} separate the density into a sum over independent components, $\rho=\sum_i\rho_i$.  Setting the present value of the scale factor, $a_0$, to unity (which will be assumed from now on, without loss of generality), and evaluating the first F-L equation at the present time ($t=t_0$), we get:
\begin{equation}
H_0^2=\frac{8\pi G}{3}\sum_i\rho_{i,0}-k+\frac{\Lambda}{3},
\end{equation}
where $H_0$ (the present-day value of the Hubble parameter) is the \textit{Hubble constant}.

We can now define dimensionless \textit{density parameters} of respective components, very useful in cosmological analyses:
\begin{equation}
\Omega_{i,0}\equiv\frac{8\pi G}{3 H_0^2}\rho_{i,0}\;,
\end{equation}
as well as the density parameter of the cosmological constant:
\begin{equation}
\Omega_{\Lambda ,0}\equiv\frac{\Lambda}{3 H_0^2}\;.
\end{equation}
The quantity $\rho_c\equiv(3 H_0^2)/(8\pi G)$ is called the \textit{critical density} (i.e.\ the density of a flat Universe with matter only, the \textit{Einstein-de Sitter model}, \citealtc{EdS}) and the present-day density parameter of non-relativistic matter, \mbox{$\Omega_{m,0}\equiv\rho_{m,0}/\rho_c$}, will from now on be simply denoted  as $\Omm$; similarly for $\Omega_\Lambda$.

\begin{figure}[!t]
\begin{center}
\includegraphics[width=0.8\textwidth]{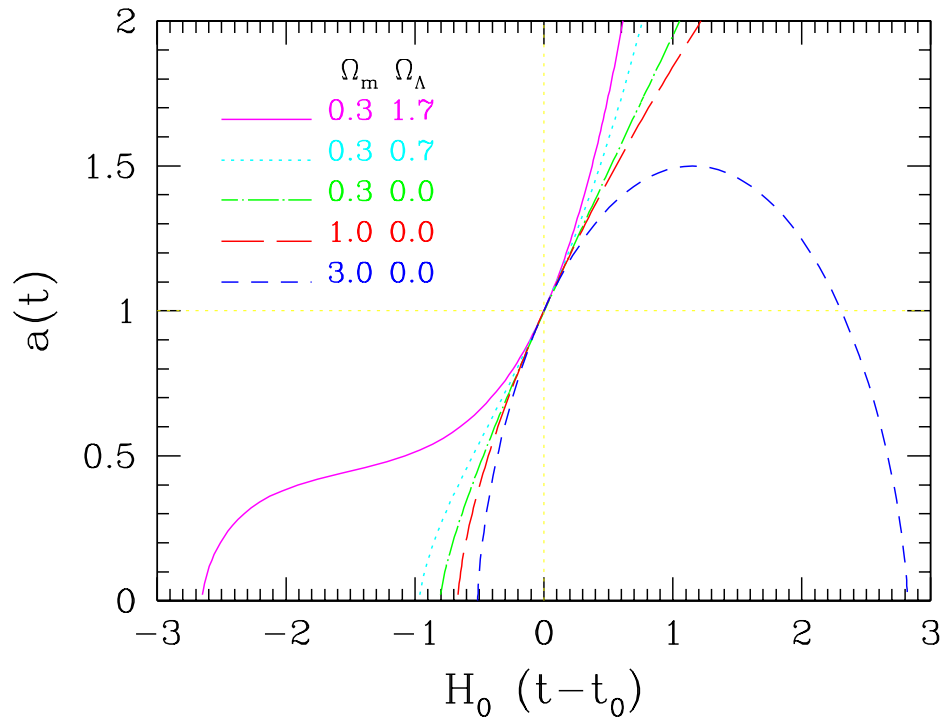}
\caption[Evolution of the scale factor]{\label{Fig:Scale factor}\small Evolution of the scale factor as a function of the cosmic time, normalized by the Hubble constant. Different curves represent different values of density parameters $\Omm$ and $\Omega_\Lambda$ as in $t_0$ (today). [Reproduced from \citec{LaSu04} by permission of the LRR.]}
\end{center}
\end{figure}

The Universe {at low redshifts} is dominated by non-relativistic matter: \textit{baryons} and collisionless \textit{cold dark matter} {and, as strongly suggested by observations mentioned at the beginning of this Chapter, the} \textit{cosmological constant} {(dubbed also \textit{dark energy}\footnote{{In general, the term \textit{dark energy} has a wider meaning and applies also to generalizations of the cosmological constant with an  equation of state different than $p=-\rho$.}}). The relevant equation} for the temporal evolution of the scale factor (expansion of the Universe) {is then}
\begin{equation}\label{Eq:a.dot}
\left(\frac{\de a}{\de t}\right)^2=H_0^2\left(\frac{\Omm}{a}+1-\Omm-\Omega_\Lambda+\Omega_\Lambda a^2\right)\;.
\end{equation}
{This means that the expansion history of the Universe in the Friedman-Lema\^{i}tre mo\-dels can be constrained if we know \textit{current} values of relevant density parameters{\footnote{{In the present Universe, both for neutrinos and radiation we have $\Omega_\nu\simeq\Omega_\gamma\simeq0$. However, Equation (\ref{Eq:a.dot}) can be modified to account for their non-zero values.}}}. Figure~\ref{Fig:Scale factor} shows $a(t)$ for several values of $(\Omm,\Omega_\Lambda)$. The currently favored \LCDM\ cosmological model is the one with $\Omm\simeq0.3$ and $\Omega_\Lambda\simeq0.7$ (light-blue dotted line).

\section{Perturbation theory}
\label{Sec:Pert.theory}
The Friedman-Lema\^{i}tre model as outlined in the previous section  explicitly assumed the homogeneity and isotropy of the Universe. {These assumptions certainly hold for the early stages of the evolution of the Universe, which is confirmed for instance by the high degree of isotropy of the CMB. There are various reasons to believe that this approximation is valid also {currently} for the largest scales (e.g.\ \citealtc{Hogg05}).} However, the mere existence of planets, stars or galaxies means that for sufficiently \textit{small} scales, our Universe is strongly inhomogeneous. Cosmology tries to answer the question how the structures such as galaxies, their clusters and superclusters, as well as voids between them (the `cosmic web' in general) came to be. The widely accepted answer says that they emerged from tiny fluctuations existing at a much earlier epoch and grew by the \textit{gravitational instability}. We will now shortly present this theory, especially for small density fluctuations, i.e.\ in the linear regime. Further details can be found in classic textbooks such as \citec{Pe80} or reviews, for instance \citec{StWi95}.

We will deal with scales much smaller than the so-called \textit{Hubble radius}, $r_H\equiv c/H_0$, and generally with small values of gravitational potential. This means that we can focus on the non-relativistic regime, where the Newtonian approximation is valid. For a fully relativistic description of the gravitational instability, the interested reader is directed to e.g.\ \citec{Mukh05}.

The basic equations for self-gravitating fluid are given by continuity, Euler and Poisson equations:
\beqa
\frac{\dep\rho}{\dep t}+\nabla\cdot(\rho\bmu) & = & 0\;, \label{eq:continuity} \\
\frac{\dep\bmu}{\dep t}+(\bmu\cdot\nabla)\bmu & = & -\frac{1}{\rho}\nabla p-\nabla\Phi\;, \label{eq:Euler} \\
\nabla^2\Phi & = & 4\pi G\rho\;. \label{eq:Poisson}
\eeqa
It is convenient to rewrite the above equations in the so-called \textit{comoving coordinates}. For that purpose we define the position $\bmx$ in the comoving frame, the \textit{peculiar velocity} $\bmv$, the \textit{density contrast} $\del$ and the gravitational potential $\phi$ in the following way:
\beqa
\bmx & = & \frac{\bmr}{a(t)}\;,\\
\bmv & = & a(t)\dot{\bmx}\;,\\
\del(t,\bmx) & = & \frac{\rho(t,\bmx)}{\bar{\rho}(t)}-1\;,\\
\phi(t,\bmx) & = & \Phi+\frac{1}{2}a\ddot{a}|\bmx|^2\;,
\eeqa
where $\bar{\rho}(t)$ is the mean density of the background at a given time. In other words, the peculiar velocity {$\bmv$} of a galaxy is the one that adds to the Hubble flow{:
\beq
\dot{\bmr}=\dot{a}\bmx+a\dot{\bmx}=H(t)\bmr+\bmv\;.
\eeq
}Equations (\ref{eq:continuity})--(\ref{eq:Poisson}) in the comoving frame take on the form:
\beqa
\frac{\dep \del}{\dep t}+\frac{1}{a}\nabla\cdot[(1+\del)\bmv] & = & 0\;, \label{eq:cont.comov} \\
\frac{\dep \bmv}{\dep t}+\frac{1}{a}(\bmv\cdot\nabla)\bmv+\frac{\dot{a}}{a}\bmv & = & -\frac{1}{\rho a}\nabla p-\frac{1}{a}\nabla\phi\;, \label{eq:Eul.comov} \\
\nabla^2\phi & = & 4\pi G \bar{\rho} a^2 \delta \label{eq:Poiss.comov}
\eeqa
(where the derivatives are now defined in the comoving coordinates). The above equations are exact in the Newtonian regime and serve as a starting point for fully-nonlinear analyses of gravitational instability. Here we will deal mainly with the linear theory, where the density fluctuations are small ($\del\ll 1$). If we now consider only dust and neglect pressure, then in the linear regime Eqs. (\ref{eq:cont.comov}) and (\ref{eq:Eul.comov}) can be linearized with respect to $\del$ and $\bmv$:
\beqa
\frac{\dep \del}{\dep t}+\frac{1}{a}\nabla\cdot\bmv & = & 0\;, \label{eq:cont.linear} \\
\frac{\dep \bmv}{\dep t}+\frac{\dot{a}}{a}\bmv + \frac{1}{a}\nabla\phi & = & 0\;. \label{eq:Eul.linear}
\eeqa
Taking now the time derivative of the continuity equation (\ref{eq:cont.linear}) and using the Poisson (\ref{eq:Poiss.comov}) end linearized Euler (\ref{eq:Eul.linear}) equations, we get:
\beq
\frac{\dep^2\del}{\dep t^2}+\frac{2\dot{a}}{a}\frac{\dep \del}{\dep t}-4\pi G\bar{\rho}\del=0\;.
\eeq
This formula for the linear evolution of the density contrast is a second-order partial differential equation in time only, so we can separate temporal and spatial dependencies and write the general solution as:
\begin{equation}
\delta=A(\bmx)D_g(t)+B(\bmx)D_d(t)\;,
\end{equation}
where $D_g$ and $D_d$ are called respectively the \textit{growing} and \textit{decaying mode}. The latter monotonically decreases as the Universe expands and eventually becomes negligibly small, so we can focus only on the growing one, written from now on simply as $D$. It is given by
\begin{equation}
D(t)=\frac{\dot{a}}{a}\intlim_0^a\frac{\de a'}{(\dot{a}')^3}\;.
\end{equation}
In the general case of a non-vanishing cosmological constant, this integral is not analytic. Closed-form expression for the both modes do exist for $\Lambda=0$, see e.g.\ \citec{Pe80}.

The growing mode can be equivalently expressed in terms of the redshift $z$:
\begin{equation}
D(z)=\frac{5\,\Omm\, H_0^2}{2}\,H(z)\intlim_z^\infty{\frac{1+z'}{H^3(z')}\,\de z'}\;.
\end{equation}
In the case of spatially-flat models with the cosmological constant, such as the \LCDM, it can be written as (e.g.\ \citealtc{LaSu04})
\begin{equation}
D(z)\propto\sqrt{1+\frac{2}{y^3}}\intlim_0^y{\left(\frac{q}{2+q^3}\right)^{3/2}\de q}\;,\qquad y\equiv\frac{2^{1/3}\left(\Omm^{-1}-1\right)^{1/3}}{1+z}\;.
\end{equation}

In the regime when the growing mode dominates, we can rewrite Eq.\ (\ref{eq:cont.linear}) as
\begin{equation}
\nabla\cdot\bmv=-a\,\del\,\frac{\dot{D}}{D}
\end{equation}
and if we define the \textit{growth factor} $f$ as
\begin{equation}
f\equiv\frac{\de \ln D}{\de \ln a}=\frac{\dot{D}}{H\, D}\;,
\end{equation}
we finally obtain the linear-theory \textit{relation between the density and peculiar velocity divergence} fields:
\begin{equation}
\nabla\cdot\bmv=-a\, H \, f\, \del\;.
\end{equation}
{In the low-redshift universe, $z\ll1$, this relation reads
\beq\label{eq:teta-delta}
\nabla\cdot\bmv=-H_0 \, f\, \del\;.
\eeq}
\ind The growth factor $f$ is a function of $\Omm$ and $\Lambda$, although the dependence on the latter is extremely weak {(e.g.\ \citealtc{LLPR91,Martel91})}. A {very accurate} approximation is the one provided by \citec{Linder05}:
\begin{equation}
f(\Omm,\Lambda)\simeq\Omm(z)^\gamma\;,
\end{equation}
with
\begin{equation}
\gamma=0.55+0.05\left(1+w|_{z=1}\right)\;,
\end{equation}
where $w(z)$ denotes the equation of state parameter of generalized dark energy. For the cosmological constant, $w\equiv-1$, so $f\simeq\Omm^{0.55}$, which reproduces the exact growth factor {in the \LCDM\ model} for any redshift to better than 0.05\% over the range $\Omm\in[0.22, 1]$ and it remains accurate to below 1\% for $\Omm \geq 0.01$. A recent paper by \citec{BB11} presents exact solutions in the case of vacuum energy that can be parametrized by a constant equation of state parameter $w$ and a very accurate approximation for the ansatz $w(a) = w_0+w_a(1-a)$.

\section{Statistical description of cosmological fields}
\label{Sec:Statistical description}
Galaxies are not distributed in space randomly, but they tend to gather in groups, clusters and superclusters, separated by huge voids. This means that the probability of finding a galaxy at location $\bmx$ is not independent of whether there is a galaxy in the vicinity of $\bmx$ or not: it is more probable to find a galaxy in the neighborhood of another one that at an arbitrary location. This is described mathematically in terms of the \textit{correlation function}, $\xi$ (e.g.\ \citealtc{Pe80}). The probability of finding a galaxy in a volume element $\de V$ at a location $\bmx$ and at the same time finding another galaxy in this volume element at a location $\bmy$ is
\beq
P_\mathrm{gg}=(\bar{n}\,\de V)^2\left[1+\xi_\mathrm{g}(\bmx,\bmy)\right]\;,
\eeq
where $\bar{n}$ is the average number density of galaxies and $\xi_\mathrm{g}(\bmx,\bmy)$ is their two-point correlation function. By analogy, the correlation function for the total matter density can be defined as
\beq
\langle\rho(\bmx)\rho(\bmy)\rangle={\bar{\rho}}\;^2\langle\left[1+\del(\bmx)\right]\left[1+\del(\bmy)\right]\rangle={\bar{\rho}}\;^2\left[1+\langle\del(\bmx)\del(\bmy)\rangle\right]=\bar{\rho}\left[1+\xi(\bmx,\bmy)\right]\;,
\eeq
because the mean (expectation) value $\langle\del(\bmx)\rangle=0$ for all locations $\bmx$. In the above {and hereafter}, angular brackets denote averaging over an ensemble of distributions that all have identical properties. Moreover, since the Universe is considered statistically homogeneous, $\xi$ can only depend on the difference $\bmx-\bmy$ and not on the individual locations. Additionally, owing to the statistical isotropy, $\xi(\bmx-\bmy)=\xi(|\bmx-\bmy|)$, i.e.\ it depends only on the separation $r=|\bmx-\bmy|$. Therefore, $\xi=\xi(r)$. The galaxy-galaxy correlation function and that of matter are in linear theory assumed to be related via the \textit{linear biasing} paradigm:
\beq\label{eq:ksi.bias}
\xi_{\mrg}(r)=b^2\,\xi_\mrm(r)
\eeq
(see e.g.\ \citealtc{StWi95}), where $b$ is the \textit{biasing parameter}. We will come back to the issue of biasing in Section \ref{Sec:Densities and velocities}.

An alternative and equivalent description of the statistical properties of {matter clustering in} the Universe is the \textit{power spectrum}, $P(k)$. It describes the level of structure as a function of the length-scale $L\simeq2\pi/k$, where $k$ is a \textit{wavenumber} in Fourier space. The larger the $P(k)$, the larger the amplitude of the fluctuations on a length-scale $2\pi/k$. Technically speaking, the density fluctuation fields is decomposed into a sum of plane waves of the form $\del(\bmx)=\sum a_\bmk\cos(\bmx\cdot\bmk)$, where $\bmk$ is the wavevector and $a_\bmk$ is the amplitude. This is a \textit{Fourier decomposition} of the density field and the power spectrum $P(k)$ describes the distribution of amplitudes with equal $k=|\bmk|$. The correlation function and the power spectrum form a Fourier transform pair, that is
\begin{equation}
\xi(r)=\frac{1}{(2\pi)^3}\int P(k)\,\mre^{\mri\bmk\cdot\bmr}\de^3\bmk\;.
\end{equation}

In principle, one can also describe the peculiar velocity field in a similar way, although this description is more complicated due to the vector nature of velocities. This however will not  be needed for the purpose of this work. We redirect the interested reader to such textbooks as \citec{Pe93} or \citec{CoLu95}.

\section{Cosmic density and velocity fields}
\label{Sec:Densities and velocities}
As we have seen in Section \ref{Sec:Pert.theory}, in the limit of small perturbations, there is a linear relation between the velocity divergence and density contrast:
\beq\label{eq:teta-delta1}
\nabla\cdot\bmv=-H_0\, f(\Omm,\Lambda)\, \del\;.
\eeq
We can now invert this Equation via the methods of electrostatics to obtain the integral relation, now in proper coordinates:
\begin{equation}\label{eq:v-delta}
\bmv(\bmr)=\frac{H_0\,f}{4\,\pi}\int{\delta(\bmr ')\,\frac{\bmr '-\bmr }{|\bmr '-\bmr |^3}\,\de^3 \bmr '}\;
\end{equation}
(the right-hand side accepts adding an arbitrary divergence-free term, which however corresponds to the decaying solution and will be neglected from now on). Writing down the \textit{peculiar acceleration} vector at a position `$\bmr$' as 
\begin{equation}\label{eq:g.theor}
\bmg(\bmr)=G\, \rho_b \int{\delta(\bmr ')\,\frac{\bmr '-\bmr }{|\bmr '-\bmr |^3}\,\de^3 \bmr '}\;,   
\end{equation}
we immediately obtain the linear relation between (peculiar) velocity and acceleration fields:
\begin{equation}\label{eq:v.and.g}
\bmv(\bmr)=\frac{H_0\, f(\Omega_\mrm)}{4\, \pi\, G\, \rho_b}\;\bmg(\bmr).  
\end{equation}

Relations (\ref{eq:teta-delta1}) and (\ref{eq:v.and.g}) can be interpreted in the following way. If we accept the gravitational instability as the mechanism of large-scale structure formation, what we obtain is that  within this framework inhomogeneities in matter distribution induce gravitational accelerations, which results in galaxies having peculiar velocities that add to the Hubble flow. These velocities in turn enhance the growth of the inhomogeneities, causing strong coupling between cosmic velocity field and large-scale matter distribution. In perturbation theory of Friedman-Lema\^{\i}tre models, in the linear regime, peculiar velocities and accelerations are aligned and proportional to each other at every point. More importantly, the proportionality coefficient of this relation is a simple function of the cosmological parameter of non-relativistic matter density, $\Omm$, and practically does not depend on the cosmological constant (nor other forms of dark energy). 

However, there are some complications. As what we observe are galaxies, we have to assume some relation between their density field and that of matter in general (including dark). {As we have mentioned in the preceding Section,} in linear theory this is usually done via the \textit{linear biasing paradigm}, here expressed in terms of density contrasts:
\beq\label{eq:biasing}
\delta_\mrg=b\, \delta_\mrm\;.
\eeq
This formula is not strictly valid as for underdensities (voids) with $\del_\mrm\gtrsim-1$ ($\rho_\mrm\simeq0$) and for $b>1$, we would obtain $\del_\mrg<-1$, which is not possible. The formal definition of linear biasing is expressed in terms of galaxy-galaxy and mass-mass correlation functions, as was given by Eq.\ (\ref{eq:ksi.bias}). The relation (\ref{eq:biasing}) follows from this, but the reverse is not true. This biasing scheme, valid for sufficiently large scales, neglects the stochasticity, as well as possible scale- and galaxy-type dependence in the relation between the two density fields. For more details on modeling the biasing, see for example the review by \citec{LaSu04} and references therein.

Including the biasing relation into Eq.\ (\ref{eq:g.theor}) and using the fact that for a spherical survey $\int\frac{\bmr '-\bmr}{|\bmr '-\bmr |^{3}}\,\de^3 \bmr '=0$, we get the following expression for the peculiar acceleration:
\begin{equation}\label{eq:g.galaxies}
\bmg(\bmr)=\frac{G}{b}\int{\rho_\mrg(\bmr ')\,\frac{\bmr '-\bmr }{|\bmr '-\bmr |^3}\,\de^3 \bmr '}\;.
\end{equation}

Owing to the nature of astronomical observations, we observe {only the} distribution {of luminous galaxies} and can measure their peculiar velocities.
Using the general framework described here, we cannot generally constrain the bias and growth factor independently. The two are thus combined into the parameter
\begin{equation}
\beta\equiv\frac{f(\Omega_\mrm)}{b}\;.
\end{equation}
Comparing now Eqs.\ (\ref{eq:v.and.g}) and (\ref{eq:g.galaxies}), we get the proportionality valid in linear theory:
\begin{equation}\label{eq:v.beta.g}
\bmv \propto  \beta\, \bmg\;,
\end{equation}
or in differential, equivalent, form:
\begin{equation}
\nabla\cdot\bmv\propto-\beta\, \delta\;.
\end{equation}
These two relations have been widely used to constrain the $\beta$ parameter for more than two decades now\footnote{The same $\beta$ parameter can be also constrained from so-called \textit{redshift space distortions}, which arise because of the effect of peculiar velocities on the shape of structures seen in redshift surveys. A more detailed description is beyond the scope of this work{, see e.g.\ \citec{Kaiser87} or \citec{Hami98}}.}. There are generally two classes of methods for this: the first one, \textit{velocity-velocity comparisons}, is to predict the velocity field from redshift surveys using Eq.\ (\ref{eq:v-delta}) and to compare it with observed peculiar velocities. The second one, \textit{density-density comparisons}, does the opposite: predicts the density field from peculiar velocity surveys using Eq.\ (\ref{eq:teta-delta}) and compares it with redshift surveys. For some recent results of using the two methods to derive the $\beta$ parameter, see e.g.\ \citec{Erdogdu06b} or \citec{Davis11} and references therein.

Both v-v and d-d comparisons have their advantages and limitations. The latter include our poor knowledge of peculiar velocities: they can be measured only if we know both the redshift and the actual distance of the galaxy. Apart from the few closest galaxies, distances estimated independently of redshifts are constrained with an accuracy no better than $10$--$20\%$, and only for several thousand galaxies with distances smaller than several dozens Mpc. This translates directly to big errors in peculiar velocities (bigger at large distances than the estimated velocities themselves). However, there is one special peculiar velocity that we know to a very good accuracy of $\sim\!5\%$ and which can be very well used for cosmological studies. It is the velocity of the system that our Galaxy belongs to: the Local Group of galaxies.

\section{Local Group as a probe: the clustering dipole}
\label{Sec:Clust.dipole}
The Local Group (hereafter `LG') is {a gravitationally bound group of galaxies with several dozen members (up to 60, \citealtc{Lees08})}, including the major players, the Milky Way and M31. Apart from internal motions of its galaxies, the whole system moves through the Universe with respect to the Hubble flow. Application of the Equation (\ref{eq:g.theor}) is simple in this case: if we take $r=0$ (the barycenter of the LG) and include biasing, we obtain
\begin{equation}\label{eq:vLG-gLG}
\vLG=\frac{H_0\, f(\Omega_\mrm)}{4\, \pi\, G\, \rho_b}\;\gLG=\frac{H_0\,\beta}{4\,\pi}\int{\delta_g(\bmr ')\,\frac{\bmr'}{|\bmr '|^3}\,\de^3 \bmr '}\;
\end{equation}
This relation could in principle be applied directly to the motion of the LG through the Universe. Consequently, comparison of the peculiar velocity and acceleration of the LG may serve as a tool to estimate the $\beta$ parameter. Independent knowledge of biasing allows to estimate the cosmological density $\Omm$. In reality, however, this procedure is not that straightforward. In order to understand why it is so, let us explain how the two quantities in this relation, LG velocity and acceleration, are measured.\\
\ind The peculiar velocity of the LG is the easier one of the two to constrain. It is known from the observed dipole anisotropy of the cosmic microwave background \citepc{Hinsh09}, interpreted as a kinematic effect, and reduced to the barycenter of the LG \citepc{CvdB99}. It equals to $v_\mathrm{_{CMB}}=622\pm35\kms$ and points in the direction $(l,b)=(272\degr\pm3\degr\!,\,28\degr\pm5\degr)$ in Galactic coordinates (Hydra constellation), where the errors both in amplitude and direction come mostly from the uncertainty of the Local Group internal dynamics. The kinematic interpretation of the CMB dipole is strongly supported by such properties of the CMB as {much smaller} amplitudes of the quadrupole and higher-order moments {of its temperature anisotropies}, and by the observed alignment of this dipole with the direction of the peculiar acceleration of the LG, although the latter is much more difficult to estimate. Constraining it requires knowledge of mass distribution in our cosmic neighborhood, and its determination had not been possible until deep all-sky galaxy catalogs became available. For that reason, the first attempts to measure the acceleration of the LG were made not earlier than about three decades ago ago \citepc{YST80, DH82}.\\
\ind Using an all-sky catalog, such measurement can be made under the assumption that visible (luminous) matter is a good tracer of the underlying density field. The general procedure is to estimate the so-called \emph{clustering dipole} of a galaxy survey and infer the acceleration of the LG. However, such inference requires several conditions to be met. First, the survey should cover the whole sky; second, the observational proxy of the gravitational force (most often the flux of the galaxy in the photometric band of the survey) should have controlled properties; and last but not least, the survey should be deep enough for the dipole to converge to the final value that we want to find. As usually one or more of these assumptions do not hold, the clustering dipole is a biased estimator of the acceleration and the estimation of the latter from the former can be done only if the mentioned effects are properly accounted for.

What is more, in reality we do not observe continuous galaxy density field, but instead discrete objects, even if in a very large number. In the following derivation, which can be found e.g.\ in \citec{ViSt87}, we model galaxies as point sources: $\rho_\mrg(\bmr) = \sum_i M_i\, \delta_D(\bmr - \bmr_i)$, where $\delta_D$ is Dirac's delta; $M_i$ and $\bmr_i$ are respectively the mass and the position of the $i$-th galaxy. Putting the coordinate system at $\bmr=0$, we obtain the acceleration of the LG as a sum of force contributions from all sources in the Universe:
\begin{equation}
\bmg=\frac{G}{b} \sum_i M_i \frac{\hat{\bmr}_i}{r^2_i}
\end{equation}
(note that as we are interested in the motion of the Local Group as an entire system, the galaxies of the LG should \emph{not} be included in the summation). This Newtonian formula\footnote{The Newtonian limit can be applied as our whole analysis concerns distances well below the Hubble radius $r_H\equiv c\slash H_0=3\, \mathrm{Gpc} / h$.} is still not useful for calculations based on observational data, as masses of individual galaxies are known very poorly, or not at all. However, if the $i$-th galaxy has an intrinsic luminosity $L_i$, we can write
\begin{equation}\label{eq:g.masses}
\bmg=\frac{4\pi G}{b} \sum_i \frac{M_i}{L_i}\frac{L_i}{4\pi r_i^2}\hat{\bmr}_i=\frac{4\pi G}{b} \sum_i \frac{M_i}{L_i}S_i\hat{\bmr}_i\;,
\end{equation}
where $S_i=L_i\slash 4\pi r_i^2$ is the flux received from the $i$-th object. Relation (\ref{eq:g.masses}) means that if we know the behavior of the \textit{mass-to-light ratio} in the band(s) of the survey, we can even try to estimate the acceleration of the LG from a \emph{two-dimensional} catalog, i.e.\ one containing astro- and photometric data only (positions and fluxes). Furthermore, if we assume that the \textit{mean} mass-to-light ratio is a universal constant, \mbox{$\Upsilon=\lan M\slash L\ran$}, we finally get
\begin{equation}
\bmg=\frac{4\,\pi\,G\,\Upsilon}{b} \sum_i S_i\hat{\bmr}_i\;.
\end{equation}
In some applications, including the present one, it is more convenient to work in terms of matter and luminosity densities. This is especially the case when the \emph{luminosity density}, $j$, is known for a given band, rather than the mean mass-to-light ratio. We have
\begin{equation}
\Upsilon=\left\lan{\frac{M}{L}}\right\ran=\frac{\rho_\mrm}{j}=\frac{3\, \Omega_\mrm\,H_0^2}{8\,\pi\, G\, j}\;
\end{equation}
which gives
\begin{equation}\label{eq:flux dipole}
\bmg=\frac{3\, \Omega_\mrm \, H_0^2}{2\,b\,j}\sum_i S_i\hat{\bmr}_i\;.
\end{equation}
The luminosity density $j$ for a particular band of the survey can be calculated for example from the luminosity function $\Phi(L)$ of galaxies in this band (e.g.\ \citealtc{Pe93})
\begin{equation}\label{eq:lum.dens} 
j=\intlim_0^{\infty}{L\, \Phi(L)\, \de L }\;.
\end{equation}
Note that using the Relation (\ref{eq:flux dipole}) in Eq.\ (\ref{eq:v.and.g}), we get the linear-theory velocity of the LG measured from the flux dipole as
\begin{equation}\label{eq:v_s}
\bmv=\beta\, \frac{H_0}{j}\sum_i S_i\hat{\bmr}_i=\beta\, \tilde{\bmg}\;,
\end{equation}
where $\tilde{\bmg}$ represents the acceleration of the LG scaled to units of velocity. The term $\sum_i S_i\hat{\bmr}_i$ is the theoretical flux dipole moment of \emph{all} sources down to the zero flux over the whole sky. The universal luminosity density $j$, measured from a fair sample of galaxies in the given band with known apparent luminosities and redshifts, is proportional to $H_0$, which means that the overall result does not depend on the Hubble constant.

A main complication to the above comes from the fact that realistic galaxy catalogs will never reach down to zero flux, irrespective of the used wavelength. Surveys are usually \emph{flux-limited}, and the number of observed sources, $N$, is finite. For that reason, in the following we will denote the flux dipole of a finite, flux-limited sample as $\bmd$:
\begin{equation}\label{eq:lim flux dipole}
\bmd=\frac{3\, \Omega_\mrm \, H_0^2}{2\,b\,j}\sum_i^N S_i\hat{\bmr}_i\;.
\end{equation}
Note that the clustering dipole calculated from the above formula may be a biased estimator of the peculiar acceleration of the Local Group, Eq.\ (\ref{eq:g.theor}). This can be overcome by extrapolating the measured dipole to zero flux \citepc{ViSt87}.

The situation becomes more favorable in the case of galaxy \emph{redshift} surveys. We can then use the redshifts as distance estimators and weight galaxies with the inverse of the selection function of the survey (e.g.\ \citealtc{YSDH91}), in order to mimic an `ideal', \textit{volume-limited} catalog. However, despite an outstanding advancement in surveying the cosmos in recent years, the deepest and densest all-sky redshift survey, the 2MASS Redshift Survey (2MRS, \citealtc{2MRS}) contains only $\sim$45,000 galaxies and has a median depth of merely $\sim\!90\Mpch$ ($z_\mrm\simeq0.03$). On the other hand, the `parent' catalog of this survey, namely the Two Micron All Sky Survey (2MASS, \citealtc{Skr06}), reaches 3 times deeper and includes about 20 times more galaxies (for details, see Chapter~\ref{Ch:Data}). Its redshift coverage, when matched with other surveys, such as the Sloan Digital Sky Survey (SDSS, \citealtc{SDSS}) or the Six Degree Field Galaxy Survey (6dFGS, \citealtc{6dF09}), is however non-uniform both on the sky and in depth (for a recent compilation, dubbed 2M++, see \citealtc{LH11}).\\
\ind For the purpose of our work, we have thus decided to sacrifice the advantages of weighting galaxies, which is possible for redshift surveys, obtaining instead an overwhelmingly greater number of sources and unprecedented depth of the survey with photometric data only. An additional motivation of using the dipole (\ref{eq:lim flux dipole}), constructed only with the use of fluxes and angular positions of individual galaxies, is the fact that it is free of any \textit{redshift-space distortions}, and in particular of the \textit{rocket effect} \citepc{Kaiser87}. The latter consists in the fact that the peculiar acceleration of the LG calculated using redshifts instead of real distances (which are mostly not known) will differ from the actual LG acceleration due to a spurious contribution from the galaxies that are in the direction of the LG motion. Here, we do not use distances measured in redshift- nor in real space, and the only possible effect of that kind would be the anisotropy modulation in the distant galaxy distribution due to the aberration effect, which is however completely negligible for our sample \citepc{Itoh10}. The only stage at which the Kaiser effect comes into play is in the measurement of the luminosity function $\Phi(L)$ and consequently the luminosity density $j$. This is addressed in the relevant papers where $\Phi(L)$ is measured, see e.g.\ \citec{6dF_Fi}.\\
\ind The machinery described in this Introduction will be applied in the following Chapters of this thesis. We start in Chapter~\ref{Ch:Nonlin} by a case study of a non-linear effect that could possibly be important for our measurements. Subsequent Chapter~\ref{Ch:Data} presents the 2MASS Extended Source Catalog and how it was prepared for our purposes. Then, in Chapters \ref{Ch:Growth} and \ref{Ch:MLE} we apply the data from the 2MASS dataset  to calculate the clustering dipole of this catalog and show how to use it to measure the $\beta$ parameter. We propose two approaches for this: one is to analyze the growth of the clustering dipole with increased depth of the sample and compare it with theoretical expectations (Chapter~\ref{Ch:Growth}); the other is to use the maximum-likelihood estimation method and optimize the window through which the measurement is done (Chapter~\ref{Ch:MLE}). We conclude and present some future prospects in Chapter~\ref{Ch:Conclusions}.



\chapter{\textsf{\textbf{Non-linear effects: a case study}}}
\label{Ch:Nonlin}
\lettrine[lraise=0.5,lines=2,findent=1pt,nindent=0em]{T}he perturbation theory described in Chapter~\ref{Ch:Intro} applies to the linear regime. Mathematically, this means that considered density contrasts must be small, $|\del|\ll 1$ (as well as should be the velocity divergences). We can thus use the linear theory only for sufficiently large scales and for that purpose, non-linear over- and underdensities are usually smoothed out in analyses of density and velocity fields. However, structures in the Universe nowadays, such as galaxy clusters or big voids, are indeed highly non-linear and they exist also in the cosmic neighborhood of the Local Group: the Virgo and Coma Clusters, the Shapley Concentration or the Local Void, to name just a few. The question then arises in the context of our work: are such objects important for the analysis of the clustering dipole as described in Section \ref{Sec:Clust.dipole}? {Can they introduce significant systematics?}\\
\ind There are at least two effects that can be significant here. One is the fact that the two vectors $\vLG$ and $\gLG$ are not parallel in reality: there is a non-zero \textit{misalignment angle} between them. This is  both known from observations (e.g.\ \citealtc{Strauss92, Schmoldt99, Maller03, Erdogdu06a}) and expected from simulations (e.g.\ \citealtc{CCK01}). However, as the observed angle is usually small, the deviations from linear-theory relation {are usually neglected} in this respect. There still remains another issue, related to observational constraints that cannot be bypassed: our inability to observe the whole Universe. First of all, every galaxy survey is limited (usually by a minimum flux). This causes effects such as the \textit{shot noise} {(e.g.\ \citealtc{StWi95}) related to the fact that only the intrinsically brightest galaxies are seen near the edge of the survey}. Secondly, irrespective of the wavelength in which the observations are performed, our Galaxy obscures a significant part of the sky, creating the so-called \textit{Zone of Avoidance} (ZoA) through which we cannot see extragalactic objects. Any `all-sky' cosmological survey will always miss some information in the Galactic plane and its vicinity, due to obscuration by dust, gas and stars of the Milky Way. This means that there can always be some objects that we are unaware of, but which have an influence on the motion of the LG. Should such a hidden object be highly non-linear {(i.e.\ very over- or underdense)} and close, it could possibly bias the calculated acceleration.\\
\ind In order to compute the clustering dipole from the available data, this fact is overcome by artificially filling the ZoA, in a more or less sophisticated way (see e.g.\ \citealtc{Lah87, LBLB89, Pl89, Maller03, PH05, Erdogdu06a, BP06, Lav10}). Still, regardless of the method one chooses, such masking of the Galactic Plane and Bulge will never completely account for any possibly existing, although unknown, large-scale structures behind the Milky Way. On the other hand, it is often claimed that the direction and amplitude of the calculated clustering dipole do not change significantly for different methods of filling the ZoA (e.g.\ \citealtc{Lah87, Erdogdu06a}). However, this does not necessarily mean that possible structures obscured by the Galaxy would have no influence on the calculation of the acceleration of the Local Group.\\
\ind This issue has been studied by \citec{LN08}, who {analyzed the dynamics of the LG by comparing the dipoles of the CMB and of the 2MRS, focusing on the lack of surveyed galaxies behind the ZoA. In order to match the two dipoles, they have inferred} excess peculiar velocity of the LG towards the ZoA and proposed a hidden nearby galaxy or a galaxy cluster as an explanation of the gravitational pull. This suggestion is however in conflict with the current observational knowledge, as already in 2004 there was confidence that all significant nearby large-scale structures behind the ZoA were known and obscuration of a big galaxy was excluded \citepc{FaLa05}. Moreover, \citec{LN08} claim that a perfect method of filling the ZoA would give `\textit{no discrepancy between the 2MRS   and the direction of the CMB dipole}' (i.e.\ between the direction of the acceleration and of the velocity of the LG). We cannot agree with this statement, since there are other sources of the misalignment angle between the two vectors, such as the scatter in the mass-to-light ratio (cf.\ \citealtc{Cr07}) and stochasticity in the non-linear relation between the velocity and acceleration of the LG (\citealtc{BChL99, CCK01, CCBCC}). As for the excess peculiar velocity of the LG, \citec{Tully.etal08} propose a different explanation (partial at least), basing on observational data of distances and velocities of nearby galaxies: motion \textit{away} from the \textit{Local Void} (LV).

In this Chapter we will present a case study of the importance of a non-linear and proximate structure, partially hidden behind the ZoA. We choose the LV for this. The Chapter is organized as follows. In Section \ref{Sec:LV} we shortly describe the Local Void. Section \ref{Sec:LV.Accel} presents our calculations of the `spurious' acceleration induced by random filling of the LV behind the ZoA. Next, Section \ref{Sec:LV.Shift} covers the issue of the directional shift of the clustering dipole due to the discussed effect. Possible corrections of the density parameter measured from density--velocity comparisons are addressed in Section \ref{Sec:LV.Omega}. Finally, in Section \ref{Sec:LV.Summ} we summarize and conclude.

\section{A case study: the Local Void}
\label{Sec:LV}

We will now analyze the possible influence of the LV on \textit{measurements} of the clustering dipole. We would like to emphasize that we do not examine here the importance of the Local Void for the very \textit{motion} of the Local Group. As was shown e.g.\ by \citec{Tully.etal08}, the push from the LV is a substantial component of the peculiar velocity of the LG. In this Chapter we are only interested in the effect of masking the intersection of the LV and the ZoA for the purpose of calculation of the clustering dipole within linear theory. Our aim is to investigate possible systematics related to the fact that a part of the LV is hidden behind the ZoA. This will be useful in assessing the total errors of the measurements presented in Chapters \ref{Ch:Growth} and \ref{Ch:MLE}.
  
\begin{figure}[!t]
\begin{center}
\includegraphics[width=\textwidth]{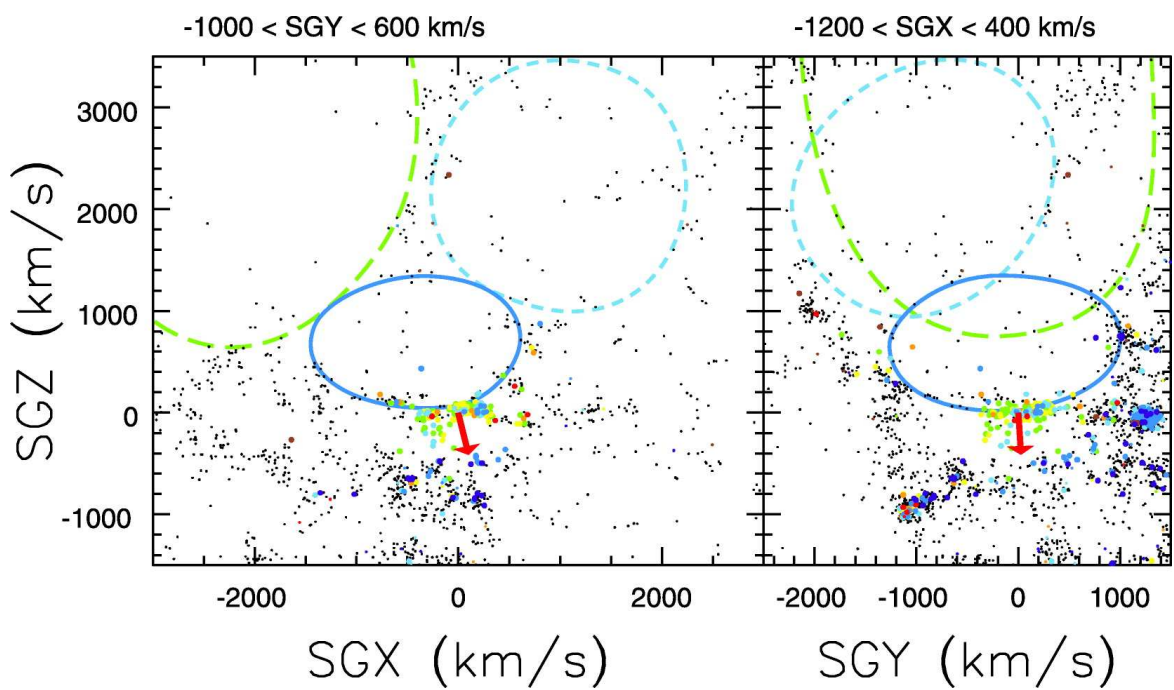}
\caption[Region of the Local Void]{\label{Fig:LV.obs}\small Region of the Local Void. The ellipses outline the three apparent sectors of the Local Void. The solid dark blue ellipses show two projections of the Inner Local Void, bounded on one edge by the Local Sheet. The North and South extensions of the Local Void are identified by the light blue short-dashed ellipses and the green long-dashed ellipses, respectively. These sectors are separated by bridges of wispy filaments. The red vector indicates the direction and amplitude of our motion away from the void. The axes are in Supergalactic coordinates in units of velocity. The Local Group is at the origin of the coordinate system, at the beginning of the red vector. [Reproduced from \citec{Tully.etal08} by permission of the AAS.]}
\end{center}
\end{figure}  
 
The Local Void is a structure first identified in the Nearby Galaxies Atlas \citepc{TF87}. Although our close cosmic neighbor, it is poorly defined because much of it lies behind the plane of the Milky Way: its center is located at Galactic coordinates $l\simeq30\degr$, $b\simeq0\degr$ (Tully 2007, private communication; but see \citealtc{NaKa11} and references therein for other determinations). It is nevertheless confirmed (by surveys in different wavelengths) that there is an underabundance, though not a total lack, of galaxies in a very large part of the sky at low redshifts (e.g.\ \citealtc{KK08}). This empty region begins at the edge of the Local Group, with the so-called \textit{Local Sheet} as a bounding surface (\citealtc{Tully08a, Tully08b, Tully.etal08}). Both the shape of the LV and its dimensions are not exactly known; however, it seems to have two components: a smaller void with a long dimension of $\sim\! 35 \, \Mpc$ enclosed within a larger void with a long dimension of the order $60\!\sim\!70 \, \Mpc$ (\citealtc{Tully07, Tully.etal08}). Figure~\ref{Fig:LV.obs} (reproduced from \citealtc{Tully.etal08}) presents the observational data in the region.\\
\ind In the framework of cosmological gravitational instability, voids are expected to form out of initial underdensities, i.e.\ regions less dense than average (e.g.\ \citealtc{HoSh82}). Their expansion is faster than the Hubble flow, which results in voids `swelling'. \citec{Tully.etal08} calculated an `effective Hubble rate' of the Local Void, under the simplest assumption of it being empty and spherical, and used this calculation, together with the peculiar velocity of the LG away from the LV, to estimate the radius of the latter as at least 16~Mpc. However, these dynamical estimates cannot exclude an effective diameter of the LV  greater even than 45~Mpc (\citealtc{Tully08a, Tully08b, Tully.etal08}).\\
\ind Such a prominent structure, partly hidden behind the Zone of Avoidance, should have an influence on the calculation of the integral in Eq.\ (\ref{eq:vLG-gLG}) and consequently, on determinations of $\beta$ and $\Omm$ from the $\vLG$ -- $\gLG$ relation. In the following Sections, using a simple model, we will analyze the amount of the systematic error it may cause for the estimation of the acceleration of the Local Group.

\section{Acceleration due to (the lack of) the Local Void}
\label{Sec:LV.Accel}

Let us define the scaled acceleration vector $\tilde{\bmg}_\mathrm{_{LG}}$ in convenient units as 
\beq
\tilde{\bmg}_\mathrm{_{LG}}=\int \frac{\delta(\bmr')(\bmr'-\bmr) }{|\bmr'-\bmr  |^3}\,\de^3\bmr'   
\eeq
(we omit at the moment the normalization, not important in the qualitative analysis that follows). We can split the integral into two parts: one over the Local Void and the other covering `the rest of the Universe':
\beq
\tilde{\bmg}_\mathrm{_{LG}}=\intlim_\mathrm{LV} (\ldots) +\intlim_{\mathbb{R}^3-\mathrm{LV}} (\ldots)\;.   
\eeq
For simplicity, we write only `LV' in this and the following formulae, although we mean in fact `LV$\cap$ZoA', as will be explained later in detail. Now, if we make a simplifying and maximizing assumption that the Local Void is completely empty ($\rho_\mathrm{LV}=0$), we have $\delta_\mathrm{LV}=-1$. On the other hand, if we filled the part of the Local Void hidden behind the ZoA by randomly chosen `galaxies', i.e.\ if we assumed average background density in this part ($\rho_\mathrm{LV}=\rho_\mathrm{b} $), we would get $\delta_\mathrm{LV}=0$. We would thus measure some acceleration that we will call \textit{spurious}, which is the difference between the calculation made with the ZoA filled randomly and the true peculiar acceleration of the Local Group:
\beqa\label{eq:g_spurious}
\lefteqn{\tilde{\bmg}_\mathrm{spur}=\tilde{\bmg}_\mathrm{cal}-\tilde{\bmg}_\mathrm{T}= } \\
&& =\intlim_\mathrm{LV}{0}\, +\,\intlim_{\mathbb{R}^3-\mathrm{LV}}{(\ldots)}\,-\,\intlim_\mathrm{LV}{\frac{(-1)\times(\bmr'-\bmr) }{|\bmr'-\bmr  |^3}\,\de^3\bmr'}\,-\,\intlim_{\mathbb{R}^3-\mathrm{LV}} (\ldots)=\intlim_\mathrm{LV} \frac{\bmr'-\bmr}{|\bmr'-\bmr  |^3}\,\de^3\bmr'\;. \nonumber
\eeqa
Let us now calculate explicitly the above integral, returning to physical units.

\begin{figure}[!t]
\begin{center}
\includegraphics[width=0.45\textwidth]{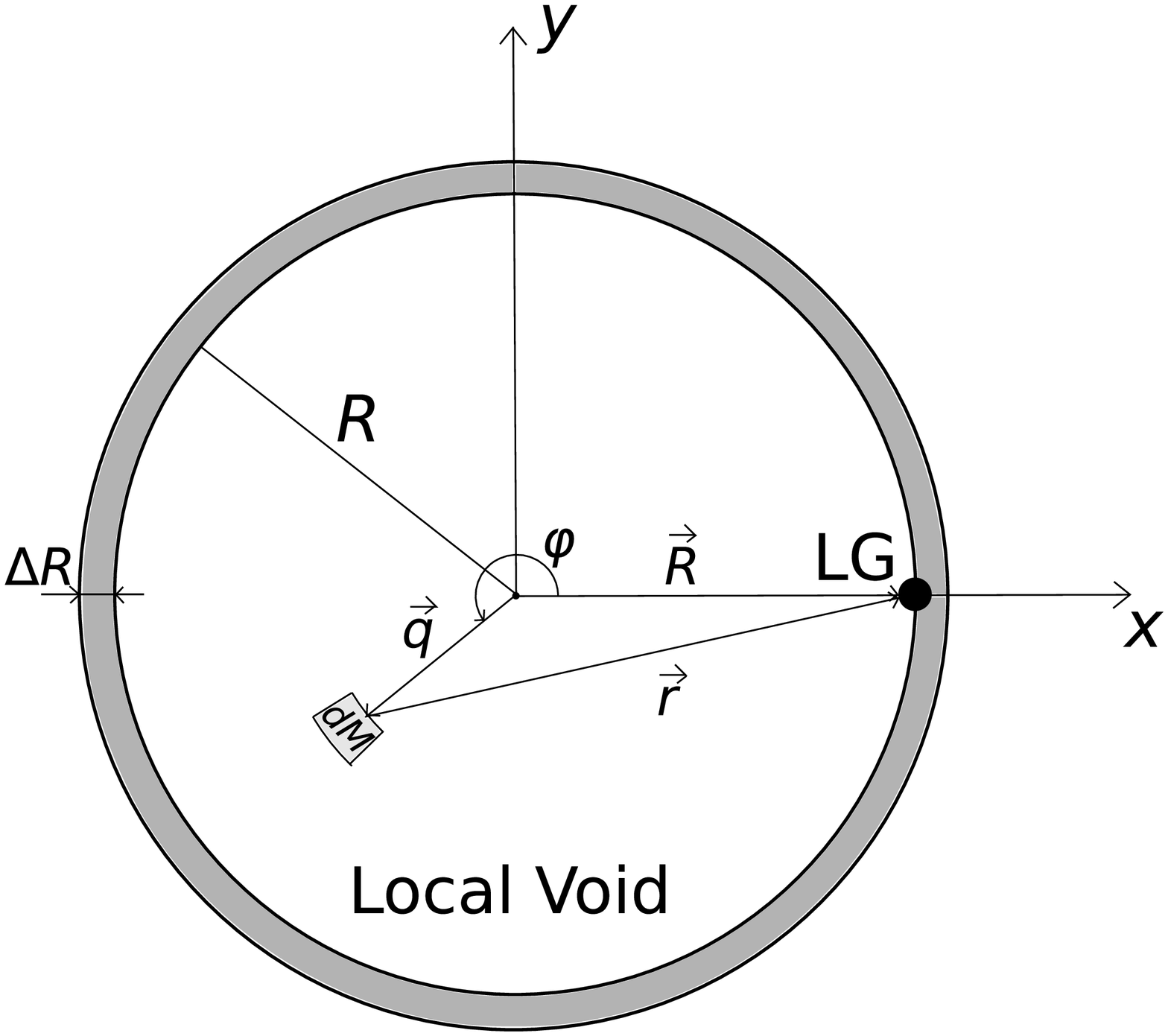}
\hspace{0.5cm}
\includegraphics[width=0.45\textwidth]{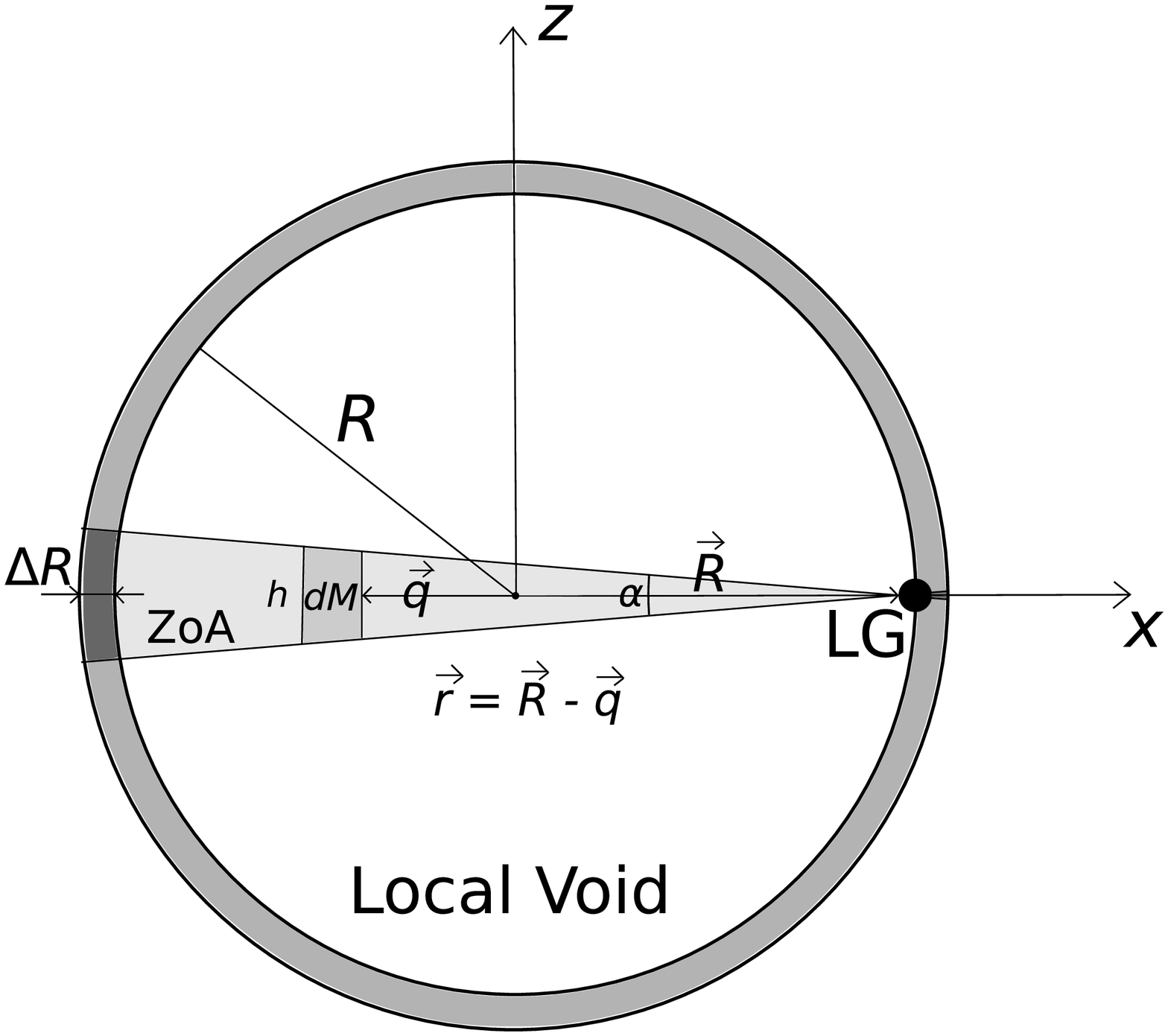} 
\caption[Schematic view of the Local Void]{\label{Fig:LV.sch}\small A schematic view of the Local Void: (left) along the $z$ axis, with the Galactic plane face-on; (right) along the $y$ axis, with the Galactic plane edge-on. On both panels, the relative sizes of the Local Group and the Local Void are not to scale.}
\end{center}
\end{figure}

\subsection{Spherical void}
For that purpose, in addition to the assumption of the Local Void being completely empty, we start by modeling it as a sphere with some radius $R$, and assume that the Local Group lies exactly at the edge of the LV (see Figure~\ref{Fig:LV.sch})\footnote{We neglect the size of the Local Group, estimated usually to about {$1.5\Mpc$ \citepc{vdB07,vdM08}}.}. We place the center of the Local Void in the plane of the Galactic equator, in accordance with observations, which additionally maximizes the studied effect. The $x$ and $y$ axes of the coordinate system lie in the Galactic plane, with the origin in the center of the LV (Figure~\ref{Fig:LV.sch}, left panel). As the latter is situated at $l_\mathrm{LV}=30\degr$, \mbox{$b_\mathrm{LV}=0\degr$}, this means that our coordinate system is shifted with respect to the Galactic one and rotated by $210\degr$ in the Milky Way plane (our $z$ axis is parallel to $z_\mathrm{Gal}$). The part missed in all-sky surveys due to the Zone of Avoidance is a spherical section bounded by two planes (Figure~\ref{Fig:LV.sch}, right panel); the angle between them, $\alpha$, depends on the survey, but for our purposes $\alpha\simeq 20\degr$ (we consider near-infrared wavelengths, as is explained hereafter). Owing to the smallness of this angle ($\alpha\simeq0.35\,\mathrm{rad}$), we treat the masked region as a thin wedge.

We additionally include a \textit{shell of compensation} at the edge of the LV, of width $\Delta R$. According to the standard picture of void formation, where such structures are grown from initial underdensities and gradually expand faster than the cosmological background, the matter expelled from inside of the void creates a layer at the edge, of density higher than the average background one (e.g.\ \citealtc{MST11} and references therein). This is supported by both simulations (see e.g.\ \citealtc{vdWSch09}) and observations (cf.\ \citealtc{Fairall98} and references therein): voids are usually bounded by `filaments' and `walls' of high density contrast. In our cosmic neighborhood, a part of such a shell is probably the Local Sheet and the Local Group is located at its inner edge (Tully 2010, private communication).

A mass element $\de M$ contributing to the spurious acceleration is
\beq
\de M=\rho_\mathrm{b}\,\de V = \rho_\mathrm{b}\,h\,\de S = \rho_\mathrm{b}\,\alpha\, r\,\de S\;,
\eeq
where $h\simeq\alpha\,r$ is the height of the wedge at a distance $r$ from the LG and $\de S$ is a surface element. The differential acceleration `measured' due to random filling of the ZoA will be
\beq
\mathitbf{\de g}=-G\,\frac{\de M(r)\, \bmr
}{r^3}=-G\,\frac{\rho_\mathrm{b}\,\alpha\,r\,\de S\,\bmr }{r^3}=
-\alpha\, G\, \rho_\mathrm{b}\,\frac{q\,\de q \, \de
  \varphi\,(\mathitbf{R}- \mathitbf{q}) }{|\mathitbf{R}- \mathitbf{q} |^2}\;,
\eeq
with $\de S = q\, \de q\, \de \varphi$, $\mathitbf{q}=(q\cos\varphi,q\sin\varphi) $, $\mathitbf{R}=(R,0) $ and \mbox{$0\leq\varphi\leq2\pi$} is the azimuthal angle. It is easy to check that from symmetry $g_y=0$. The $x$-component of the spurious acceleration is given by
\beq
g_x=-\alpha\,G\,\rho_\mathrm{b}\, R\, \mathcal{I}\;,
\eeq
where
\beq
\mathcal{I}=\intlim^{2\pi}_0 \de \varphi\intlim^1_0 \de \xi \;
\frac{\xi-\xi^2\cos\varphi}{\xi^2+1-2\xi\cos\varphi}=\pi\;.
\eeq
Thus the value of the peculiar velocity caused by the spurious acceleration is (in linear theory)\footnote{\label{biasing}Here and later in this Chapter, we neglect the biasing (cf.\ Eq.\ \ref{eq:biasing}), or in other words, we set it equal to unity. Its exact value does not affect the conclusions of this Section.}
\beq\label{eq:v_spurious}
v_\mathrm{spur}
=\frac{H_0\,f(\Omm) }{4\pi\,G\,\rho_\mathrm{b}
}\,g=\frac{1}{4} \alpha\,H_0\,f(\Omm)\,R\;.
\eeq
It is interesting to compare this value with the one induced by a sphere of radius $R$ and density $\rho_\mathrm{b}$. The peculiar acceleration at the surface of the sphere is then $g_\bullet=\frac{4}{3}\pi G \rho_\mathrm{b} R$, which gives the linear peculiar velocity of
\beq
v_\bullet=\frac{1}{3}H_0\,f(\Omm)\,R\;.
\eeq
We thus have
\beq
\frac{v_\mathrm{spur}}{v_\bullet}=\frac{3}{4}\, \alpha\;,
\eeq
{which shows that the effect is expected to be small.}

The above calculation applied to an isolated void without any shell of compensation. In order to verify the effect of such a shell, let us first assume that the average matter density inside the layer is constant, $\rho_\mathrm{sh}>\rho_b$, and its thickness is $\Delta R$. The Local Group, of negligible size, is still located at the edge of the Local Void, as in Figure~\ref{Fig:LV.sch}. Putting now the mass element inside the intersection of the shell with the ZoA, we can see that the differential spurious acceleration will read
\beq
\mathitbf{\de g}_\mathrm{sh}=\alpha\, G\, (\rho_\mathrm{sh}-\rho_\mathrm{b})\,\frac{q\,\de q \, \de   \varphi\,(\mathitbf{R}- \mathitbf{q}) }{|\mathitbf{R}- \mathitbf{q} |^2}\;,
\eeq
with $\mathitbf{q}$ and $\mathitbf{R}=(R,0)$ defined as before, but now \mbox{$R\leq q \leq\,R+\Delta R$}. The $y$-component of the acceleration vanishes from symmetry, and along the $x$ direction we have
\beq
g^\mathrm{(sh)}_x=\alpha\,G\,(\rho_\mathrm{sh}-\rho_\mathrm{b})\,R\, \mathcal{I}_\mathrm{sh}\;,
\eeq
with $\mathcal{I}_\mathrm{sh}$ different from $\mathcal{I}$ only by the limits of integration in $\xi$ and equal to
\beq
\mathcal{I}_\mathrm{sh}=\intlim^{2\pi}_0 \de \varphi\intlim^{1+\frac{\Delta R}{R}}_1 \de \xi \; \frac{\xi-\xi^2\cos\varphi}{\xi^2+1-2\xi\cos\varphi}=0\;.
\eeq

\noi This result means also that the spurious acceleration of the LG due to the part of the compensating shell hidden behind the ZoA will equally vanish if the density distribution $\rho_\mathrm{sh}$ is not constant, but depends on the distance from the center of the LV (as we can divide the shell into infinitesimally thin layers of constant density each). The bottom-line is that a compensating shell with radial density distribution will \textit{not} affect the spurious acceleration of the Local Group provided that the latter is located at the inner edge of the shell, as observational constraints suggest.\footnote{In fact, the spurious acceleration from the hidden part of the shell would vanish even if the LG was \textit{inside} the LV, as long as the center of the latter was coplanar with the Galactic equator.} On the other hand, if the LG was placed in the interior of the compensating layer, or at its outer edge, the discussed spurious acceleration from the LV would be \textit{diminished}, by up to $2\slash 3$ in the limit of an infinitesimally thin shell. This means that the calculations presented so far concern a maximizing case.

Having calculated the amplitude of the spurious acceleration in the model, we can apply our results to observational data. As an example, we take the analysis of \citec{Maller03} of the clustering dipole of the Two Micron All Sky Survey\footnote{We will come back to the issue of the 2MASS clustering dipole when presenting our own measurements (Chapters \ref{Ch:Growth} and \ref{Ch:MLE}). Here however we want to obtain an estimate of systematics that can be used for our calculations, hence we use externally provided earlier results.}, since these are the same data that will be used in our analysis of Chapters \ref{Ch:Data}--\ref{Ch:MLE}. \citec{Maller03} used the measurements in the near-infrared $K_\mathrm{s}$ band and defined the Zone of Avoidance as the region with $|b|<7\degr$ for $l>230\degr$ or $l<130\degr$ and $|b|<12\degr$ for $l>330\degr$ or $l<30\degr$. Let us take a `mean value':
\beq
\alpha=2\,\Delta b=\frac{1}{2}\,(2\times7\degr+2\times12\degr)=19\degr\simeq0.332\,\mathrm{rad}.
\eeq
We now use Eq.\ (\ref{eq:v_spurious}) and for consistency with \citec{Tully.etal08}, we apply \mbox{$H_0=74\kmsMpc$}, $\Omm=0.24$ and $R=16\,\mathrm{Mpc}$, to obtain an approximate value of the additional, spurious velocity of the Local Group, measured if the Local Void is not properly accounted for:
\beq
v_\mathrm{spur} \simeq 45\kms\;.  
\eeq
When compared to the velocity of the LG relative to the CMB reference frame, $v_\mathrm{_{CMB}}=622\pm35\kms$, one can see that this effect is of the same order as the error in the measurement of $v_\mathrm{_{CMB}}$. Note also that due to the proportionality of $v_\mathrm{spur}$ to the radius of the LV (Eq.\ \ref{eq:v_spurious}), increasing its diameter to $45\,\mathrm{Mpc}$, while preserving sphericity, would cause the spurious velocity to raise significantly to $\sim\!60\kms$. However, owing to the geometry of the problem, this would not largely affect general conclusions of our analysis (see Secs.\ \ref{Sec:LV.Shift} and \ref{Sec:LV.Omega}). Moreover, observational constraints point rather to some degree of elongation of the LV than to a larger size of the whole structure. We will now address this issue.

\subsection{Elongated void}
The calculations so far assumed a simplistic model of a spherical void. However, we can see for example in Figure~\ref{Fig:LV.obs} that the Local Void should be possibly modeled by a more sophisticated structure, such as an ellipsoid. Current observational data suggest that the LV is elongated in the Supergalactic $\mathrm{SGY}=0$ plane, roughly coincident with the Galactic plane and the $x$ axis of our coordinate system. One should bear in mind however that this effect could be a manifestation of the existence of the ZoA itself: lack of observed galaxies in this region of the sky may be simply due to obscuration. Nevertheless, in case the elongation is real, for completeness of our analysis let us examine this possibility.

For that purpose we assume that the section of the LV in our $xy$ plane is an ellipse with semiaxes $B\equiv R$ and $A\equiv\kappa\,R $, where elongation \mbox{$\kappa>1$}. The major axis of the ellipse is placed along the $x$ axis of the coordinate system. As in the spherical case, owing to symmetries of the problem, the only non-vanishing component of the spurious acceleration `acting' on the LG is the $x$ one. It is easy to check that the relevant formula for $g_x$ is now
\beq
g_x=-\alpha\,G\,\rho_\mathrm{b}\, \kappa \, R \, \mathcal{I}_e\;,
\eeq
with the integral $\mathcal{I}_e$ given by
\beq
\mathcal{I}_e=\intlim^{2\pi}_0 \de \varphi\intlim^{\Xi(\varphi)}_0 \de \xi \;
\frac{\xi-\xi^2\cos\varphi}{\xi^2+1-2\xi\cos\varphi}\;.
\eeq
Here, $\Xi(\varphi)\equiv1\slash\left(\kappa\sqrt{1-e^2 \cos^2\varphi}\right)$ is an appropriately normalized equation of the ellipse in polar coordinates with the eccentricity \mbox{$e\equiv\sqrt{1-\kappa^{-2}}$}.

The spurious velocity of the LG due to such elongation of the LV will be larger in comparison to the spherical case by a factor
\beq
\tau\equiv\frac{v_\mathrm{spur} ^{(\mathrm{ell})}}{v_\mathrm{spur} ^{(\mathrm{sph})}}=\kappa\,\frac{\mathcal{I}_e}{\mathcal{I}}=\kappa\,\frac{\mathcal{I}_e}{\pi}\;.
\eeq
Obviously, for $\kappa=1$, we have $\tau=1$. What is important here is that a linear increase in $\kappa$ results only in a slower-than-linear raise of the $\tau$ factor: for example $\kappa=2$ gives $\tau=4\slash 3$ and if $\kappa=3$, then $\tau=3\slash 2$. This means that for the observationally allowed elongation of $\kappa\simeq2$ and minor semiaxis $B\simeq15\Mpc$ \citepc{Tully.etal08}, the spurious velocity of the LV would raise by $1/3$ to \mbox{$\sim\!60\kms$}, which is the same as in the previously discussed case of the enlargement of a spherical and empty void. Note however that observations clearly show that the Local Void is not completely empty (cf.\ Figure  \ref{Fig:LV.obs}) and this high value of the spurious velocity will be an upper limit for our considerations.

\section{Shift of the clustering dipole}
\label{Sec:LV.Shift}
Knowing the amplitude of the spurious velocity induced by randomly-filled intersection of the Local Void and the Zone of Avoidance, we would like to check the shift of the \textit{direction} of the measured clustering dipole when the effect of the LV is accounted for. From Eq.\ (\ref{eq:g_spurious}), the true velocity of the LG (proportional to its acceleration in linear theory) is related to the calculated (`measured') one and the spurious component via
\beq
\mathitbf{v}_\mathrm{T}=\mathitbf{v}_\mathrm{cal}- \mathitbf{v}_\mathrm{spur}\;.
\label{eq:vel_rel}
\eeq

The vector $\mathitbf{v}_\mathrm{spur}$ is directed to the center of the Local Void;\footnote{Note that although in our coordinate system the only non-vanishing component of the spurious velocity is the $x$ one, when projected on Galactic coordinates it has $x_\mathrm{Gal}$ and $y_\mathrm{Gal}$ components of comparable values, equal respectively to $0.87\,v_\mathrm{spur}$ and $0.5\,v_\mathrm{spur}$.} as $v_\mathrm{T}$ we adopt $v_\mathrm{_{CMB}}=622\kms$. For the measured clustering dipole we choose the dipole of the 2MASS survey, as given in \citec{Maller03}  for random filling of the ZoA: $l_\mathrm{cal}=266\degr$, $b_\mathrm{cal}=47\degr$. Using these values altogether, after some calculations we find that the direction of the dipole is shifted down by $5\degr$ in $l$ and $2\degr$ in $b$. Thus, the `true' direction of the 2MASS dipole would be 
\beq 
l_\mathrm{T} \simeq 261\degr\,, \qquad b_\mathrm{T} \simeq 45\degr\,.  
\eeq
The shift is small, but as a systematic effect, it should be in principle accounted for in the measurement of the clustering dipole. However, random filling is not the only possible, nor the most optimal, way to deal with the ZoA. A better method, preferred by us (see Chapter~\ref{Ch:Data}) is to clone the sky below and above the ZoA, which has the advantage of approximately tracing  structures through the ZoA. For the latter method, \citec{Maller03} obtained $l_\mathrm{cal}=263\degr$, $b_\mathrm{cal}=40\degr$, so differences between the two methods give $\Delta l_\mathrm{cal}=3\degr$, $\Delta b_\mathrm{cal}=7\degr$. Therefore, the differences in the direction of the 2MASS dipole resulting from distinct methods of treating the ZoA are comparable to, or even greater than, the effect of the LV. These conclusions do not change significantly even if we include the elongation of the LV: for $v_\mathrm{spur}=60\kms$, we have a shift by $\Delta l=6\degr$ and $\Delta b=2\degr$ towards $l_\mathrm{T}=260\degr$, $b_\mathrm{T}=45\degr$.

The shift of the direction of the 2MASS clustering dipole changes the misalignment angle with respect to the CMB dipole. However, the amplitude of the shift is comparable to the uncertainty of the CMB dipole direction ($3\degr$ and $5\degr$ respectively for $l$ and $b$). Moreover, {owing to the specific geometry of the problem}, the calculated change of the misalignment angle turns out to be smaller than $1\degr$ even for high (but reasonable) values of $v_\mathrm{spur}$. We thus conclude that masking the LV has negligible impact on the misalignment angle between the 2MASS and CMB dipoles.

\section{Correcting the density parameter measurement}
\label{Sec:LV.Omega}

Application of Equation~(\ref{eq:vLG-gLG}) serves as a method to measure the cosmological parameter $\Omm$, in principle by comparing the velocity of the LG (equal to $\bmv_{\mathrm{_{CMB}}}$) to its gravitational acceleration inferred from an all-sky galaxy survey (but see the discussion in Section \ref{Sec:Clust.dipole}). From Eq.~(\ref{eq:vLG-gLG}) it follows that $v_{\mathrm{_{CMB}}} = \Omega_{\mrm}^{0.55}\, \tilde{g}_{\mathrm{_{LG}}}$ (with $b=1$, cf.\ footnote \ref{biasing}), where $\tilde{g}_{\mathrm{_{LG}}}$ is the gravitational acceleration of the LG expressed in units of velocity. The acceleration measured without the LV accounted for results in the `calculated' value of $\Omm$, such that $\Omega_{\mathrm{cal}}^{0.55} = v_{\mathrm{_{CMB}}}/\tilde{g}_{\mathrm{cal}}$. If the LV is taken into account, we find the `true' value of $\Omm$, i.e.\ $\Omega_{\mathrm{T}}^{0.55} = v_{\mathrm{_{CMB}}}/\tilde{g}_{\mathrm{T}}$. When we divide $\Omega_{\mathrm{cal}}^{0.55}$ by $\Omega_{\mathrm{T}}^{0.55}$, all the scaling factors relating the velocity to acceleration in linear theory {(including the biasing parameter)} cancel out. Therefore,
\beq \label{eq:ratioLV}
\frac{\Omega_{\mathrm{cal}}^{0.55}}{\Omega_{\mathrm{T}}^{0.55}} =
\frac{v_{\mathrm{T}}}{v_{\mathrm{cal}}} \,.
\eeq
The velocity $\bmv_{\mathrm{T}}$ is related to $\bmv_{\mathrm{cal}}$ by Equation~(\ref{eq:vel_rel}), where $\bmv_{\mathrm{spur}}$ is a small correction. Therefore, we can expect that the relative change in the value of $\Omega_{\mrm}$ will be small. We thus write $\Omega_{\mathrm{cal}} = \Omega_{\mathrm{T}} + \Delta\Omega$ and expand the expression $\Omega_{\mathrm{cal}}^{0.55} / \Omega_{\mathrm{T}}^{0.55}$ to first order in $\Delta\Omega/\Omega$. The result is
\beq
\frac{\Omega_{\mathrm{cal}}^{0.55}}{\Omega_{\mathrm{T}}^{0.55}} = 1 + 0.55\,\frac{\Delta\Omega}{\Omega}\;.
\eeq
The right-hand-side of Eq.~(\ref{eq:ratioLV}), calculated using Formula~(\ref{eq:vel_rel}) with $v_\mathrm{spur}=45\kms$, is $1.029$. Hence finally 
\beq
\frac{\Delta\Omega_{\mrm}}{\Omega_{\mrm}} = 0.053 \,.  
\eeq
In other words, not accounting for the existence of the LV in measurements of the clustering dipole biases the estimated value of $\Omega_{\mrm}$ by about 5\% for the radius of the spherical LV equal to $16\Mpc$. If we allow for non-sphericity, this bias rises to some 7\%. This means that the influence of the LV (as discussed here) on the determination of the cosmic density parameter from the comparison of the velocity and acceleration is small. Indeed, typically the uncertainty of the degenerate combination  $\beta=f( \Omega_{\mrm})\slash b$ (with $b$ not necessarily unity), amounts to at least $10\!\sim\!20\%$ (for determinations using 2MASS data see e.g.\ \citealtc{PH05,Erdogdu06a} or \citealtc{Davis11}). Owing to our ignorance of the exact value of $b$, which typically has errors as big as $20\%$ \citepc{Maller05}, we can conclude that the total relative error in $\Omega_{\mrm}$ is likely to be higher than in $\beta$ and the inclusion of the effect of the Local Void would not contribute largely to the final error budget, although it possibly should be included as a systematic effect.

The above analysis, although concerning a non-linear effect, was performed within the linear theory, in which the estimated peculiar velocity, compared to the observed one, is inferred directly from the scaled peculiar acceleration (the clustering dipole). However, a completely empty void is a non-linear structure (e.g.\ \citealtc{BiCh08}) and the actual spurious velocity of the LG, generated by the LV, will be greater than the corresponding scaled spurious acceleration. It is known that for a void with $\delta=-1$ (i.e.\ $\rho=0$), the relation is $v_\mathrm{spur}\simeq1.5\,g_\mathrm{spur}$ (\citealtc{BChL99, BiCh08}). Application of this result would increase the systematic effects considered in this Chapter by roughly 50\% and enhance their significance. Nevertheless, in order to make this approach self-consistent, one would have to take into account non-linear effects from all other sources, especially those nearby, both over- and underdense. This would be a very difficult task {to perform analytically}, if not impossible, and is beyond the scope of this analysis, which deals with a simple model of the Local Void surrounded only by a shell of compensation.

\section{Summary and conclusions}
\label{Sec:LV.Summ}
A serious problem plaguing determinations of the clustering dipole is that every survey called `all-sky' misses a significant amount of galaxies due to obscuration by dust, gas and stars in the disk and bulge of the Milky Way (the Zone of Avoidance). To overcome this problem, in order to calculate the clustering dipole of the given survey, the ZoA is filled with mock galaxies. Their properties are chosen in a way to mimic the true, although unknown, galaxy distribution in the obscured part of the sky, extrapolating the one known from the rest of the celestial sphere.

A part of the ZoA intersects with a nearby void region, the Local Void. When the existence of such a structure is not accounted for in the calculation of the acceleration of the LG, a spurious term is generated. In this Chapter we have calculated both the amplitude and the direction of this spurious acceleration. For simplicity, we have first assumed that the LV is spherical and for its size we have adopted the value estimated by \citec{Tully.etal08}. We have also made the assumption that the LV is completely empty. Even then the amplitude of the spurious component amounts only to $45\kms$ in units of velocity. Including the observed elongation of the LV enhances this value by $1/3$. On the other hand, possible presence of massive structures inside the LV, hidden behind the ZoA, could only lower this value.

This artificial acceleration changes also the direction of the calculated clustering dipole. We have shown that this change is comparable to the uncertainty in the direction of the peculiar velocity of the LG, determined from the dipole component of the CMB temperature distribution, reduced to the barycenter of the LG. Moreover, by chance it points almost perpendicularly to the misalignment vector (i.e.\ the difference between the vectors of the velocity and acceleration of the LG). This results in a negligible shift of the misalignment angle, by less than $1\degr$.

The final effect that we considered was the error in the inferred value of the non-relativistic matter density $\Omm$ resulting from the negligence of the LV. We have estimated the relative error in this parameter due to this effect as approximately $5\%\!\sim\!7\%$. Therefore, up to this accuracy the influence of the Local Void on the determination of $\Omm$ from velocity--density comparisons can be neglected. On the other hand, this additional systematic should be taken into account in the total error budget of the density parameter determined by such a method.\\
\ind We would like to reiterate that our results do not negate the dynamical influence of the Local Void on the Local Group; on the contrary, \citec{Tully.etal08} have shown that this influence is significant. It is only the effect of masking the intersection of the LV and the ZoA that seems to be of little importance for the purpose of calculation of the clustering dipole within the linear theory. This partially supports the claims that the Zone of Avoidance is not a crucial issue in determinations of the peculiar acceleration of the LG from all-sky surveys, especially such as 2MASS, where Galactic extinction is much weaker than in optical wavelengths. We will include the effect discussed here as a systematic in the error budget of the 2MASS clustering dipole analysis, presented in subsequent Chapters. This analysis will start with a presentation of the data that we used.

\chapter{\textsf{\textbf{The data: Two Micron All Sky Survey (2MASS)}}}
\label{Ch:Data}

\lettrine[lraise=0.5,lines=2,findent=1pt,nindent=0em]{T}he analysis of the clustering dipole requires a catalog of galaxies that covers the whole sky, and desirably has uniform properties, such as photometry and astrometry. These are generally demanding requirements and only several catalogs of that type have been obtained so far. Apart from the attempts to gather all-sky galaxy data by compiling several sources (e.g.\ \citealtc{Lah87} or \citealtc{Hudson93}), two approaches are possible. The first one is to use a single instrument in space; the second is to have two identical telescopes on the two Earth hemispheres. The former approach was taken by the \textit{Infrared Astronomical Satellite} (\textit{IRAS}) team, which resulted in the first really uniform all-sky galaxy catalogs (e.g.\ \citealtc{IRAS}), widely used afterwards for cosmological measurements, including those of the clustering dipole and LG acceleration (see the list of references in the beginning of Chapter~\ref{Ch:Growth}). The second method to obtain extragalactic data for the whole sky, which consists in using two ground-based instruments on both hemispheres, was the key idea behind the Two Micron All Sky Survey.

\section{2MASS Extended Source Catalog}
\label{Sec:2MASS XSC}
The Two Micron All Sky Survey (2MASS, \citealtc{Skr06}) is the first near-infrared survey of the whole sky (covering 99.998\% of the celestial sphere), and was performed in the period 1997--2001 in the $J\, (1.25\, \mu\mathrm{m})$, $H\, (1.65\, \mu\mathrm{m})$ and $K_s\, (2.16\, \mu\mathrm{m})$ bands, with the use of twin 1.3-m ground-based telescopes, one at Mount Hopkins (Arizona, USA) and the other at Cerro Tololo Inter-American Observatory (Chile). All the data from the survey are available through the NASA/IPAC Infrared Science Archive.\footnote{\texttt{http://irsa.ipac.caltech.edu/Missions/2mass.html}} The main outcome of this project are two photometric catalogs: of point sources (PSC), containing about 471 million objects (mainly stars and some quasars), and of extended ones, with more than 1.6 million objects, majority of which are galaxies ($>98\%$) with additionally some diffuse Galactic sources \citepc{Jar04}. The Extended Source Catalog (XSC), which was used for the purpose of our analysis, is complete for sources brighter than {$K_s\simeq13.6$} mag ($\sim\!2.7\,\mathrm{mJy}$) and resolved diameters larger than $\sim\!10$ -- $15''$. The near-infrared flux is particularly useful for the purpose of large-scale structure studies as it samples the old stellar population, and hence the bulk of stellar mass, and it is minimally affected by dust in the Galactic plane \citepc{Jar04}. An additional advantage of using 2MASS data, especially in the context of calculating the flux dipole (Eq.\ \ref{eq:lim flux dipole}), for which \emph{apparent} magnitudes are used, is the global photometric uniformity of the catalog, which was enforced by nightly photometric calibration to an extensive set of standard star fields. {This procedure assured equally that there is no bias or offset between the photometry or astrometry obtained with the two telescopes used for the survey \citepc{Skr06}.} On the other hand, as any survey, 2MASS is not perfect. It is biased against optically blue and low surface brightness galaxies, such as dwarfs, but sensitive to the early type, bulge-dominated ones. However, as the former have very small luminosities and masses, their possible underrepresentation in the catalog should not influence significantly our results. What is more, these biases are much less severe than it was in the case of \textit{IRAS} which, due to longer wavelengths (middle and far infrared), was sensitive mainly to starburst galaxies.
\begin{figure}[!ht]
\begin{center}
\includegraphics[width=0.89\textheight, height=0.92\textwidth, angle=90]{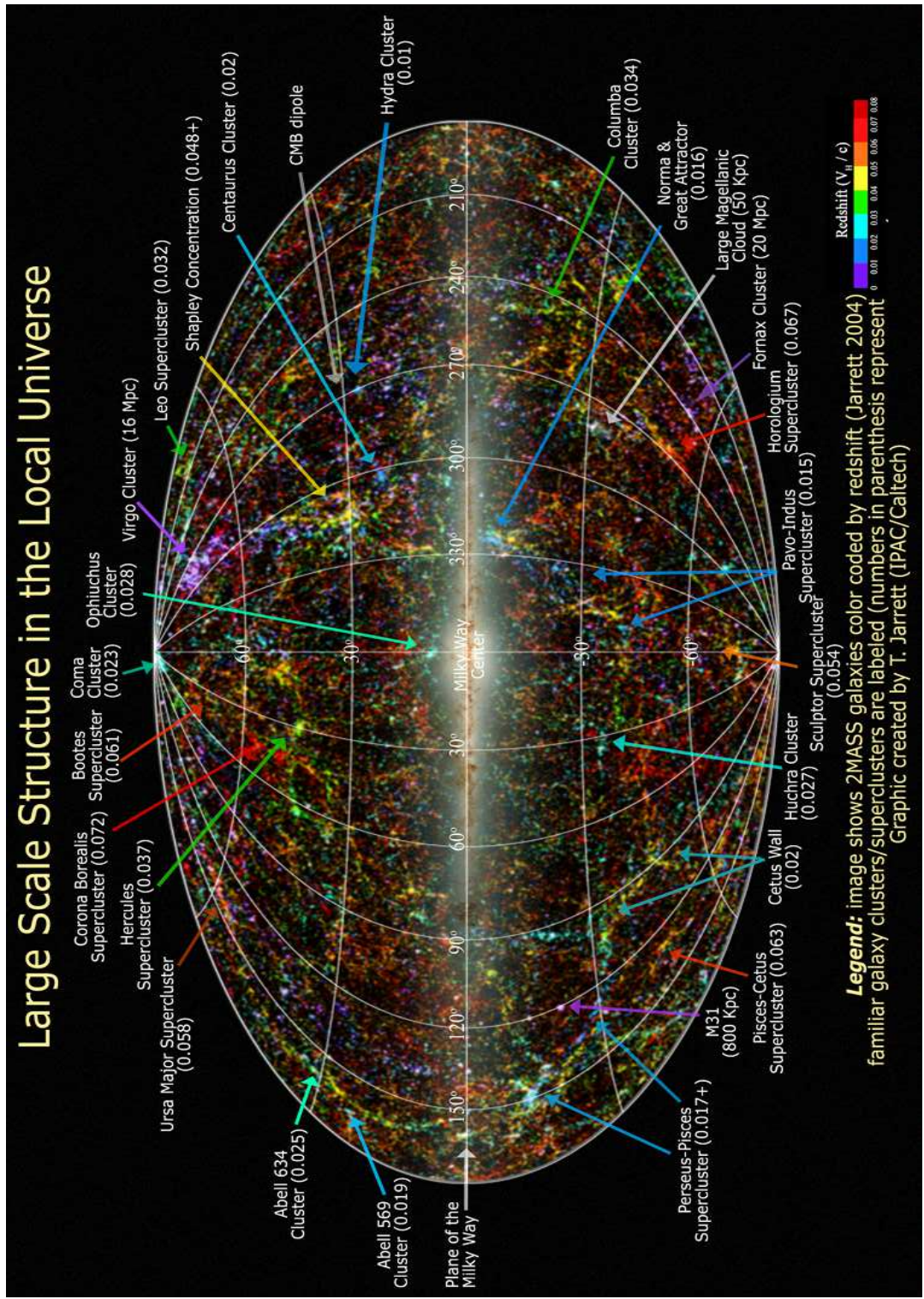}
\end{center}
\caption[Distribution of 2MASS galaxies on the sky]{\label{Fig:LSS}\small Distribution of galaxies from the 2MASS Extended Source Catalog.}
\end{figure}

Figure~\ref{Fig:LSS}, courtesy of Thomas Jarrett\footnote{\texttt{http://web.ipac.caltech.edu/staff/jarrett/lss/index.html}\vspace{2mm}}, presents an Aitoff (equal-area) projection of the 2MASS XSC galaxy distribution in Galactic coordinates, color-coded by redshifts or their estimates from \citec{Jar04}. Several local large-scale structures are  clearly visible, {including the Supergalactic plane (right of the 0-th meridian)}. The Milky Way galaxy, creating the Zone of Avoidance superimposed on the `cosmic web', covers only a small fraction of the sky when compared to surveys in other bands. {Note that the sample gathered by \citec{Jar04}, which is a promising first step towards deriving photometric redshifts for all those galaxies in the 2MASS XSC that do not have spectroscopic ones measured, is not useful for the purposes of our work. Those redshift estimates are only preliminary, do not have controlled errors and their accuracy depends strongly on the position of the galaxy on the sky.}

\section{Data preparation}
\label{Sec:Data.prep} 
The 2MASS photometry offers several types of `magnitudes' for extended objects, depending for instance on the type of aperture used. Throughout the whole analysis we use the $20\,\mathrm{mag\slash{}sq.''}$ isophotal fiducial elliptical aperture magnitudes, which are defined as magnitudes inside the elliptical isophote corresponding to a surface brightness of $\mu_\mathrm{band}=20\,\mathrm{mag\slash{}sq.}''$. We prefer those to the Kron ones as the latter use large and noisy apertures, prone to contamination, resulting in systematic overestimation. Our choice is additionally supported by the considerations in the appendix of \citec{Kochanek01}. However, we must remember to correct the values used by adding an offset of $\Delta=-0.2$~mag when converting to flux, in order to compensate for the flux lost outside the aperture (typically $\sim\!10\%$ -- $20\%$, \citealtc{LGA}). We have checked that this offset is roughly equal to the one between isophotal fiducial elliptical aperture magnitudes and the `total' ones, obtained from fit extrapolation {(the latter are also noisy and not recommended for such analyses as ours)}. The magnitude correction by a constant factor certainly introduces some scatter in total flux estimates, as it may depend on galaxy morphology. The latter is very hard to constrain from 2MASS data, we will thus treat this scatter as a systematic effect that needs to be included in the error budget. Our tests have shown that this error is of the order of a couple percent.

In order to prepare the data for our purposes, we have proceeded as follows. First of all, we applied the extinction correction from \citec{SFD98}, by calling the procedure \texttt{dust\_getval}\footnote{\texttt{http://www.astro.princeton.edu/\textasciitilde{}schlegel/dust/dustpub/CodeC/README.C}\vspace{2mm}}  for Galactic coordinates of each of the objects. The procedure yielded values of \mbox{$E(B-V)$}, which were subtracted from the original magnitudes with appropriate multiplicative factors $R_V$ taken from \citec{CCM89}: $0.902$ for $J$, $0.576$ for $H$ and $0.367$ for $K$. We performed the subtraction for objects with \mbox{$|b|>5\degr$} due to the statement of \citec{SFD98} that for \mbox{$|b|<5\degr$} the predicted reddenings should not be trusted (the sources in the ZoA were not included in our catalog apart from several brightest; see below). Moreover, for some minor parts of the sky, the extinction correction gave unreliably high reddenings, which resulted in some objects becoming unrealistically bright (with negative magnitudes) and eventually deleted. At this stage, we have also identified and removed the following sources (with some found in more than one category):
\begin{itemize}
\item artifacts: flag \texttt{cc\_flg=a} in the 2MASS XSC (122 objects);
\item sources with \texttt{NULL} or unreliable $K$ magnitudes (as described above) (718 objects);
\item non-extended sources: flag \texttt{vc=2} in the 2MASS XSC (7383 objects);
\item Local Group galaxies, taken from the list of \citec{Lees08} (31 objects);\footnote{Not all the objects from the Local Group were found in the database. These were some dwarf galaxies of low mass and near-IR luminosity, hidden behind the Galaxy or with surface brightness below the threshold of 2MASS.}
\item some of the Milky Way sources, taken from a list of 4454 such objects, separately identified earlier in the 2MASS XSC by Thomas Jarrett (private communication).
\end{itemize}

\section{Removal of Milky Way sources}
\label{Sec:MW.removal}
\begin{figure}[!t]
\begin{center}
\includegraphics[angle=-90,width=\textwidth]{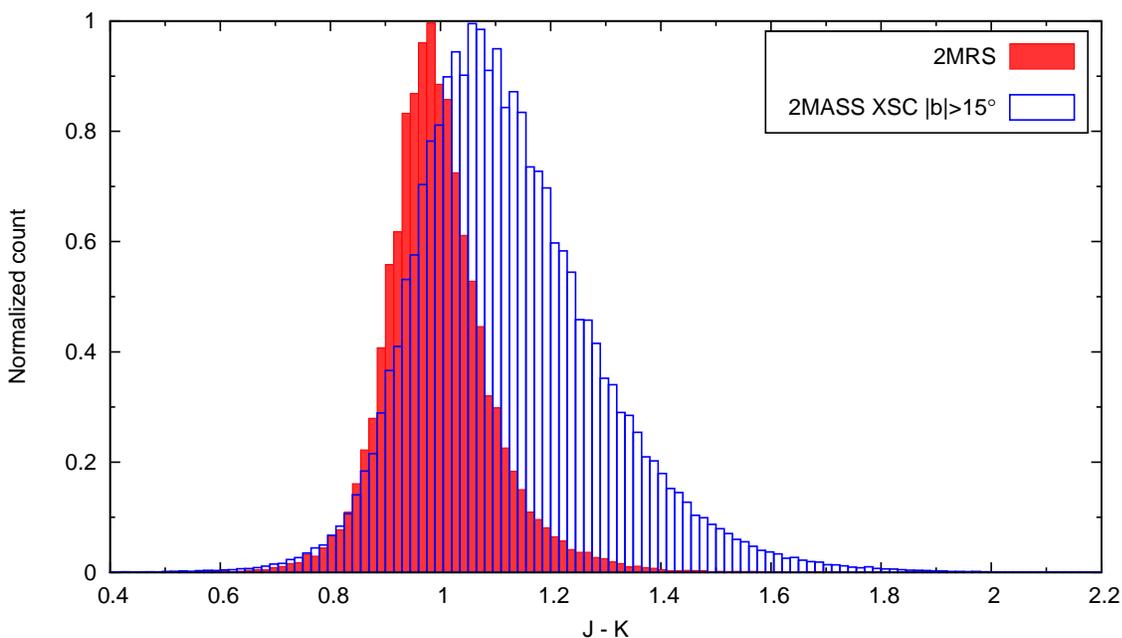}
\end{center}
\caption[Histogram of the $J-K$ color for 2MRS and 2MASS XSC]{\label{Fig:Colors}\small  Histogram showing the $J-K$ color distribution of the 2MRS galaxies (red filled bars) and of the 2MASS extended objects with $|b|>15\degr$ (blue open bars). The counts are normalized to a common value for clarity.}
\end{figure}
The 2MASS Extended Source Catalog contains mainly galaxies; however, it is also comprised of Milky Way entities, such as stellar clusters, planetary nebulae, HII regions, young stellar objects and so on. In order to keep our analysis reliable, these objects had to be removed from the catalog. This was partially done for the 4454 sources mentioned above. However, owing to the size of the catalog, any further `manual' procedure of Galactic object removal was impossible and only a method based on some general properties could be applied. A useful one in this regard is the color, i.e.\ difference of magnitudes in two bands. In their analysis, \citec{Maller03} made a cross-correlation with galaxies spectroscopically confirmed from the Sloan Digital Sky Survey (SDSS) and excluded extended sources brighter than \mbox{$K=12$ mag} with colors \mbox{$J-K<0.75$} or \mbox{$J-K>1.4$}; at fainter magnitudes only those objects with \mbox{$J-K<0.5$} were removed. We have checked these conditions by examining the \mbox{$J-K$} distribution of galaxies in the 2MASS Redshift Survey {(which by construction contains only extragalactic sources)}. We have found that indeed galaxies are clustered around \mbox{$J-K=1$} {(cf.\ Fig.\ \ref{Fig:Colors})}; however, we have decided to alter the limits given by \citec{Maller03}. Analyzing additionally the distribution of XSC objects with \mbox{$K <13.5$ mag} and \mbox{$|b|>15\degr$}, among which there are mainly non-Milky Way sources, apart from some molecular clouds (private communication of Janusz Ka{\l}u\.zny), we have decided to keep in our catalog those objects that have \mbox{$0.6<J-K<2.0$} (see also \citealtc{NIRGMA}). {Figure~\ref{Fig:Colors}, showing the histogram of the $J-K$ color for the 2MRS sample and for the above mentioned 2MASS objects off the Galactic plane, supports our choice}. We use our criterion for all sources, independently of magnitude, as we think that a differentiation as in \citec{Maller03} could lead to a bias in the sample. An additional visual verification of the 100 brightest objects which pass this filter off the Galactic plane confirms that indeed all of them are galaxies and that only one galaxy with an extreme value of $J-K$ is removed by this procedure up to $K_s\sim\!7.5$~mag.

\section{Zone of Avoidance}
\label{Sec:ZoA}

As was already discussed in Chapter~\ref{Ch:Nonlin}, an important issue in the calculation of the clustering dipole is the Zone of Avoidance (ZoA), i.e.\ the region of the sky with small Galactic latitudes $b$, which obscures galaxies behind the Galactic plane and bulge. Although the Galactic extinction is much lower in the near infrared than in visible bands \citepc{CCM89} and this applies equally to the ZoA \citepc{Jar00}, the 2MASS XSC is still incomplete near the Galactic equator, mainly due to high stellar density in this region of the sky \citepc{KKJ05}. For that reason, and owing to inapplicability of the extinction maps of \citec{SFD98} for \mbox{$|b|<5\degr$}, we have masked out the Galactic plane and bulge in the following way. For the shape of the mask we have chosen the one proposed in \citec{Erdogdu06a}, i.e.\ we have skipped all the objects with \mbox{$|b|<5\degr$} (plane) and \mbox{$|b|<10\degr$} for \mbox{$l<30\degr$} or \mbox{$l>330\degr$} (bulge). Then we have filled the resultant gap by cloning the adjacent strips, with mirror-like reflections: for instance, objects with $10\degr<b<20\degr$ were copied to the bulge by assigning \mbox{$b_\mathrm{new}:=20\degr-b_\mathrm{old}$} and keeping other parameters unchanged (such as the longitude $l$ and magnitudes). An analogous procedure was used for the negative latitudes and for the Galactic plane.  {Figure~\ref{Fig:ZoA} illustrates schematically this procedure.}  Such cloning has the advantage over random filling (considered both in \citealtc{Maller03} and \citealtc{Erdogdu06a} and discussed in Chapter~\ref{Ch:Nonlin}) that it extends the structures from above and below the ZoA; moreover, the only artificial discontinuity of the galaxy distribution created in this procedure is at the Galactic equator and at the edges of the box masking the bulge. We have also tried other masks and methods of filling the ZoA, and found no special importance for the results of the analysis presented here. This will be shortly addressed in Section~\ref{Sec:Growth}.
\begin{figure}[!t]
\begin{center}
\includegraphics[angle=0,width=0.65\textwidth]{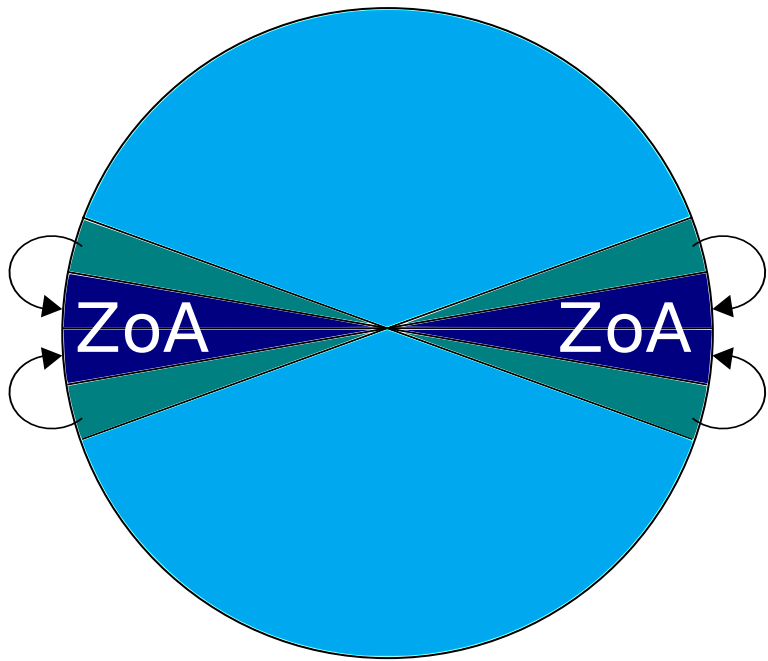}
\end{center}
\caption[Masking and filling the ZoA]{\label{Fig:ZoA}\small Schematic illustration of masking the Zone of Avoidance and cloning the adjacent strips. Galaxies from the navy areas were removed and replaced by those from above and below, with their latitudes mirrored using the edges of the ZoA as the axes of symmetry. The observer is located at the center of the circle.}
\end{figure}

\section{Large Galaxy Atlas}
\label{Sec:LGA}
Once the ZoA has been masked and filled, we have added to our catalog several galaxies that were not present in the 2MASS XSC but could be found in the 2MASS Large Galaxy Atlas (LGA, \citealtc{LGA}). This atlas\footnote{Accessible through IPAC Infrared Science Archive (IRSA) at\\ \texttt{http://irsa.ipac.caltech.edu/applications/2MASS/LGA/}} contains the $\sim\!600$ largest galaxies as seen in the near-infrared, of which around 50 are not present in the XSC or are located in the ZoA, $|b|<5\degr$ (17 sources). Among the latter, three are of particular importance for the Local Group motion, namely Maffei~1, Maffei~2 and Circinus. We will discuss their influence on our results later in the text. Note that this addition of LGA galaxies does not spoil the photometric uniformity of the resulting sample because all the galaxies from the LGA present also in the XSC were assigned the magnitudes from the former catalog when the final version of the latter one was constructed.

In case of those LGA galaxies that were present in the ZoA, the \citec{SFD98} maps are known to \textit{overestimate} the Galactic extinction by roughly 15\% (e.g.\ \citealtc{Schroder}); we have thus decreased the $E(B-V)$ by that amount there. The exceptions are Maffei~1 and Maffei~2, for which we used \textit{exact} values of extinction, given in \citec{Maffei2}, as well as Circinus with $E(B-V)=0.677$ (For et al., in preparation). Note however that apart from those three galaxies, which have extinction-corrected magnitudes below $K=5$~mag, all the remaining ones added from LGA are much fainter, by 2~mag or more, and possible misestimation of their extinction does not largely influence our analysis.

\section{The final catalog}
\label{Sec:catalog}
The final catalog contained 1,464,028 galaxies up to the 14th magnitude in the $K_S$ band. Among these, 108,023 were clones copied to the ZoA. Appendix \ref{App:Brightest} provides information about the 50 brightest galaxies of the sample (excluding the Local Group).


\chapter{\textsf{\textbf{Growth of the 2MASS dipole}}}
\label{Ch:Growth}
\lettrine[lraise=0.5,lines=2,findent=1pt,nindent=0em]{I}n this Chapter we will examine the clustering dipole of galaxies from the Two Micron All Sky Survey Extended Source Catalog  as a function of increased depth of the sample. This analysis will allow us to constrain the $\beta$ parameter from this dipole and consequently to estimate the density parameter $\Omm$. Our method however is not to directly compare the peculiar velocity and acceleration of the Local Group; instead, we use the observed growth of the dipole to obtain these constraints.\\
\ind In the past, many different datasets have been applied to calculate the clustering dipole. Generally speaking, there is no consistency on the \emph{amplitude}, the \emph{scale of convergence} of the dipole, and even on the \textit{convergence itself}. The pioneering works used the revised Shapley-Ames catalog \citepc{YST80} and the CfA catalog \citepc{DH82}. A great advancement came with the launch of the far-infrared \emph{IRAS} satellite and catalogs obtained thanks to this mission. The LG dipole from \emph{IRAS} was studied first from two-dimensional data only \citepc{YWRR86, MD86, HLM87, ViSt87}, then with redshifts included \citepc{Strauss92, Schmoldt99, RR2000, DMSa04, BP06} and with optical data added \citepc{LRRLB88, KaLa89}. Samples with optical data only were also used \citepc{Lah87, Hudson93}, as well as galaxy clusters \citepc{PliVal91,Tini95, KE06}. Among the most recent analyses one finds those directly related to the study presented here, which used the data from the Two Micron All Sky Survey. \citec{Maller03} used the 2MASS Extended Source Catalog, concluded convergence of the clustering dipole from flux data only and used it to calculate the average mass-to-light ratio in the near-infrared $K_s$ band and to estimate the linear biasing parameter $b_K$.\footnote{{From now on we denote the general biasing parameter as $b$ and the one in the $K_s$ band as $b_K$.}} \citec{Erdogdu06a} studied the acceleration of the LG from the 2MASS Redshift Survey, {claimed convergence already at $\sim\!60\Mpch$  and} estimated the $\beta$ parameter by comparing the dipole with the LG velocity. In a more recent work, \citec{Lav10} used an orbit-reconstruction algorithm to generate the peculiar velocity field for the 2MRS, extended it to larger radii, and observed no convergence of the clustering dipole up to at least $120\Mpch$.\\
\ind Let us shortly reiterate some considerations of Section \ref{Sec:Clust.dipole}. Since both the gravitational force and the flux of a galaxy are inversely proportional to its distance squared, the clustering dipole can be calculated by assuming some average value of the mass-to-light ratio and summing only flux `vectors' of galaxies (with positions on the sky used as angular coordinates). Such a dipole does not use distance nor redshift information of particular objects, which hinders the direct estimation of the depth up to which the dipole is measured and can bias the measurement of LG acceleration from it. This drawback is partially removed if a galaxy \emph{redshift} survey (with redshifts of galaxies used as proxies of their distances) is used for the analysis; however, up to date the densest and deepest \emph{all-sky} redshift surveys do not reach farther than to $\sim\!100\Mpch$ and contain no more than several dozen thousand sources. Owing to some recent claims of large-scale flows in the local Universe (e.g.\ \citealtc{WFH09,KABK08}), it is important to estimate the dipole from as deep catalogs as possible, even if they do not contain redshift measurements. We  have thus decided to trade the advantages of redshift measurements for a huge number of galaxies and a much greater depth of a photometric-only catalog. The angular dipole thus obtained is additionally free from any redshift distortions, in particular from the rocket effect \citepc{Kaiser87}.\\
\ind In this Chapter, we focus on the issue of the convergence of the dipole, by analyzing its growth with increased depth of the sample. For that purpose we use the data from the 2MASS XSC, similarly to what has been presented in \citec{Maller03}. One of our goals is to expand and refine the latter work. We do it by modifying the criteria of removing Galactic objects from the sample (Section \ref{Sec:MW.removal}) and masking and filling the Galactic plane and bulge (the Zone of Avoidance, Section \ref{Sec:ZoA}), but keeping in the ZoA the brightest and closest galaxies of big influence for the local motion (Section \ref{Sec:LGA}). We analyze the growth of the dipole and check how its direction changes on the sky. Next, we compare the observed growth of the dipole with theoretical expectations, using the condition of known peculiar velocity of the Local Group, in a similar manner as was done in \citec{Lav10}. From this comparison we evaluate the $\beta$ parameter and, by using externally provided value of bias, constrain the cosmic density $\Omm$.

The Chapter is organized as follows. Section \ref{Sec:Growth} focuses on the observed growth of the 2MASS clustering dipole. Next, Section \ref{Subsec:Angle} deals with the position of the 2MASS dipole on the sky. In subsequent Section \ref{Sec:Discussion} we discuss the results: first, in Subsection \ref{Subsec:Theory} we present the theoretical basis to analyze the growth for known LG velocity; Subsection \ref{Subsec:Windows} focuses on the observational window of our measurement; finally, in Subsection \ref{Subsec:Compare} we compare our results with theoretical expectations and use this comparison to estimate the $\beta$ parameter from the growth of the dipole alone. We shortly summarize and conclude this Chapter in Section \ref{Sec:Gr.SummConcl}, where we also provide the obtained value of the $\Omm$ parameter.

\section{Observed growth of the 2MASS clustering dipole}\label{Sec:Growth}
In this section we will use the data prepared as has been described in Chapter~\ref{Ch:Data} to calculate the clustering dipole of the galaxies from the 2MASS XSC and analyze its growth. For that purpose we start with the formula for the flux dipole of the survey given by Eq.\ (\ref{eq:lim flux dipole}) and change units of $\bmd$ into$\kms$, defining the \emph{scaled dipole} as
\begin{equation}\label{eq:g_v} 
\tilde{\bmd} =\frac{2\,b}{3\, H_0\, \Omega_\mrm}\,\bmd =\frac{H_0}{j}\sum_i^N S_i\hat{\bmr}_i\;.
\end{equation}
The flux of each galaxy is calculated from its magnitude $m_i$ as
\begin{equation}\label{eq:S_i}
S_i=S_0\,10^{-0.4m_i}\,,
\end{equation}
where $S_0$ is the flux for a 0-magnitude object. As was already stated, we consider magnitudes in the  \mbox{$K_s$} band, which was the main (`target') band of the 2MASS survey (for simplicity of notation, we sometimes skip the `$s$' subscript). Then Eq.\ (\ref{eq:g_v}) takes on the form:
\begin{equation}\label{eq:g_v_Ks}
\tilde{\bmd}=\frac{H_0}{j_K}\,S(0\,\mathrm{mag}) \sum_{i}^{N} 10^{-0.4(K _i+\mathrm{ZPO})} \hat{\mathbf{r}}_i\;,
\end{equation}
where \mbox{$S(0\,\mathrm{mag})=1.122\times10^{-10}\,\mathrm{W\,m^{-2}}$ $(\pm1.685\%)$} and the zero point offset $\mathrm{ZPO}=0.017\pm0.005$ \citepc{CWM03}. The $K_i$ magnitudes in Eq.\ (\ref{eq:g_v_Ks}) include also a negative offset of $\Delta=-0.2$ mag added due to the underestimation of total fluxes by the isophotal magnitudes in the 2MASS XSC. The quantity $j_K$ is the luminosity density in the $K$ band. It is obtained for example from the integral (\ref{eq:lum.dens}) using the luminosity function in this band. The value of $j_K$ has been estimated by many authors in the recent decade: \citec{Kochanek01, Cole01, Bell03, Eke05, 6dF_Fi, SLC09, LH11} {and the results from one work to another do not always agree within the errors quoted. Here, we use the determination that} we consider the most reliable, calculated by \citec{6dF_Fi} from the luminosity function of more than 60,000 galaxies in the 6-degree Field Galaxy Survey (6dFGS, \citealtc{6dF09}): $j_K=(5.9\pm0.6) \times 10^8\, h\, \mathrm{L}^K  _\odot \, \mathrm{Mpc}^{-3}$ (where $\mathrm{L}^K _\odot\simeq6.8\times10^{24}\,\mathrm{W}$, \citealtc{Rieke08}\footnote{$\mathrm{L}^K _\odot$ is the Solar luminosity in the $K$ band, calculated based on table 7 of \citec{Rieke08},\\ \texttt{http://www.iop.org/EJ/article/1538-3881/135/6/2245/aj271287\_mrt7.txt}.}). Note that as the luminosity density depends {linearly} on the Hubble constant, the $h$ factor cancels out and Eq.\ (\ref{eq:g_v_Ks}) may be rewritten as 
\begin{equation}\label{eq:dim-less dipole}
\tilde{\bmd}=\mathcal{C} \, \sum_{\Kmin}^{\Kmax} 10^{-0.4K_i}\hat{\mathbf{r}}_i\;,
\end{equation}
where $\mathcal{C}\simeq 2620\,\kms$ for parameter values as given above. The limits of the summation are now the minimum and maximum $K$ magnitudes of galaxies in the sample. The lower limit in Eq.\ (\ref{eq:dim-less dipole}) is the magnitude of the brightest object in the catalog (excluding LG galaxies) and the upper one will be increased, as will be discussed later in the text. To retain the reliability of the analysis, we must note that the results remain trustworthy only up to the completeness limit of the catalog: the sample becomes incomplete for objects with \mbox{$K\gtrsim 13.6$ mag} \citepc{Jar04}.\\
\ind {The growth of the dipole was calculated by incrementing $\Kmax$ in the sum given in Eq.\ (\ref{eq:dim-less dipole}). Results are illustrated in Figures~\ref{Fig:growth.N} and \ref{Fig:growth.r}. Figure~\ref{Fig:growth.N} shows the growth of the dipole as a function of the number of galaxies used for the calculation and presents also Galactic Cartesian components of the acceleration ($d_x$ points towards the Galactic center, $l=0\degr$, $b=0\degr$; $d_y$ is perpendicular to it in the Galactic plane; $d_z$ is perpendicular to the Galactic plane). The similarity of this plot to figure~1 of \citec{Maller03} is intentional and allows us to draw the conclusion that although our data analysis was slightly different to that of \citec{Maller03} (altered shape and filling of the ZoA, different removal of Galactic objects), the two approaches are qualitatively the same. {Nonetheless, we want to stress that such a presentation of the results may be misleading. {First of all, as the discussed scales can be treated as Euclidean, for a uniform distribution the number of sources in a given radius is $N(r<R)\propto R^3$, so a linear scale in the number of galaxies on the abscissa `compresses' the left-hand side of the plot, while `stretching' the right-hand one. This may suggest the convergence of the dipole, which was actually concluded by \citec{Maller03}. However, local inhomogeneities complicate this $N(r<R)$ relation and finally a linear scaling in the number of galaxies up to a given flux threshold is not related in a simple manner to a scale expressed in distances nor in cutoff magnitudes.} For that reason, what we need in order to draw proper conclusions about the growth and possible convergence of the dipole, is a linear scale \emph{in distance} on the abscissa.}\\
\begin{figure}[!t]
\centering\includegraphics[angle=270,width=\textwidth]{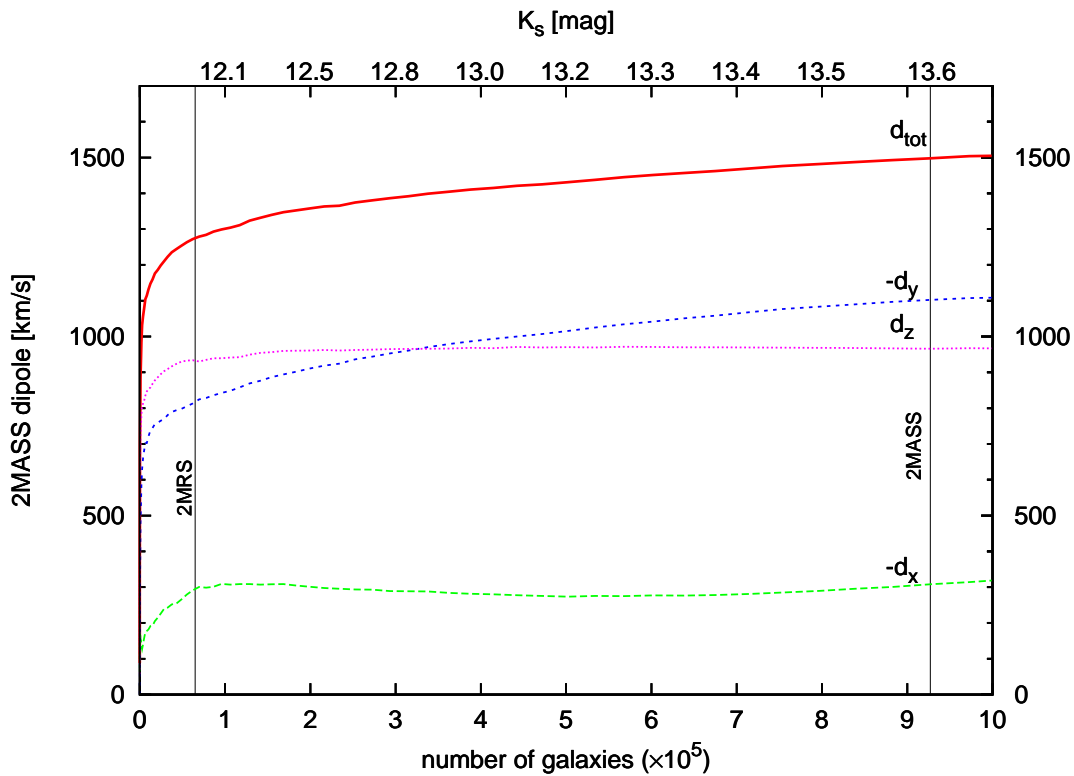}
\caption[Growth of the 2MASS dipole as a function of galaxy number]{\label{Fig:growth.N}\small Growth of the 2MASS clustering dipole as a function of the number of galaxies used for the calculation (bottom axis), ordered by their $K_s$ magnitudes (top axis). The thick red line is the amplitude of the dipole; the thin dotted and dashed lines (green, blue and magenta) are the Cartesian components (in Galactic coordinates). Two vertical lines illustrate the limits of the 2MRS  $K_s\leq11.75$~mag   sample and completeness of the 2MASS XSC   ($K_s=13.6$~mag).   Such data presentation could suggest that the 2MASS dipole has converged within sample limits, which is \textit{not} the case.}
\end{figure}
\begin{figure}[!t]
\centering\includegraphics[angle=270,width=\textwidth]{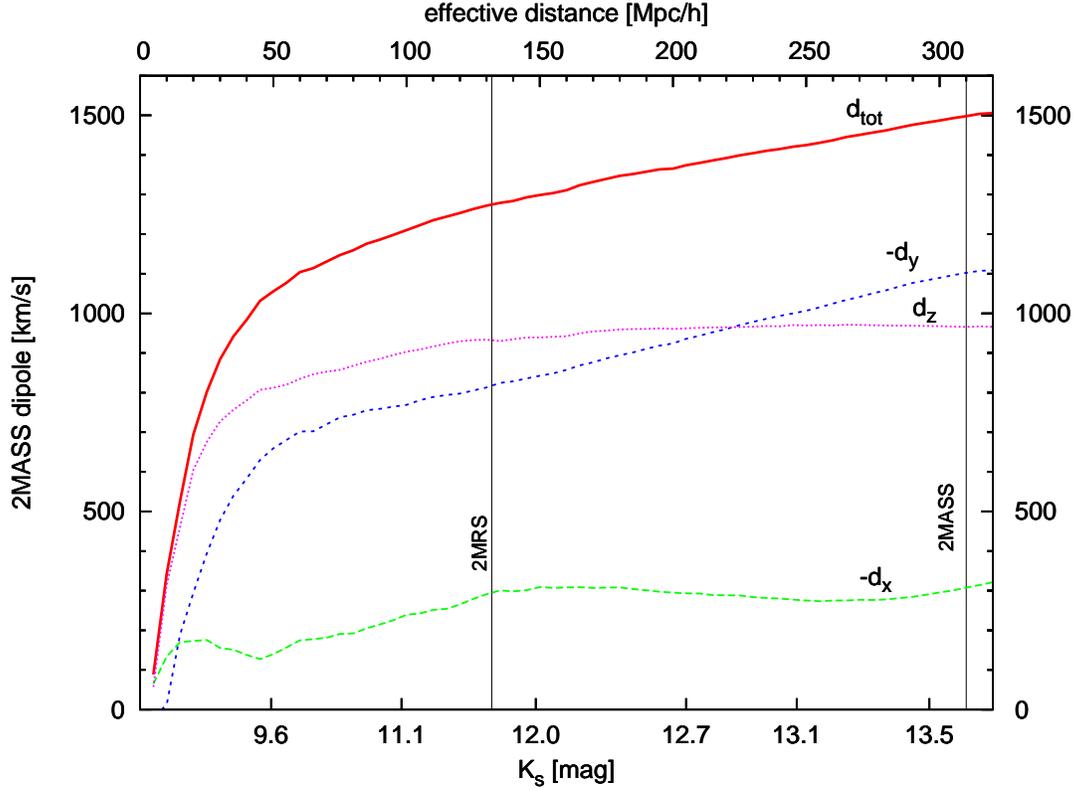}
\caption[Growth of the 2MASS dipole as a function of effective depth]{\label{Fig:growth.r}\small Growth of the 2MASS clustering dipole as a function of increased maximum $K_s$ magnitude of the subsample (bottom axis). Corresponding  effective distance is given at the top. The thick red line is the amplitude of the dipole; the thin dotted and dashed lines (green, blue and magenta) are the Cartesian components (in Galactic coordinates). Two vertical lines illustrate the limits of the 2MRS $K_s\leq11.75$~mag sample and completeness of the 2MASS XSC  ($K_s=13.6$~mag).   The lack of convergence of the dipole is clearly visible. Note also the steady growth of the amplitude as well as of the Galactic $y$ component of the dipole.}
\end{figure}
\ind Neither distances, nor even redshifts (including photometric ones) are currently measured for the whole 2MASS XSC (although some attempts are being made regarding photo-$z$'s, see \citealtc{Jar04} and \citealtc{FraPea10}). We thus {decided to define} effective distances of galaxies {based on} their fluxes, with the use of the luminosity function (LF) in the $K$ band. If all the galaxies had the same luminosity, say $L_*$, the relation between the observed flux $S$ and distance $r$  would be straightforward: $r=\sqrt{L_*\slash 4\pi S}$. However, galaxies have different {intrinsic} luminosities, and obviously their LF is not a Dirac's delta (on the contrary, it is very broad). Therefore an estimated distance of a galaxy with a given flux must have a scatter. Constructing the estimator, a first choice could be the conditional mean, i.e.\ the expectation value for $r$ given $S$. Instead, we think that it is better to choose the conditional \emph{median} for $r_\mathrm{eff}$ (a median value of distance given the flux). We consider it being more adequate to our problem: the same number of galaxies with a given flux have distances smaller and greater than the median, by definition. {However, as we show in Appendix \ref{App:r_eff}, the two estimators differ by only a few percent.} For the $K$-band LF as given by \citec{6dF_Fi}, i.e.\ a \citec{Sche76} function
\begin{equation}
\Phi(M)\,\de M\propto10^{0.4(1+\alpha)(M_*-M)}\exp\left[-10^{0.4(M_*-M)}\right]
\end{equation}
with $M_*=-23.83+5\log h\pm0.03$ and $\alpha=-1.16\pm0.04$, this effective distance for the magnitude $K$ equals to
\begin{equation}\label{eq:r_eff}
r_\mathrm{eff}\simeq0.59\times10^{0.2K}\Mpch \;
\end{equation}
(detailed calculations are presented in Appendix \ref{App:r_eff}; see also \citealtc{Pe93}). This proxy of distance is used in Figure~\ref{Fig:growth.r}, which differs from Figure~\ref{Fig:growth.N} by a different scaling of the abscissa. The growth of the clustering dipole up to the completeness limit of the 2MASS XSC is now evident. Additionally, note that the growth has an essentially constant slope for $r_\mathrm{eff}>150\Mpch$, i.e.\ $K_s>12$ mag.

An interesting feature is the behavior of the Galactic Cartesian components of the dipole. The $x$ and $z$ ones are virtually constant for $r_\mathrm{eff}>150\Mpch$; however, the $y$ component still grows even at the limit of the catalog, similarly as does the total amplitude. This could point to some systematic effect, related to masking and filling of the Zone of Avoidance. We have however checked that the same qualitative behavior of the three components is observed for different shapes of the mask and the way it is filled; what is more, the effect exists even if we calculate the dipole having removed from the catalog all the galaxies with $|b|<10\degr$ (leaving the resulting strip completely devoid of galaxies). {We thus conclude that the effect is real and must be due to the large-scale matter distribution in our cosmic neighborhood. An indirect confirmation comes from recent measurements of the local bulk flow, i.e.\ the volume-averaged peculiar velocity of galaxies in a several-dozen megaparsec sphere around us. Different authors (e.g.\ \citealtc{WFH09, ND11}) who do not necessarily agree on the \textit{amplitude} of the flow, still find consistent results when it comes to the \textit{direction} (namely, $l_\mathrm{bf}\simeq280\degr$, $b_\mathrm{bf}\simeq10\degr$, i.e.\ roughly along the Galactic $y$-axis). The simplest explanation of the pull could be a single supermassive structure at a large distance, hidden behind the Galactic plane (as there is no known supercluster in this region just above the ZoA, see Figure~\ref{Fig:LSS}). However, an alternative and perhaps more probable option is the summed effect of several sources, such as the Shapley Concentration and the Horologium Supercluster. This issue still awaits quantitative studies.}

We have also observed that adding to the sample the LGA galaxies that were not present in our {basic} catalog has virtually no influence on the amplitude of the dipole and only slightly changes the values of particular components (see also below).

\section{Misalignment angle}
\label{Subsec:Angle}
\begin{figure}[!t]
\centering\includegraphics[angle=270,width=\textwidth]{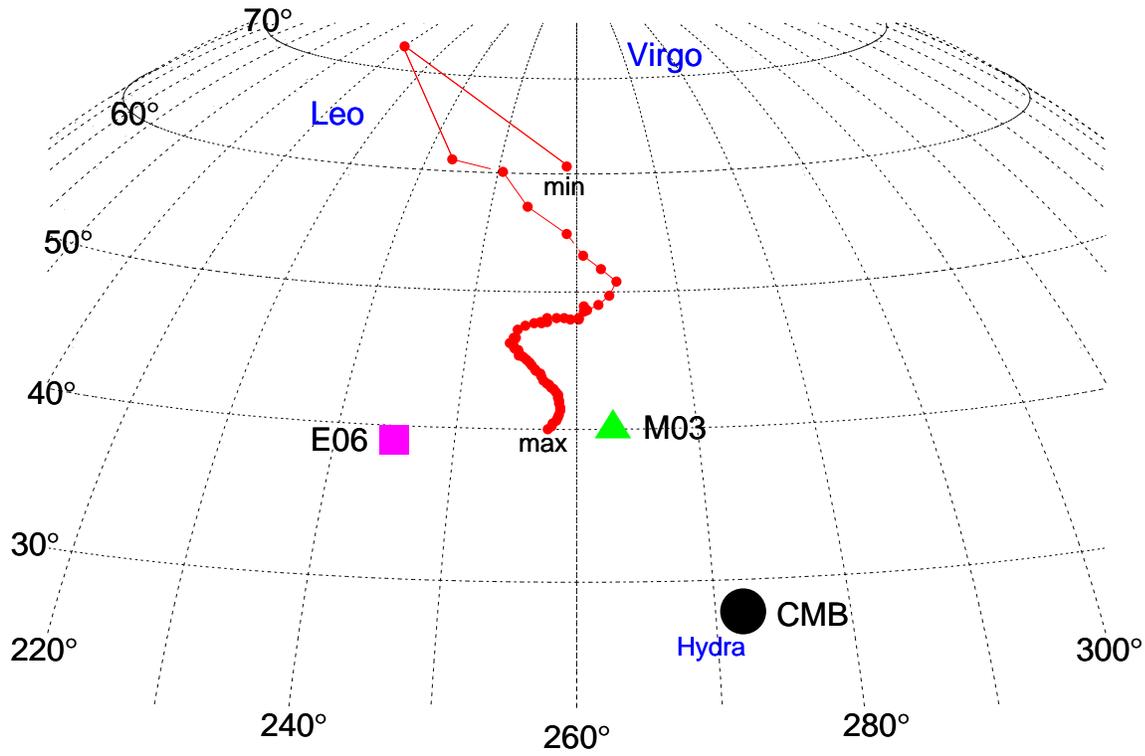}
\caption[Position of the 2MASS dipole on the sky]{\label{Fig:on.the.sky}\small Position of the 2MASS clustering dipole on the sky as a function of increased depth of the sample. The grid shows Galactic coordinates: longitude $l$ (labels on bottom) and latitude $b$ (left). The min/max labels refer to the minimum and maximum depth of the sample for which the dipole was calculated, {equivalent to effective distances of} $5\Mpch$ and $310\Mpch$, respectively. The black disk marked by `CMB' is the direction of the peculiar velocity of the Local Group ($l=272\degr$, $b=28\degr$). The green triangle labeled `\textsc{\textcolor{kolor_cyt}{M03}}' shows the 2MASS clustering dipole of \citec{Maller03}: $l=263\degr$, $b=40\degr$ (with cloning in the ZoA). The magenta square marked by `\textsc{\textcolor{kolor_cyt}{E06}}' is the direction of the flux-weighted 2MRS dipole in the CMB frame of \citec{Erdogdu06a}: $l=245\degr$, $b=39\degr$. Blue labels indicate important structures in the local Universe: the Virgo Cluster ($D\simeq17\Mpc$), the Hydra Cluster ($D\simeq47\Mpc$) and the Leo Supercluster ($z\gtrsim0.031$).\vspace{3cm}}
\end{figure}
\begin{figure}[!t]
\centering\includegraphics[angle=270,width=\textwidth]{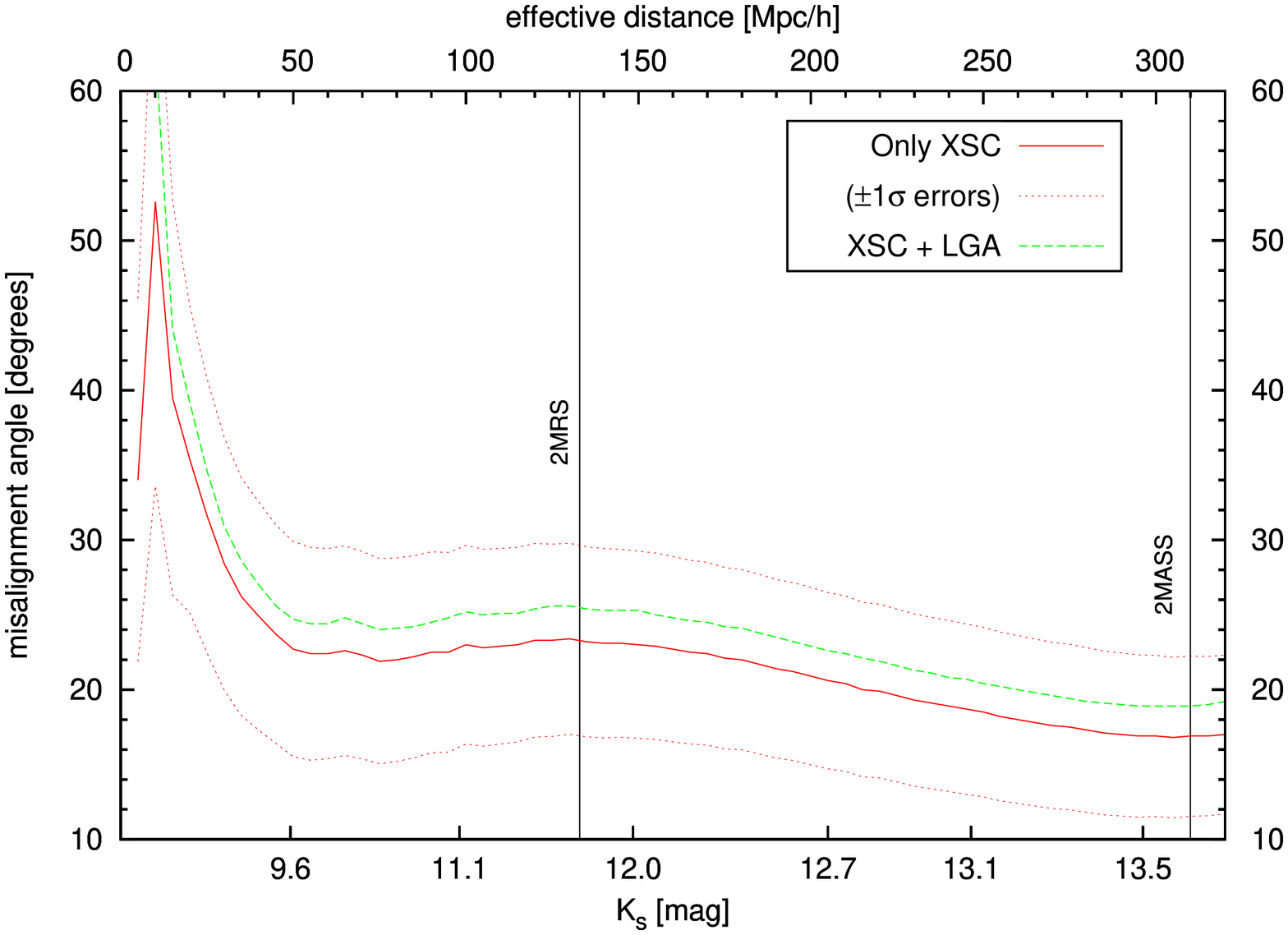}
\caption[Misalignment angle between the 2MASS dipole and the velocity of the LG]{\label{Fig:misal}\small Misalignment angle between the 2MASS clustering dipole and the peculiar velocity of the Local Group as a function of increased maximum $K_s$ magnitude of the subsample (bottom axis). The corresponding effective distance is given at the top. Two vertical lines illustrate the limits of the 2MRS $K_s\leq11.75$~mag  sample and completeness of the 2MASS XSC  ($K_s=13.6$ mag).  The red solid and green dashed lines are shown to illustrate the effect of adding the Large Galaxy Atlas to the sample. {Dotted lines show $1\sigma$ errors (only for the XSC sample without the LGA, to avoid overcrowding of the plot).}}
\end{figure}
The linear theory relation (\ref{eq:vLG-gLG}) between the peculiar velocity and the acceleration of the Local Group predicts that the two vectors should be parallel. In reality, a non-zero misalignment angle between them is expected, due to various reasons (cf.\ Chapter~\ref{Ch:Nonlin}). Observed values of this angle are usually of the order of {$10\degr\sim25\degr$} \citepc{Strauss92, Schmoldt99, Maller03, Erdogdu06a} and we have obtained similar results, which confirms the validity of the linear approximation for the scales of interest. Figure~\ref{Fig:on.the.sky} shows how the dipole direction on the sky changes as the sample depth increases from `min', i.e.\ $5\Mpch$, to `max', equal to $310\Mpch$ {(expressed here in terms of the effective distance)}. The black disk shows the CMB dipole. For comparison, we present also two other results of 2MASS data analysis: the 2MASS dipole with `cloning' the ZoA, from \citec{Maller03} (green triangle labeled \textsc{\textcolor{kolor_cyt}{M03}}) and the flux-weighted 2MRS dipole in the CMB frame of \citec{Erdogdu06a} (magenta square, \textsc{\textcolor{kolor_cyt}{E06}}). We increase the Galactic longitude $l$ from left to right for easier comparison with relevant figures in these two papers (respectively fig.\ 2 of \textsc{\textcolor{kolor_cyt}{M03}} and fig.\ 7 of \textsc{\textcolor{kolor_cyt}{E06}}). {The direction that we find at the limit of the sample lies somewhere in between those two results.}

In Figure~\ref{Fig:misal} we plot the misalignment angle between the 2MASS clustering dipole and the CMB dipole direction, as a function of growing depth of the sample. The two central curves illustrate the effect of adding the LGA galaxies to the catalog, and in particular of the three bright galaxies located in the ZoA (Maffei 1 \& 2 and Circinus). A specific `tug-of-war' between these structures, located on almost opposite sides of the Galactic plane, results in a general raise of the angle by $\sim\!1\degr${, which is well within $1\sigma$ errors of the measurement (estimated via bootstrap resampling, see Subsec.\ \ref{Subsec:Compare} and Appendix~\ref{App:Bootstrap})}. As can be equally seen, the brightest (and presumably the closest{, cf.\ Appendix~\ref{App:Brightest}}) galaxies have the most influence on the value of the angle. Note also that already for an effective distance as small as $50\Mpch$ the misalignment angle reaches a value of $\sim\!22\degr$ ($\sim\!0.35$~rad), with a minimum of $17\degr\pm5\degr$ for $r_\mathrm{eff}\simeq300\Mpch$.

One can propose several reasons for this misalignment. The first possibility are influences from extremely large scales. We cannot address this issue directly here; note only that the claims of \citec{WFH09} or \citec{KABK08} of large-scale bulk flows with values much larger than expected in \LCDM\ have been recently put in doubt by \citec{NBD11.BF}, \citec{ND11} and \citec{Osborne}: the matter is thus far from being settled. The second option could be {nearby} unsuspected structures in the Zone of Avoidance. This issue was partly addressed in Chapter~\ref{Ch:Nonlin}, where the influence of one such partially obscured object -- the Local Void -- was analyzed. As was shown, the improper filling of the ZoA in the LV region cannot bias the angle by more than $~1\degr$. Additionally, already in 2004 it was quite certain that thanks to surveys in various wavelengths (such as HI or X rays) there should be no unknown nearby large-scale structures hidden behind the Milky Way (e.g.\ \citealtc{FaLa05}). Finally, such a big value of the misalignment angle can be due to improper accounting of very local influences; we have tested this possibility by adding the LGA galaxies to our sample.\\
\ind Note however that our statistical assumption that the flux is a good proxy of the gravitational force, which may be appropriate for the whole sample, can fail for individual galaxies. In particular, we have used a constant mass-to-light ratio for all galaxies, which is probably valid when averaged over many of them, especially since we use the $K_s$-band luminosity, which is known to be a better tracer of stellar mass than optical and other NIR wavebands. Still, the $M/L_K$ ratio should vary somewhat with morphology and luminosity. For galaxies with smallest magnitudes, these variations may significantly affect their contribution to the direction of the clustering dipole. The same was noted by \citec{Maller03}, who also concluded that the $16\degr$ difference between their 2MASS dipole and the CMB one was caused by the brightest galaxies. We obtain a similar result here and interpret it as due to a specific non-linear effect: a few sources with the largest fluxes have a large influence on the misalignment.\\
\ind The high significance of the most luminous (in terms of the observed flux) galaxies for the direction of the dipole confirms theoretical predictions of \citec{CCBCC}: the misalignment angle could be lowered if the observational window used was the optimal one, which would be the case if we \textit{removed} those brightest galaxies from the sample (as was also tried both by \citealtc{Maller03} and \citealtc{Erdogdu06a}). This issue will be addressed in the following Chapter~\ref{Ch:MLE}, concerning the optimized estimation of $\beta$ from the 2MASS dipole.

\vspace{3cm}

\section{Discussion and estimation of the $\bm\beta$ parameter}
\label{Sec:Discussion}
Our results showing that the clustering dipole of 2MASS galaxies still grows even at the completeness limit of the sample, taken at face value, are consistent with what has been obtained by several authors, who used various datasets and methods: redshift survey of \emph{IRAS} galaxies, \citec{Strauss92}; \emph{IRAS} PSC\emph{z}, \citec{Schmoldt99} and reanalysis by \citec{BP06}; X-ray selected clusters, \citec{KE06}; reconstructed velocity field of 2MRS, \citec{Lav10}. On the other hand, they contradict claims of convergence at scales even as small as $60\sim100\Mpch$: optical sample of \citec{Hudson93}; redshift sample of Abell/ACO clusters, \citec{PliVal91,Tini95}; \emph{IRAS} PSC\emph{z}, \citec{RR2000}; \emph{IRAS} PSC\emph{z} and BTP, \citec{DMSa04}; 2MRS, \citec{Erdogdu06a}. Our analysis also suggests a different interpretation of the results of \citec{Maller03} --- data presentation as in our Figure~\ref{Fig:growth.r} instead of figure~1 therein would possibly point to similar lack of convergence. On the other hand, in order to be able to \textit{directly} compare our results with those of \citec{Erdogdu06a}, we would have to apply the same weighting as was done for the 2MRS sample, namely by the inverse of the flux-weighted selection function. We are unable to do it, not knowing distances nor redshifts for the whole sample.\\
\ind Apart from galaxy weighting, the discrepancies between the above listed results most probably stem from the different nature of catalogs and methods used for the calculation, and in particular may be due to distinct \emph{observational windows}. Such a window for a given survey describes the sample: it may be interpreted as a filter (in real or Fourier space) through which we observe the Universe. Knowledge of the observational windows, necessary to correctly confront results as those given above, is also essential if we want to make comparisons with theoretical expectations.

\subsection{Theoretical framework}\label{Subsec:Theory}
We would now like to check if the behavior of the 2MASS flux dipole is consistent with the predictions of the currently favored cosmological model, namely Lambda-Cold-Dark-Matter (\LCDM). We start by presenting the theoretical framework for such a comparison. It was first derived in the context of then-popular models like cold-dark-matter and isocurvature baryon, as described in detail in two classic papers: \citec{JVW90} and \citec{LKH90}. More recently, this approach was taken by \citec{Lav10}, who reconstructed the local peculiar velocity field (up to $\sim\!150\Mpch$), applying the data from the 2MRS. The basic quantity for these comparisons is the joint probability distribution function for $\bmv$ and $\bmg$, assumed to be a multivariate Gaussian (for more on this p.d.f., see next Chapter).

In our case, we want to find the expectation value for the amplitude of the acceleration of the Local Group \emph{knowing} its peculiar velocity. As was already mentioned, the latter equals to $v_\mathrm{_{CMB}}=622\pm35\,\mathrm{km\slash s}$ in the direction $(l,b)=(272\degr\pm3\degr\!,\,28\degr\pm5\degr)$. In the analysis of the present Chapter we do not use the directional constraint: the misalignment angle is integrated out (c.f.\ \citealtc{JVW90}). In the next Chapter, concerning the maximum-likelihood determination of $\beta$ from 2MASS XSC, we \textit{will} use the fact that the misalignment angle is known. Here, the theoretical prediction gives us the conditional velocity, $\bmv_\mrc$ (in units of $\kms$), which is related to the acceleration via the linear-theory Relation (\ref{eq:v_s}). The relevant formula for its amplitude has the form given by eq.\ (8a) of \citec{JVW90} (note a typo therein):
\begin{eqnarray}\label{eq:v_cond}
v_\mrc & \equiv & \langle v_m\vert v_\mathrm{_{CMB}}\rangle = \\
&&=\sigma_m \frac{1-\varrho^2+\varrho^2 u^2}{\varrho u}\,\mathrm{erf}\left\{\frac{\varrho u}{\left[2(1-\varrho^2)\right]^{1\slash 2}}\right\}+ \sigma_m\sqrt{\frac{2}{\pi}(1-\varrho^2)}\,\exp\left[-\frac{\varrho^2 u^2}{2(1-\varrho^2)}\right]\;, \nonumber
\end{eqnarray}

\noi{}where we have used the following quantities:
\begin{itemize}
\item
$\bmv_m$ is the velocity induced by a single realization of the density field {$\del(\bmr)$} given the assumed power spectrum of density fluctuations:
\begin{equation}\label{eq:v_m}
\bmv_m=\frac{H_0\, f(\Omm)}{4\,\pi}\int{\delta(\bmr)\,\frac{\bmr}{r^3}\,W(\bmr)\,\de^3 \bmr}\;,
\end{equation}
\item[]with $W(\bmr)$ the window of the survey, included to mimic the velocity measured from a given survey;
\item
$\sigma^2_m$ is the predicted (ensemble average) {1-D} variance of the velocity measured from the survey:
\begin{equation}\label{eq:sigma_m}
\sigma^2_m=\frac{H_0^2\, f^2(\Omm)}{6\,\pi^2}\intlim_0^\infty{P(k) \tilde{w}^2(k)\,\de k}\;,
\end{equation}
\item[]with $P(k)\equiv \langle\vert\delta_\bmk\vert^2\rangle$ being the {aforementioned} power spectrum, $\bmk$ the wavevector and $\tilde{w}(k)$ the observational window in Fourier space;
\item $\bmu\equiv\bmv_m\slash\sigma_m\;;$
\item
$\varrho$ is the cross-correlation coefficient of $v_m$ and $v_\mathrm{_{CMB}}$:
\begin{equation}\label{eq:rho}
\varrho=\frac{H_0^2}{6\,\pi^2\,\sigma\,\sigma_m}\intlim_0^\infty{P(k) \tilde{w}(k)\,\de k}\;,
\end{equation}
\item[]where $\sigma$ is given by Eq.\ (\ref{eq:sigma_m}) with $\tilde{w}(k)\equiv 1$. 
\end{itemize}
The second moment of the conditional velocity is given by eq.\ (8b) of \citec{JVW90}:
\begin{equation}
\mu_\mrc\equiv\langle v^2_m\vert v_\mathrm{_{CMB}} \rangle=3\sigma_m^2(1-\varrho^2)+\varrho^2 u^2 \sigma_m^2\;,
\end{equation}
hence the variance is equal to
\begin{equation}\label{eq:sigma_cond}
\sigma^2_\mrc=\mu_\mrc-v^2_\mrc\;.
\end{equation}

The presence of the factor $f(\Omm)$ in Eqs.\ (\ref{eq:v_m})--(\ref{eq:sigma_m}) and the fact that we also have to include biasing $b$ mean that in order to properly compare the conditional velocity $v_\mrc$ given by Eq.\ (\ref{eq:v_cond}) with {the amplitude of the} dipole, $d$, measured from the data (Eq.\ \ref{eq:g_v}), the latter has to be rescaled, using a best-fit $\beta$ parameter, as in Eq.~(\ref{eq:v_s}). In fact, this rescaling enables us to estimate $\beta$ in a straightforward way; we come back to this point a few paragraphs later.

\subsection{Observational window}\label{Subsec:Windows}
In this subsection we discuss the proper observational window for our measurement. We start by noting that when calculating the peculiar gravitational acceleration of the Local Group from the dipole of the all-sky galaxy distribution, in general there are two schemes to postulate a relation between this distribution and that of the underlying mass (see e.g.\ \citealtc{Erdogdu06a}). The first scheme, called \emph{number weighting}, assumes that the mass is distributed in the Universe as a continuous density field, which is sampled by galaxies in a Poissonian way. The second prescription, namely \emph{flux weighting}, uses the assumption that all the mass in the Universe is locked to the mass of the halos of luminous galaxies. The 2MASS dipole as calculated in Section \ref{Sec:Growth} is of the flux-weighted type. 

The observational window $W(r)$ of a flux-limited survey is just its selection function $\varphi(r)$. In the number-weighted scheme, this function measures what fraction of galaxies located in the distance interval $(r,\,r+\de r)$ are included in the survey:
\begin{equation}
\varphi(r)=\frac{\intlim_{4\pi r^2 S_\rmmin}^\infty \Phi(L)\, \de L}{\intlim_0^\infty \Phi(L)\, \de L}\;,
\end{equation}
where $\Phi(L)$ is the luminosity function and $S_\rmmin$ is the limiting flux of the survey (e.g.\ \citealtc{DH82, YSDH91}). In our case of the flux-weighted dipole, as given by Eq.~(\ref{eq:lim flux dipole}), the appropriate window is the flux-weighted selection function, $\Psi(r)$, defined in such a way that $1-\Psi(r)$ is the percentage of light from a distance $r$ which is not visible for the survey \citepc{Erdogdu06a, CCBCC}. It is given by
\begin{equation}\label{eq:W_S_r}
W_S(r)=\frac{\intlim_{L_\rmmin}^\infty L\, \Phi(L)\, \de L}{\intlim_0^\infty L\, \Phi(L)\, \de L}\;,
\end{equation}
where $L_\rmmin=4\pi r^2 S_\rmmin$. The window $W_S(r)$, where the subscript $S$ emphasizes the dependence of the window on the flux limit of the catalog, smoothly decreases to zero with increased distance from the observer. For a detailed discussion see \citec{CCBCC}.

On the other hand, `ideal' surveys would be distance- (or volume-) limited: in this case there is no loss of signal up to a limiting distance $R$. Then the window function is simply unity for $r<R$ and zero otherwise (the so-called \textit{top-hat window}). It has thus a form of a Heaviside step function; for a spherical survey we have
\beq\label{eq:W.top-hat}
W_R(r) = \Theta_{\mathrm H}(R - r)
\eeq
(e.g.\ \citealtc{JVW90}). In order to better reconstruct the dipole of the galaxy distribution from a flux-limited catalog, volume-weighting of the survey is commonly mimicked by weighting individual galaxies by the inverse of the selection function, at the expense of increasing shot noise from large scales \citepc{Strauss92} and at a price of a possibly large rocket effect \citepc{Kaiser87}. The top hat is then the relevant window. In our case however, as we are dealing with angular (photometric) data only, we cannot estimate the selection function even in redshift space. Therefore, we do not weight galaxies in the sample (or in other words, we assign unit weights to them) and the relevant window function is given by (\ref{eq:W_S_r}).

{The Fourier-space counterpart of the observational window, $\tilde{w}(k)$, is obtained for the clustering dipole from the real-space one as} 
\begin{equation}\label{eq:W_Four}
\tilde{w}(k)\equiv k\intlim_0^\infty{W(r)\,j_1(kr)\,\de r}
\end{equation}
\citepc{KaLa89, LKH90, JVW90}, where $j_1$ is the spherical Bessel function of the first kind of order 1. Note that this window is \emph{not} the Fourier transform of $W(r)$. In our case of $W_S(r)$ given by (\ref{eq:W_S_r}), the Fourier-space window reads (eq.\ 59 of \citealtc{CCBCC})
\begin{equation}\label{eq:2MASS window}
\tilde{w}_s(k)=1-\frac{8\pi S_\rmmin}{k\int_0^\infty{L\,\Phi(L)\,\de L}}\intlim_0^\infty{\sin(kr)\,L_\rmmin\,\Phi(L_\rmmin)\,\de r}\;,
\end{equation}
where $S_\rmmin$ and $L_\rmmin$ are as above. The window $\tilde{w}_s$ is a function of both the wavenumber $k$ and the minimum flux $S_\rmmin$ (hence the subscript $s$). On the other hand, it does \emph{not} depend on the distance, as the latter is integrated out, together with absolute luminosities (with the use of the luminosity function). This means that when comparing observations to theoretical expectations, we never use the effective distance (\ref{eq:r_eff}) for the calculations of the conditional velocity $v_\mrc$ (\ref{eq:v_cond}) and its variance $\sigma_\mrc$ (\ref{eq:sigma_cond}). Our observable is the minimum flux of the sample, related to the maximum magnitude via $S_\rmmin=S_0\,10^{-0.4\Kmax}$. For clarity however, in the relevant plots we prefer to show the results in terms of linear scaling in the effective distance. 

\begin{figure}[!t]
\centering\includegraphics[width=\textwidth]{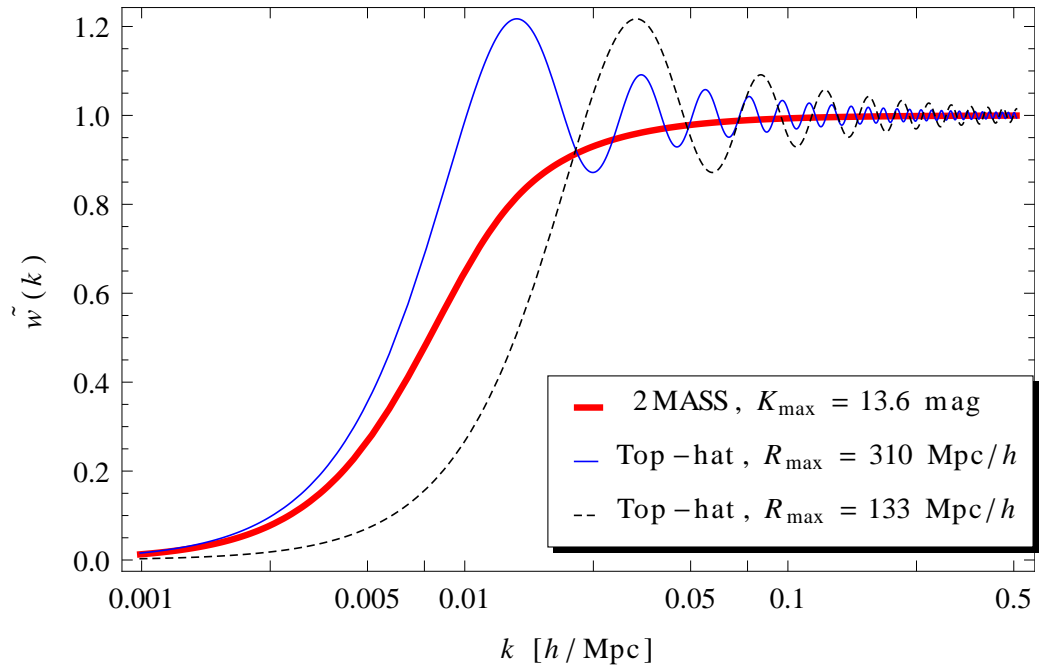}
\caption[Observational windows in Fourier space]{\label{Fig:windows}\small Comparison of three observational windows in Fourier space: the one of the 2MASS flux-limited survey (thick red line) for a maximum magnitude of $13.6$  mag and two for distance-limited surveys (top-hat) with  $R_\rmmax=310\Mpch$   (blue thin line) and  $R_\rmmax=133\Mpch$   (black dashed line). The former radius is the effective distance of galaxies at the limit of the 2MASS XSC; the latter corresponds to   $K=11.75$    mag, the limit of the 2MRS.}
\end{figure}

In Figure~\ref{Fig:windows} we plot the 2MASS flux-limited window with  $\Kmax=13.6$   mag and compare it with two top-hat windows for distance-limited surveys: one with $R_\rmmax=310\Mpch$ and the other with $R_\rmmax=133\Mpch$, which are respectively the effective distances of galaxies at the limit of the 2MASS and 2MRS $\Kmax=11.75$~mag sample. The Fourier form of the top-hat window {for the clustering dipole} is (e.g.\ \mbox{\citealtc{JVW90}):}
\begin{equation}
\tilde{w}_\mathrm{th}(k)=1-j_0(k R_\rmmax)\;
\end{equation}
with $j_0\equiv\sin x\slash x$ being the spherical Bessel function of the first kind of order 0. The oscillating behavior of this window in Fourier space is the result of rapid truncation in real space. As we can see, the top-hat window with $R_\rmmax=310\Mpch$, which asymptotically behaves in the same way as the 2MASS one with $\Kmax=13.6$ mag, passes clearly more large-scale signal ($k\lesssim0.02\hMpc$) than the latter does. This will result in much faster convergence of the theoretical dipole measured through the top-hat window than through the 2MASS flux-weighted one (see below). On the other hand, the top-hat window with a cutoff equivalent to the limit of 2MRS at  $133\Mpch$ blocks almost all the signal already for $k\lesssim0.01\hMpc$, as expected.

\subsection{Comparison with theoretical predictions}\label{Subsec:Compare}
We start by reminding that the analysis in this Chapter is performed within the linear theory. For that reason, we do not include non-linear effects in the cross-correlation coefficient $\varrho$ (Eq.\ \ref{eq:rho}). {Mathematically, this means that we set to unity two functions: the ratio of power spectra of density and scaled velocity divergence, $\calR(k)$, and the coherence function, $C(k)$. In the next Chapter we relax these assumptions and the two functions will be discussed in greater detail there.}

Our goal here is to compare the observational data to the expectations of the \LCDM\ concordance model. Hence the power spectrum used in Eqs.\ (\ref{eq:sigma_m}) and (\ref{eq:rho}) is the spectrum of cold dark matter, with baryon effects included, as given by \citec{EH98}. In this framework, $P(k)\propto k^{n_s} T^2(k)$, where $n_s$ is the spectral index of primordial fluctuations and the transfer function $T(k)$ depends on the parameters $h$ (Hubble constant divided by $100\kmsMpc$), $\Omm$, $\Omb$ (density parameter of baryons) and $\sigma_8$ (present value of root-mean-square density contrast of matter fluctuations within a sphere of $8\Mpch$). In our calculations we use the following set of parameters obtained from \emph{WMAP} seven-year observations \citepc{Larson11}: $h=0.71$, $\Omm h^2=0.1335$, $\Omb h^2=0.02258$, $n_s=0.963$ and $\sigma_8=0.801$.

\begin{figure}[!t]
\centering\includegraphics[angle=270,width=\textwidth]{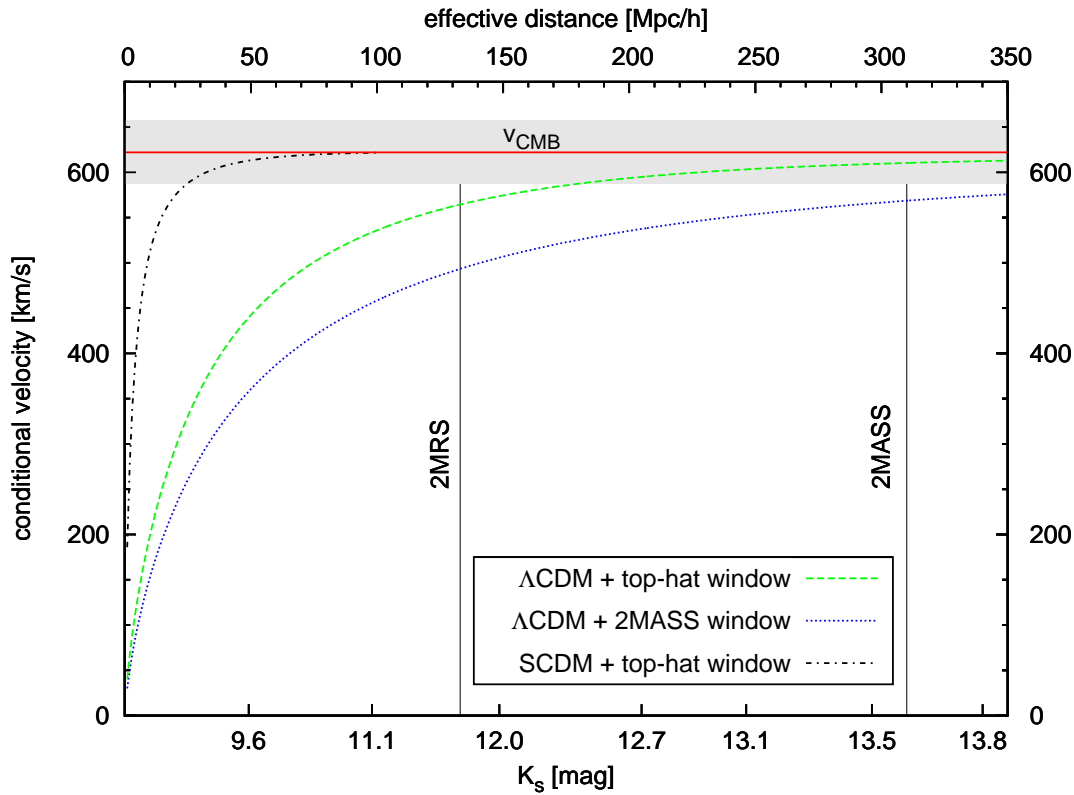}
\caption[Theoretically predicted growth of the conditional velocity of the LG]{\label{Fig:predicted}\small Theoretically predicted growth of the conditional velocity of the Local Group for known $v_\mathrm{_{CMB}}$ using two different observational windows: the one of 2MASS (blue dotted line) and top-hat (green dashed line), both calculated for the \LCDM\ as the underlying cosmological model. The horizontal solid line is the observed velocity of the Local Group with respect to the CMB. For comparison, the prediction for the currently disfavored Standard CDM model with the top-hat window is shown with the black dot-dashed line.}
\end{figure}
\ind Before applying the observational data, in Figure~\ref{Fig:predicted} we compare the expected growth of the conditional velocity for the 2MASS window and for the top-hat case, both calculated with the above \LCDM\ model parameters. As we can see, for such a flux-limited survey as 2MASS, the predicted velocity (blue dotted line) is far from converging to the limit of $v_\mathrm{_{CMB}}=622\,\mathrm{km\slash s}$ (horizontal line) even for $r_\mathrm{eff}\simeq350\Mpch$. In the top-hat case (green dashed line), the expected convergence is much faster, as could have been deduced already from the comparison of the windows, presented in Figure~\ref{Fig:windows}. However, it should be noted that even for all-sky catalogs that include redshifts, for which we can effectively model the dataset with a top-hat window (like in the case of the 2MRS), the convergence of the dipole is not likely before some $200\Mpch$, opposite to the results of \citec{Erdogdu06a}, where it is claimed that the contribution from structure beyond $6000\,\kms\,(=60\Mpch)$ is negligible. Figure~\ref{Fig:predicted} shows that for the latter distance, the conditional velocity for the top-hat window has reached less than $75\%$ of its final value. On the other hand, for the 2MASS window, $v_c$ attains $0.95\,v_\mathrm{_{CMB}}$ no sooner than for $r\simeq470\Mpch$ ($\sim\!14.5$ mag in the $K_s$ band, not shown in the plot), far beyond the completeness of the 2MASS XSC. As an instructive  case, in Figure~\ref{Fig:predicted} we additionally plot the prediction for the once-popular `Standard CDM' model, currently strongly disfavored by observations ($h=0.5$, $\Omm=1$, $\Omb=0.05$, $n_s=1$, $\sigma_8=1$), with the top-hat window (black dot-dashed line). {Note how rapidly the dipole is expected to grow in this case. We suppose this may be one of the reasons why fast convergence of the observed dipole is still taken for granted by some authors also in the framework of the \LCDM\ model.}\\
\begin{figure}[!t]
\centering\includegraphics[angle=270,width=\textwidth]{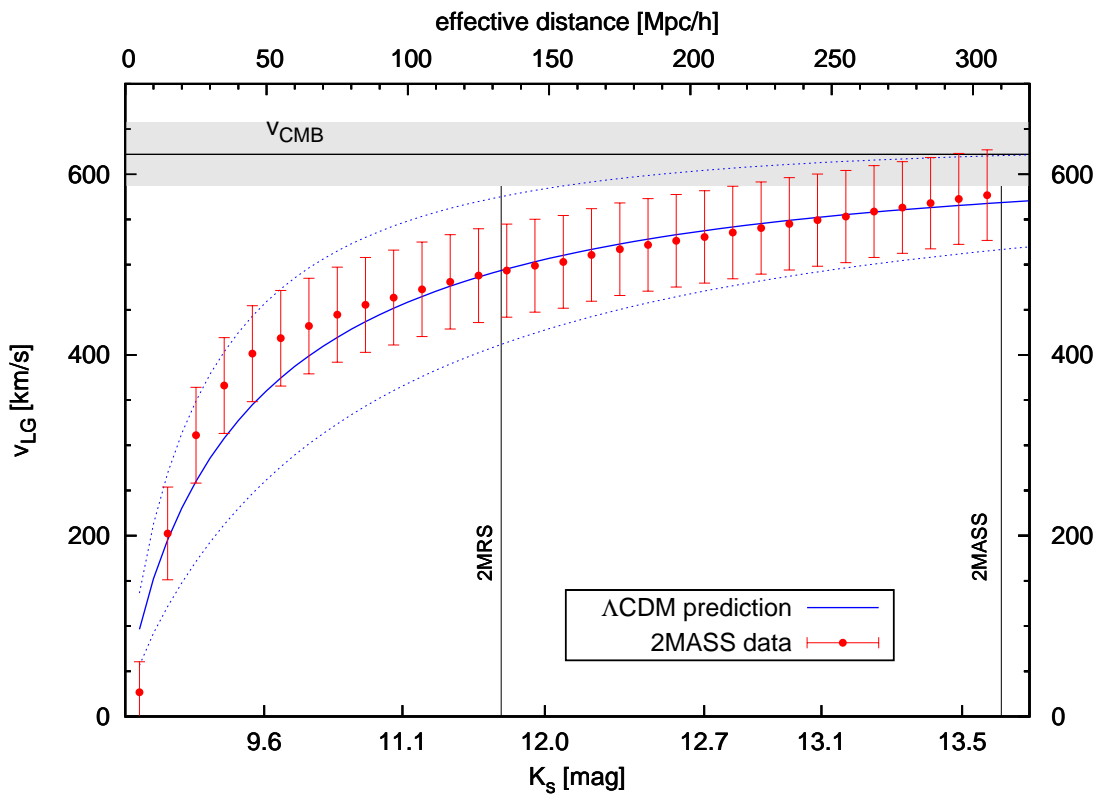}
\caption[Growth of the 2MASS dipole compared with the theoretical expectation]{\label{Fig:g_m and v_c}\small Growth of the 2MASS clustering dipole (red points with 1$\sigma$ errorbars), compared with the theoretical expectation for the conditional LG velocity in the \LCDM\ model (solid blue line with dotted blue lines showing 1$\sigma$ variances). The datapoints were rescaled by the $\beta$ parameter (see text). The horizontal solid black line is the Local Group velocity with respect to the CMB. Vertical lines give the limits of the 2MRS and 2MASS XSC catalogs.}
\end{figure}
\ind The most important result of this Chapter is presented in Figure~\ref{Fig:g_m and v_c}. It shows the observed growth of the 2MASS dipole compared with the conditional velocity calculated from Eq.\ (\ref{eq:v_cond}). As was already discussed at the end of Subsection \ref{Subsec:Theory}, we rescale observational data by a best-fit parameter $\beta\equiv\Omm^{0.55}\slash b$. The errorbars of the measured dipole were obtained from 1000-fold bootstrap resampling of the catalog {(for details see Appendix \ref{App:Bootstrap})}. The $\beta$ parameter was fitted by a minimum-$\chi^2$ procedure, including both observational errors (from bootstrapping) and the theoretical (cosmic) variance. Note that formally we are `over-fitting', as the datapoints are correlated with each other, due to integral (cumulative) nature of the measured dipole. {The analysis could in principle be redone for the differential dipole (shells in `flux space'). For that purpose one would need to develop the theoretical part, which may not be straightforward especially for the second moment and so far has not been presented in any work that we are aware of.}\\
\ind The observed growth, once rescaled, is well within $1\sigma$ range of the theoretical prediction (apart from the datapoints for the smallest distances, where the number of galaxies is very small, hence the measurement is noisy). Remarkably, the best agreement is obtained in the range $100\Mpch\lesssim r_\mathrm{eff}\lesssim300\Mpch$, i.e.\ roughly between the extent of the 2MRS catalog and the 2MASS XSC completeness limit. The result of the fitting gives $\beta=0.38$ with $1\sigma$ confidence intervals of $\pm0.02$. Owing to the considerations of Section \ref{Sec:Data.prep} concerning the scatter introduced by the offset added when passing from isophotal fiducial elliptical aperture magnitudes to total fluxes, as well as due to findings of Chapter~\ref{Ch:Nonlin} on the bias introduced by the fact of possibly improper masking of the Local Void, we double this formal error. Hence our estimate is
\begin{equation}
\beta=0.38\pm0.04\;,
\end{equation}
{precise within 11\%. This measurement} is in agreement with the analysis of \citec{Erdogdu06a}, where also data from 2MASS were used, although from much smaller scales. The clustering dipole was calculated there from the 2MASS Redshift Survey, then complete up to $\sim\!105\Mpch$ (maximum magnitude in the $K_s$ band equal to $11.25$~mag). Their result, based on claimed convergence of the flux-weighted dipole, was $\beta=0.40\pm0.09$. {We note that with our method, using only `2MRS-like' data, i.e.\ up to $K_s=11.25$~mag, the best-fit value would be $\beta=0.35\pm0.06$, which is smaller, but still consistent with our measurement from the whole sample and the one by \citec{Erdogdu06a}.} A somewhat larger value was obtained by \citec{PH05}, where the peculiar velocity field within $65\Mpch$ predicted from 2MASS photometry and public redshift data was compared to three independent peculiar velocity surveys based on Type Ia supernovae, surface brightness fluctuations in elliptical galaxies, and Tully-Fisher distances to spiral galaxies. The best-fit from this comparison was $\beta=0.49\pm0.04$ {[although close inspection of tables 1--3 from \citec{PH05} shows that these errors are possibly underestimated].} On the other hand, our value of $\beta$ agrees within the errors with that obtained by \citec{Davis11}, who reconstructed the cosmological large scale flows in the nearby Universe using the SFI++ sample of Tully-Fisher measurements of galaxies and compared it with the whole sky distribution of galaxies in the 2MRS to derive $0.28<\beta<0.37$ (68.3\% confidence). Finally, in a recent paper, \citec{NBD11.GR} used the galaxy distribution in the 2MRS to solve for the peculiar velocity field, estimated absolute magnitudes of galaxies and constrained $\beta$ by minimizing their spread. They found a value in excellent agreement with our determination: $\beta=0.35\pm0.1$.

\section{Summary and conclusions}
\label{Sec:Gr.SummConcl}
In this {Thesis}, as the estimator of the LG acceleration we use the clustering dipole of the galaxies from the 2MASS Extended Source Catalog, which contains positions and fluxes of almost one million sources. As we want to reach as far from the Local Group as possible with the state-of-the-art observations, we have decided to trade the advantages of redshift surveys for the huge number of objects and unprecedented sky coverage of the 2MASS XSC. This allowed us to measure the dipole up to $310\Mpch$, which is more than 2 times farther than for the deepest existing all-sky survey of galaxies with measured redshifts, the 2MASS Redshift Survey. The price to pay was our inability to weight galaxies with the inverse of their selection function (which would require redshifts to be known), so we could not directly mimic an `ideal', distance-limited survey, with our data. On the other hand, the flux dipole does not suffer from redshift-space distortions, and in particular from Kaiser's rocket effect.\\
\ind We have focused mainly on the issue of convergence of the dipole, which has been a subject of debate for more than two decades now. We have shown that the flux dipole of the 2MASS XSC does \textit{not} converge up to the completeness limit of the sample, which is $\sim\!13.6$~mag in the $K_s$ band. Moreover, we find that beyond an effective distance of some $150\Mpch$, the growth is induced mainly in the negative Galactic Cartesian $y$ direction. We have checked that this behavior is not due to systematic effects related to masking out and artificially filling the Zone of Avoidance. {We thus suppose that the effect is related to large-scale matter distribution in our cosmic neighborhood.} Additionally, the misalignment angle between the measured dipole and the velocity of the LG is found to be of the order of $20\degr$. We also find that the misalignment comes mainly from some nearby galaxies, bright in the near infrared. This is a hint that their removal could optimize the analysis, as was already theoretically investigated by \citec{CCBCC}. We come back to this issue in the next Chapter.\\
\ind{The observed lack of convergence of the dipole reflects the fact that for such a survey as the 2MASS XSC, we measure the growth using only angular and photometric data. As the luminosity function of galaxies is very broad, the information coming from a given distance is `spread' onto a wide interval of measured fluxes. However, by comparing this observed growth with theoretical expectations for known $\bmv_\mathrm{_{CMB}}$, in the framework introduced by \citec{JVW90} and \citec{LKH90}, we find consistency with the predictions of the \LCDM\ model, once the proper observational window of the 2MASS flux-limited catalog has been included. On the other hand, in the case of the top-hat window, which is applicable for a perfect, distance-limited survey, the theoretically expected convergence of the dipole is much faster, and takes place at a scale of $\sim\!200\Mpch$. This is considerably larger than the $60\Mpch$ claimed by \citec{Erdogdu06a}, where galaxies from the 2MASS Redshift Survey were weighted by the inverse of their selection function to mimic such a volume-limited catalog.}

{For the 2MASS XSC, we do not have redshifts for the whole sample and cannot define the selection function. Our inability to weight galaxies and to mimic a distance-limited survey, together with the lack of convergence of the dipole, do not allow us to measure the parameter $\beta$ from a direct $\vLG$ -- $\gLG$ comparison. However, we were still able to estimate this parameter due to the consistency of the observed growth with theoretical predictions; the result is $\beta=0.38\pm0.04$. This value is in agreement with and has better precision than the one from the analysis of \citec{Erdogdu06a}, who found $\beta=0.40\pm0.09$ from the clustering dipole of the 2MRS, which has a depth of $\sim\!100\Mpch$.}


Having estimated the value of $\beta$, we can go further and derive the density parameter $\Omm$. {By definition of $\beta$, we have $\Omm=(\beta\,b)^{1/0.55}$, we can thus express our measurement as
\beq
\Omm=b^{1.82}\times(0.17\pm0.03)\;.
\eeq
We can also constrain $\Omm$ alone,} provided that we know the linear bias, $b$. The latter {is dependent on the wavelength and} was estimated for the $K_s$ band  by \citec{Maller05}, who calculated the angular correlation function of galaxies in 2MASS and inverted it using singular value decomposition to measure the three-dimensional power spectrum. A fit of CDM-type power spectra in the linear regime allowed them to constrain the linear bias as $b_K=1.1\pm0.2$. Using this value, we obtain the following estimate:
\begin{equation}
\Omm=0.20\pm0.08\;,
\end{equation}
{with a relative error of $\sim$40\%, much larger than for $\beta$ alone due to the big uncertainty in the biasing parameter. Still,} this result, supporting the fact that the density of non-relativistic matter in the Universe is well bellow the critical one, roughly agrees with other independent dynamical estimates of $\Omm$, such as e.g.\ that by \citec{Feld03}, who used mean relative peculiar velocity measurements for pairs of galaxies and obtained $\Omm=0.30^{+0.17}_{-0.07}$. {Our findings are also perfectly consistent with the results derived from galaxy cluster data. For instance, \citec{MADE10} used observations of the growth of massive, X-ray flux-selected galaxy clusters and derived, for spatially flat models with a constant dark energy equation of state, $\Omm=0.23\pm0.04$.} {Last but not least, the density parameter derived here agrees within the errors with the one estimated from the analyses of the angular power spectrum of the cosmic microwave background temperature anisotropies, {for instance those} from \emph{WMAP} seven-year observations \citepc{Larson11}: $\Omm=0.267\pm0.029$.}\\
\ind The conclusions of this Chapter can be verified if new data become available. In particular, one could examine the growth of the 2MASS dipole with redshifts of galaxies as their distance estimates, using proper weights. In the near future we cannot however hope for a uniform and sufficiently deep sample of spectroscopic redshifts for the catalog, even if the 2MASS Redshift Survey is continued (as it was planned to reach $K_s=12.25$~mag, see \citealtc{Huchra05}). The calculation may be thus feasible only for photo-\textit{z}'s, if they are available for the whole 2MASS XSC. Some first steps towards this goal have been already taken \citepc{Jar04} and a promising direction may be to cross-correlate 2MASS with the data from the Wide-field Infrared Survey Explorer (WISE, \citealtc{WISE}) that are already available \citepc{WISErelease}. We can also take a different approach to estimate $\beta$ from LG velocity-acceleration comparison. By maximizing the correlation between observed velocity and acceleration, we can use the maximum likelihood method to try to obtain an \textit{optimal} measurement of $\beta$. Such study is presented in the next Chapter{, where we use the same 2MASS data but refine the method of statistical analysis}.

\chapter{\textsf{\textbf{Optimized estimation of $\bm\beta$ parameter}}}
\label{Ch:MLE}
\lettrine[lraise=0.5,lines=2,findent=1pt,nindent=0em]{T}he analysis presented in Chapter~\ref{Ch:Growth} shows that the growth of the 2MASS clustering dipole is consistent with \LCDM\ predictions. This consistency allowed us to constrain the $\beta$ parameter by comparing the observed growth with the theoretically predicted one. However, it is still possible to optimize our measurement. First of all, owing to the integral nature of the dipole, the points in Figure~\ref{Fig:g_m and v_c} are strongly correlated. Fitting a theoretical curve gives then a minimum $\chi^2$ much less than 1. Secondly, the window function $W_S(r)$ as given by Eq.\ (\ref{eq:W_S_r}) is not the optimal one, which results in a significant value of the misalignment angle between the peculiar velocity of the LG and the 2MASS clustering dipole. As was shown e.g.\ by {\citec{Strauss92} and followed by} \textsc{\textcolor{kolor_cyt}{Chodorowski et al.\ }}(\textsc{\textcolor{kolor_cyt}{2008}}; hereafter `\CCBCC'), in order to {obtain the best estimates of}  cosmological parameters by comparing the LG velocity and acceleration{, one should preferably use} the maximum likelihood approach {and} optimize the window through which the {clustering dipole} is measured. {As one of the indications of this optimization,} the misalignment angle is then expected to decrease.\\
\ind In this Chapter we will confront the theoretical predictions of \CCBCC\ with observational data, optimize the window and use the maximum likelihood method to constrain the parameter $\beta$. The Chapter is organized as follows. Section \ref{Sec:Likelihood} presents the analytical description of the likelihood function for parameter estimation from the LG velocity-acceleration comparison. In Section \ref{Sec:Optimize} we apply 2MASS data to the theoretical framework of \CCBCC\ in order to determine the optimal window for the measurement. Section \ref{Sec:Opti.beta} deals with the optimized measurement of the $\beta$ parameter. Finally, in Section \ref{Sec:Opti.Summary} we summarize the Chapter.

\section{Analytical description of the likelihood}
\label{Sec:Likelihood}
Let us come back to the linear-theory Equation (\ref{eq:vLG-gLG}) relating the peculiar velocity of the LG with its acceleration:
\beq\label{eq:vLG-beta-gLG}
\vLG=\beta\,\bmg_\mathrm{_{LG}}\;.
\eeq
For a number of reasons, such as nonlinear effects, observational windows through which the velocity and acceleration of the LG are observed or shot noise {from the edges of the catalog}, this linear relationship is spoiled. Therefore, even if we had a perfect estimator of $\gLG$, in order to measure $\beta$ from this relation we could not simply equate the two dipoles, but a more sophisticated approach would be needed. Commonly adopted is the \textit{maximum likelihood} approach, which provides a way to account for these effects and to compute errors of estimated cosmological parameters. We will now shortly sketch the analytical description of the method. For details see \citec{CiCh04}, denoted hereunder as `\textsc{\textcolor{kolor_cyt}{CC04}}', or \CCBCC. Note that for simplicity we will express both $\vLG$ and $\gLG$ in units of km/s and as the theoretical quantity for the former (and its divergence) we will use the \textit{scaled} velocity $\bmv_\mathrm{sc}\equiv f^{-1}(\Omm)\bmv_\mathrm{obs}$, where $f$ is the growth factor. We will come back to  the `observed' (measured) velocity of the LG when dealing with the expression for the likelihood. Finally, we will often refer to the LG acceleration as `LG gravity' (or simply, `gravity') and skip the subscript `LG' in both gravity and velocity.

Let $p(\bmg,\bmv)$ denote the joint probability distribution function (p.d.f.) for the LG gravity and peculiar velocity. In a Bayesian approach, we ascribe {\it a priori} equal probabilities to values of unknown parameters, which allows us to express their likelihood function, given $\bmv$ and $\bmg$ of the LG, via the p.d.f.\ for $\bmv$ and $\bmg$:
{\begin{equation}
\label{eq:likeli}
\mathcal{L}({\rm param.}|\bmg,\bmv) = p(\bmg,\bmv|\rm param.) \,.
\end{equation}}
\ind It is a standard practice to approximate the p.d.f.\ by a multivariate Gaussian \citepc{Strauss92,Schmoldt99} and this assumption was already made implicitly in Section \ref{Sec:Discussion}. Numerical simulations \citepc{Kofman94,CiCh03} show that non-gaussianity of fully nonlinear $\bmg$ and $\bmv$ is indeed small. This is not surprising since, e.g.\ gravity is an integral of density over a large volume (Eq.~\ref{eq:g.theor}), so the central limit theorem should at least partly be applicable. 

Using statistical isotropy of $\bmg$ and $\bmv$, their joint p.d.f.\ can be simplified to the form:
\beq \label{eq:p.d.f.} 
p(\bmg,\bmv) = \frac{(1 - \varrho^2)^{-3/2}}{(2 \pi)^{3}
\sig_\bmg^{3} \sig_\bmv^{3}} \exp\left[- \frac{x^2 + y^2 - 2\, \varrho\, \mu\,
x\, y}{2(1 - \varrho^2)}\right]
\eeq
\citepc{JVW90,LKH90}, where $\sig_\bmg$ and $\sig_\bmv$ are the {theoretical} r.m.s.\ values of a single Cartesian component of gravity and velocity, respectively. From isotropy, $\sig_\bmg^2 = \lan \bmg\cdot\bmg \ran/3$ and $\sig_\bmv^2 = \lan \bmv\cdot\bmv\ran/3$, where $\lan \cdot \ran$ denote the ensemble averaging. Next, $(\bmx,\bmy) = (\bmg/\sig_\bmg,\bmv/\sig_\bmv)$, and $\mu = \cos\theta$, with $\theta$ being the misalignment angle between $\bmg$ and $\bmv$. Finally, $\varrho$ is the cross-correlation coefficient of $g_m$ with $v_m$, where $g_m$ and $v_m$ denote an arbitrary Cartesian component of respectively $\bmg$ and $\bmv$. This coefficient was already introduced for a specific case in Eq.\ (\ref{eq:rho}). Here we will provide its general form. From isotropy,
\beq \label{eq:err}
\varrho = \frac{\lan \bmg \cdot \bmv \ran}{\lan g^2 \ran^{1/2} \lan v^2
\ran^{1/2}} \,.
\eeq
Also from isotropy,
\beq \label{eq:cross}
\lan x_m\, y_n \ran = \varrho \, \del_{mn} \,,
\eeq
where $\del_{mn}$ denotes the Kronecker delta. In other words, there are no cross-correlations between different spatial components. 

For a given all-sky galaxy survey, the LG gravity is measured effectively through the window of the survey, $W_\bmg$ (cf.\ Subsection \ref{Subsec:Windows}):
\begin{equation}\label{eq:gint}
\bmg = \int \frac{{\rm d}^3 r}{4 \pi} \del(\bmr) W_\bmg(\bmr) \frac{\bmr}{r^3}
\,.
\end{equation}
In contrast, the LG velocity is not estimated from a velocity survey (i.e., from a catalog of peculiar velocities of galaxies), but measured directly from the dipole anisotropy of the CMB. Still, to relate it to theoretical quantities, we write:
\begin{equation}\label{eq:vint}
\bmv = \int \frac{{\rm d}^3 r}{4 \pi} \te(\bmr) W_\bmv(\bmr)
\frac{\bmr}{r^3} \,.
\end{equation}
The quantity $\te \equiv - \nabla \cdot \bmv$ is the minus scaled velocity divergence and we assume that the velocity field is irrotational. Kelvin's circulation theorem assures that the cosmic velocity field is vorticity-free as long as there are no shell crossings.  \textit{N}-body simulations (e.g.\ \citealtc{PiBe99,PuSc09}) show that the vorticity of velocity is small in comparison to its divergence even in the fully nonlinear regime. Thus, similarly to $\bmg$, also $\bmv$ can be expressed as a Coulomb (Newton) integral over its source, i.e.\ the field of the velocity divergence. Here we do {\em not\/} assume that we know the latter from observations, but we know from theory its statistical relation to the density field.  This is sufficient for our purposes in this Chapter. Since $\bmv$ is directly measured from the CMB dipole, the effective velocity window, $W_\bmv$, which we have introduced in Equation~(\ref{eq:vint}), is essentially unity. This was implicitly assumed in Section \ref{Subsec:Theory}. Here we modify slightly this form of the window to reflect the finite size of the LG. We adopt
\begin{equation}\label{eq:W_v}
W_\bmv = \left\{ \begin{array}{ll}
0, & r < r_{\scriptscriptstyle \rm LG} \,, \\
1, & \mbox{otherwise}\,,
\end{array} \right. 
\end{equation}
which has a small-scale cutoff, {$r_{\scriptscriptstyle \rm LG} = 1\Mpch$ \citepc{vdB07,vdM08}.} A specific form of the gravity window, $W_\bmg$, of the 2MASS survey was given in Subsec. \ref{Subsec:Windows}. We will come back to the issue of its optimization later.

In Fourier space, relations~(\ref{eq:gint}) and~(\ref{eq:vint}) read:
\begin{equation}
\label{eq:g_Four}
\bmg_{\bmk}=\frac{i\bmk}{k^2}\delta_{\bmk}\tilde{w}_{\bmg}(k),
\end{equation}
\begin{equation}
\label{eq:v_Four}
\bmv_{\bmk}=\frac{i\bmk}{k^2}\te_{\bmk}\tilde{w}_{\bmv}(k),
\end{equation}
where the subscript $\bmk$ denotes the Fourier transform and $\tilde{w}$ are windows in Fourier space. From relation (\ref{eq:W_Four}) we obtain for the velocity window 
\beq
\tilde{w}_\bmv(k) = j_0(k r_{\scriptscriptstyle \rm LG}) \,,
\label{eq:W_v(k)}
\eeq
where $j_0$ represents the spherical Bessel function of the first kind of order 0.

From equations~(\ref{eq:g_Four}) and~(\ref{eq:v_Four}) we have
\beq \label{eq:g^2}
\lan \bmg \cdot \bmg \ran = \frac{1}{2 \pi^2} \int_0^\infty
\tilde{w}_\bmg^2(k) P(k) \de k \,,
\eeq
and
{\beq
\lan \bmv \cdot \bmv \ran = \frac{1}{2 \pi^2} \int_0^\infty
\tilde{w}_\bmv^2(k) \calR(k) P(k) \de k \,,
\label{eq:v^2}
\eeq
with
\beq\label{eq:calR}  
\calR(k) \equiv \frac{P_{\te}(k)}{P(k)} \,
\eeq
being the ratio of the power spectra of scaled velocity divergence and of density.} In Chapter~\ref{Ch:Growth} we have assumed this ratio to be unity; here, as we want to include possible non-linear effects, we relax this assumption. In particular, we will use the results obtained by \textsc{\textcolor{kolor_cyt}{CC04}} based on numerical simulations. They found the following fit:
\beq\label{eq:pvpgfit_fix}
{\calR}(k) = [1+(7.071k)^4]^{-\alpha} \,,
\eeq
where
\beq 
\alpha = -0.06574 + 0.29195\,\sigma_8 \qquad 
\mathrm{for}~0.3<\sigma_8<1 \,,
\eeq
with $\sig_8$ being the present value of the root-mean-square density contrast of matter fluctuations within a sphere of $8\Mpch$. \textsc{\textcolor{kolor_cyt}{CC04}} argued that the ratio of the power spectra practically does not depend on the background cosmological model. This ratio is unity in the linear regime ($k \ll 1$) but decreases in the nonlinear regime, because the velocity grows slower than it would be expected from the linear approximation.

Furthermore,
\beq\label{eq:g-v}
\lan \bmg \cdot \bmv \ran = \frac{1}{2 \pi^2} \int_0^\infty
\tilde{w}_\bmg(k)\, \tilde{w}_\bmv(k)\, C(k)\, 
P_{\te}^{1/2}(k)\, P^{1/2}(k)\, \de k ,
\eeq
where $C(k)$ is the coherence function (CF), or the correlation coefficient of the Fourier components of the gravity and velocity fields:
\beq\label{eq:coh_def} 
C(k) \equiv \frac{\lan \bmg_\bmk \cdot \bmv_\bmk^\ast \ran}{\lan
|\bmg_\bmk|^2 \ran^{1/2} \lan |\bmv_\bmk|^2 \ran^{1/2}} 
= \frac{\lan \del_\bmk \te_\bmk^\ast \ran}{\lan |\del_\bmk|^2 
\ran^{1/2} \lan |\te_\bmk|^2 \ran^{1/2}}
\eeq
\citepc{Strauss92}. Again, as in the case of the ratio of the power spectra, deviations of this function from unity were neglected in Section \ref{Sec:Discussion}. It is however not constant and, similarly as was in the case of $\calR$, depends on the wavevector $k$ and the $\sig_8$ parameter. This function was analyzed in detail by \citec{ChCi02} and refined by \textsc{\textcolor{kolor_cyt}{CC04}}, where the following fit was found, also based on numerical simulations:
\begin{equation}\label{eq:CF_fit}
C(k) = \left [ 1+(a_0 k - a_2 k^{1.5} + a_1 k^2)^{2.5} \right]^{-0.2},
\end{equation}
with the coefficients given by scaling relations in $\sigma_8$:
\begin{eqnarray}
a_0 &=& 4.908\ \sigma_8^{0.750} \nonumber ,\\
a_1 &=& 2.663\ \sigma_8^{0.734} , \\
a_2 &=& 5.889\ \sigma_8^{0.714} \nonumber .
\end{eqnarray}
The fit was calculated for $k\in \langle 0, 1\rangle \hMpc$ and $\sigma_8 \in \langle 0.1,1\rangle$, with the imposed constraint $C(k=0)=1$. This constraint assures that for sufficiently large, linear scales, the relation between the gravity and the velocity is deterministic and linear (Eq.~\ref{eq:vLG-beta-gLG}). \citec{ChCi02} investigated numerically also the dependence of the CF on $\Omega_m$ and found it to be extremely weak.

Hence, we obtain finally for the cross-correlation coefficient of velocity and gravity:
\beq \label{eq:rho_full} 
\varrho = \frac{\int_0^\infty \tilde{w}_\bmg(k)\, \tilde{w}_\bmv(k)\,
C(k)\, \calR^{1/2}(k)\, P(k)\, \de k}{\left[\int_0^\infty \tilde{w}_\bmg^2(k) \,
P(k) \, \de k\right]^{1/2} \left[\int_0^\infty \tilde{w}_\bmv^2(k) \,
\calR(k)\, P(k)\, \de k\right]^{1/2}} \,.
\eeq
Equations~(\ref{eq:g^2}), (\ref{eq:v^2}) and~(\ref{eq:rho_full}) specify all the parameters (the variances and the cross-correlation coefficient) that determine the joint p.d.f.\ for $\bmg$ and $\bmv$, Equation~(\ref{eq:p.d.f.}), {\em in the absence of observational errors}. The deviation of the cross-correlation coefficient from unity is then due to different windows through which the gravity and the velocity of the LG are measured, and due to nonlinear effects. The latter are described by the two functions: the CF and the ratio of the power spectra. Later on we will describe the corrections to the cross-correlation coefficient if observational errors are non-zero, as is in the case of our analysis. {As in the preceding Chapter, when calculating the theoretical quantities that require the power spectrum of density perturbations, we use the framework of \citec{EH98} with cosmological parameters derived from \emph{WMAP} seven-year observations \citepc{Larson11}: $h=0.71$, $\Omm h^2=0.1335$, $\Omb h^2=0.02258$, $n_s=0.963$ and $\sigma_8=0.801$.}\\
\ind The final quantity necessary to calculate the joint p.d.f.\ of LG velocity and gravity is the observational window of the latter. It was derived and discussed in detail by \CCBCC. In Chapter~\ref{Ch:Growth} we have already presented its form when all the galaxies brighter than some limiting flux $S_\rmmin$ are included. However, as was shown by \CCBCC, in order to optimize the measurements which use the p.d.f., one should remove from the dataset the objects \textit{brighter} than an appropriate minimum magnitude $\Kmin$. The reason for that procedure is that we need to mitigate nonlinear effects as much as possible. For large Fourier-space wavenumber $k$, the coherence function of velocity with gravity (Eq.~\ref{eq:CF_fit}) drops significantly below unity, decreasing the value of the cross-correlation coefficient (Eq.~\ref{eq:rho_full}). Suppressing then the gravity window for large $k$ has almost no effect on the cross-term (which is the numerator of Eq.~\ref{eq:rho_full}), while it decreases the gravity variance, the square root of which appears in the denominator of this equation. This manipulation on the gravity window helps therefore to achieve the best possible correlation between the LG velocity and gravity. However, when one suppresses the gravity window for scales which are linear enough so that the CF is close to unity, one worsens the correlation again. This is so because even for linear fields (CF and the ratio of power spectra equal to unity) the cross-correlation coefficient decreases for increasingly different windows of velocity and gravity. As a result, for some value of $K_{\rm min}$ (or equivalently, a maximum flux $S_{\rm max}\propto10^{-0.4 K_{\rm min}}$){, at which we should cut the sample by removing bright galaxies,} the cross-correlation coefficient will have a maximum.

The observational window of the 2MASS survey including all the galaxies between the maximum flux $S_\rmmax$ and the minimum one $S_\rmmin$ (the latter given now by the completeness of the catalog) is 
\begin{equation}\label{eq:W_r_opti}
W_\bmg(r)=\frac{\intlim^{L_{\rm max}}_{L_{\rm min}} 
{L\,\Phi (L)\, \de L}}{\intlim^{\infty}_{0}{L\,\Phi (L)\, \de L}}
\end{equation}
(eq.\ 47 of \CCBCC) where $L_{\rm max} = 4 \pi r^2 S_{\rm max}$ and $\Phi(L)$ is the luminosity function of galaxies. In Fourier space this translates into 
\beqa \label{eq:optimal window}
\tilde{w}_\bmg(k)  =  \frac{8\pi S_\rmmax}{k\int_0^\infty{L\,\Phi(L)\,\de L}}\intlim_0^\infty{\sin(kr)L_\rmmax\,\Phi(L_\rmmax)\,\de r}  - \qquad \nonumber \\
\qquad -\,  \frac{8\pi S_\rmmin}{k\int_0^\infty{L\,\Phi(L)\,\de L}}\intlim_0^\infty{\sin(kr)L_\rmmin\,\Phi(L_\rmmin)\,\de r} 
\eeqa 
(eq.\ 54 of \CCBCC). Note however that the flux (or gravity) window does not appear in Equation~(\ref{eq:v^2}) for the velocity variance, and in Equation~(\ref{eq:rho_full}) for the cross-correlation coefficient it appears in such a way that its absolute normalization cancels out. In other words, the p.d.f.\ is sensitive only to the {\em shape\/} of the gravity window. 

\CCBCC\ calculated the cross-correlation coefficient as a function of the minimum magnitude $K_{\rm min}$ for the window (\ref{eq:optimal window}) and with the LF taken from \citec{Bell03}. They found that $\varrho$ has a maximum ($\simeq0.97$) for $K_{\rm min}=4.5$ mag, under the assumption of negligible shot noise (from dilute sampling by distant galaxies of the underlying mass density field). Additionally, if the shot noise can be neglected, the minimum of the cross-correlation coefficient was shown to exactly correspond to the minimal expectation value of the misalignment angle between the LG velocity and acceleration. On the other hand, should the shot noise prove to be significant, the optimal value of $K_{\rm min}$ was predicted to increase.

{In order to obtain the final expression for the likelihood of cosmological parameters as a function of observables and theoretical quantities, based on Relation (\ref{eq:likeli}), we must account for observational errors. Let us then assume that for the LG acceleration they are given by a 3D vector, $\bm{\eps}$. The variance of a {\em single\/} spatial component of measured gravity, $\tsig_\bmg^2$, is a sum of the 1D cosmological component, $\sig_{\bmg,c}^2$, and errors, $\eps^2/3$. Because the gravity is inferred from the observations of a galaxian, rather than mass, density field, we must include biasing: $\sig_{\bmg,c}^2 = b^2 \lan g^2 \ran /3$. Thus,
\begin{equation}\label{eq:sigma_g}
\tsig_\bmg^2 = \frac{b^2 \lan g^2 \ran + \eps^2}{3} \,.
\end{equation}}
\ind{As for the measured velocity of the LG, known from the CMB dipole, we follow \textsc{\textcolor{kolor_cyt}{CC04}} and \CCBCC\ and neglect its observational errors for the purpose of our analysis. Its 1D theoretical (cosmic) variance can be then calculated by plugging Eq.\ (\ref{eq:pvpgfit_fix}) into the integral (\ref{eq:v^2}). There is no biasing in this case, but in order to compare it with observables, we return to `physical' variables, inserting the growth factor $f(\Omm)$ squared. The 1D velocity variance is thus a function of $\Omm$:
\beq
\tsig_\bmv^2 = \frac{\Omm^{1.1}\lan v^2 \ran }{3} \,,
\eeq
where ${\lan v^2 \ran}^{1/2}=988\kms$ is constant (depends only on the velocity window and the power spectrum, which are fixed).}

{As was shown by \textsc{\textcolor{kolor_cyt}{CC04}}, including the {observational} errors has the effect of lowering the value of the cross-correlation coefficient. 
The \textit{effective} cross-correlation coefficient will have the form
\begin{equation}\label{eq:rho_prim}
\tro  = \varrho \left(1+ \frac{\eps^2}{b^2 \lan g^2 \ran} \right)^{\! -1/2}
 \end{equation}
(\textsc{\textcolor{kolor_cyt}{CC04}}; \CCBCC), where  $\varrho$ is given by Equation~(\ref{eq:rho_full}).}

{We can now proceed to the expression for the likelihood function. The \textit{observables} (`data') are the measured values of the LG velocity and gravity, $v_{\rm m}$ and $g_{\rm m}$, respectively, as well as $\mu_{\rm m}$, i.e.\ the cosine of the misalignment angle (where $g_{\rm m}$ and $\mu_{\rm m}$ vary when the window is changed). The \textit{model parameters} $\tsig_\bmg$ and $\tro$ depend on the biasing $b$, the gravity errors $\bm{\eps}$ and the gravity window $W_\bmg$; $\tilde{\sigma}_\bmv$ depends only on $\Omm$. Because of the relation $\Omm^{0.55}=\beta\,b$, the minus log-likelihood function can be cast to the following form, depending explicitly on the biasing and $\beta$:}
{\beqa \label{eq:loglike}
-\ln\mathcal{L} = 3 \ln\beta + 3 \ln\left[b\, \tsig_\bmg \, \sig_\bmv (1-\tro^2)^{1/2}\right] +  \qquad \qquad \qquad \qquad \qquad \qquad  \qquad \qquad \quad   \\
 \qquad \qquad \qquad \qquad \qquad +\,  \frac{1}{2(1-\tro^2)} \left( \frac{g_{\rm m}^2}{\tsig_\bmg^2} + \frac{v_{\rm m}^2}{\beta^2\, b^2 \sig_\bmv^2} - \frac{2\, \tro\,\mu_{\rm m}\, g_{\rm m} v_{\rm m}}{\beta \, b \, \tsig_\bmg \,  \sig_\bmv} \right)+ \mathrm{const} \nonumber 
\eeqa
(\textsc{\textcolor{kolor_cyt}{CC04}}; \CCBCC), where $\sig^2_\bmv\equiv{\lan v^2 \ran}/3$ is the theoretical 1D variance of scaled velocity}. In order to obtain the most likely values for particular parameters, one has to find the {minimum} of the function given above with respect to them{, given the observational data in hand}.

\section{In quest for the optimal window}
\label{Sec:Optimize}
We will now apply the 2MASS data and try to find the optimal window for our measurements. The catalog was prepared and analyzed in the same way as described in Chapter~\ref{Ch:Data}. The difference with respect to the analysis of the growth of the flux dipole presented in Chapter~\ref{Ch:Growth} is that now the measured dipole has a constant cutoff at the high-magnitude side, $K_{\rm max}=13.6$~mag, reflecting the completeness of the sample, and it is the minimum magnitude of the sample that is increased:
\begin{equation}\label{eq:dipole opti}
\tilde{\bmd}_{\rm opt}=\mathcal{C}\, \sum_{\Kmin}^{13.6} 10^{-0.4K_i}\hat{\mathbf{r}}_i\;,
\end{equation}
where $\mathcal{C}\simeq 2620\,\kms$ as explained in Section \ref{Ch:Growth}. {The amplitude of this dipole for a given $\Kmin$ should be understood as the `measured gravity' $g_\mrm$ referred to in this Chapter. Note that removing the brightest (in terms of received flux) galaxies from the sample means that the estimated amplitude of the dipole will usually be smaller than the one measured from the whole dataset. However, this loss of signal will be compensated for by effectively `weighting' gravity by its theoretical r.m.s.\ (\ref{eq:g^2}) with a relevant window.}

\begin{figure}[!t]
\centering\includegraphics[angle=270,width=\textwidth]{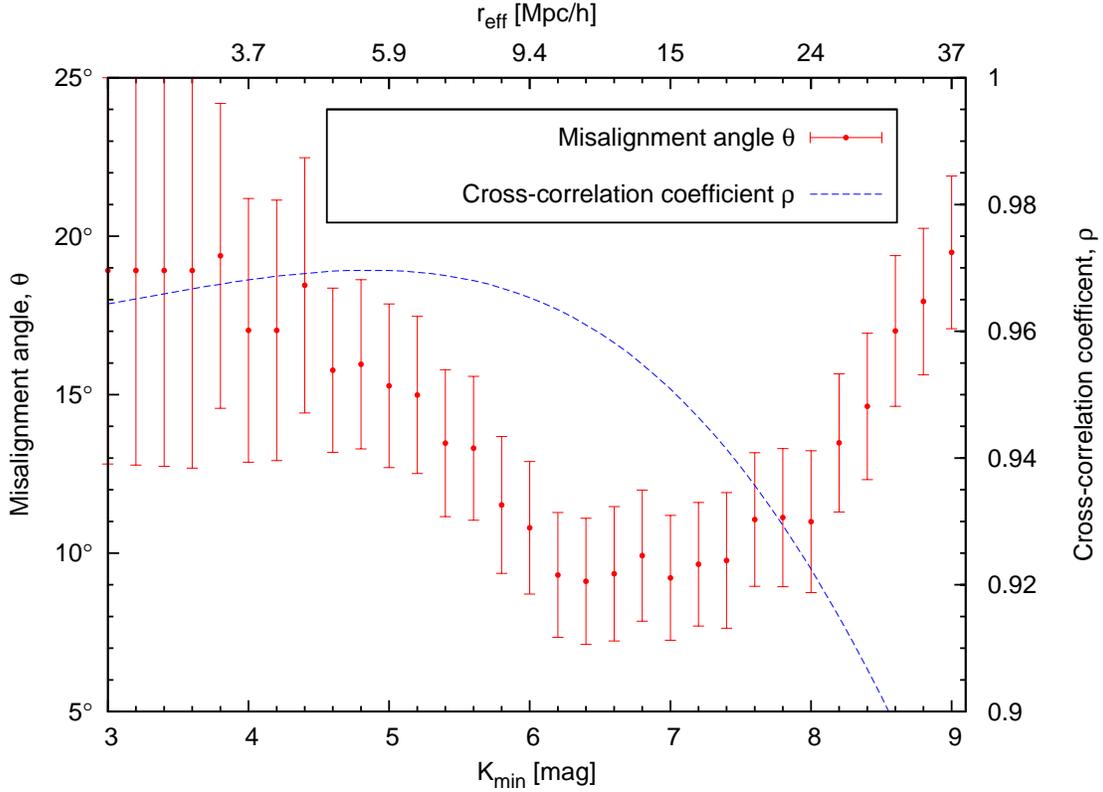}
\caption[Cross-correlation coefficient and the misalignment angle]{\label{Fig:teta_rho}\small Comparison of the theoretically predicted cross-correlation coefficient $\varrho$ between LG gravity and velocity (blue dashed line, right-hand-side scaling) with the observed misalignment angle between LG velocity and the 2MASS clustering dipole (red points with errorbars, left-hand-side scaling), both as a function of increased minimum magnitude of the sample, and for a constant maximum one, $\Kmax=13.6$~mag. {The top axis gives effective cut-off distances related to $\Kmin$.}}
\end{figure}

We will now try to find the optimal value of $\Kmin$ for which we can make the maximum-likelihood estimation of $\beta$. We start by examining the theoretically predicted cross-correlation coefficient, $\varrho$, as given by Eq.\ (\ref{eq:rho_full}){, which would be applicable in the absence of observational errors}. It depends on the shape of the gravity window and we plot it as a function of increased $\Kmin$ in Figure~\ref{Fig:teta_rho}. The cross-correlation coefficient reaches its maximum $\varrho_\rmmax=0.971$ for $\Kmin=5$~mag (at an effective distance of $5.9\Mpch$). This magnitude is somewhat larger then the \mbox{$\Kmin=4.5$~mag} predicted by \CCBCC\ and {in our opinion the main reason for that slight discrepancy is} the form of the luminosity function used to calculate the gravity window (\ref{eq:W_r_opti}). \CCBCC\ used the \citec{Bell03} LF, i.e.\ a \citec{Sche76} function with $\alpha=-0.77$ and $M^*=-23.29+5\log h$, while we take the updated LF by \citec{6dF_Fi}, where $\alpha=-1.16$ and $M^*=-23.83+5\log h$. Note however that our value of $\varrho_\rmmax$ itself is very close to the one {found} by \CCBCC.\\
\ind{\CCBCC\ emphasize that statistically, the maximum of the cross-correlation coefficient corresponds to the minimum of the misalignment angle (m.a.) \textit{provided that} observational errors, such as the shot noise, are negligible. However, owing to a broad p.d.f.\ of the expected angle (cf.\ figure~3 of \CCBCC), the scatter around this relationship is quite significant.} As can be seen in Figure~\ref{Fig:teta_rho}, indeed if we vary the observational window (increase the minimum magnitude of the sample), the m.a.\ gradually decreases from the original $\sim\!19\degr$ to a minimum value. However, the smallest m.a.\ is attained for $\Kmin$ significantly larger than 5~mag. Analysis of the data shows that we have {a relatively wide `plateau' of small values of the angle with $\theta_\rmmin=9\degr\pm2\degr$ for $\Kmin$ around $6.3$~mag ($r_\mathrm{eff}=10.8\Mpch$). Applying such a cutoff  would be equivalent to removing some 50 brightest galaxies from the sample.  The $1\sigma$ errorbars here, as well as in Figure~\ref{Fig:teta_rho} and hereafter, were estimated from 1000-fold bootstrap resampling of the catalog, similarly to what was done in Chapter~\ref{Ch:Growth} (see Appendix \ref{App:Bootstrap} for details)}.\\
\begin{figure}[!t]
\centering\includegraphics[angle=270,width=\textwidth]{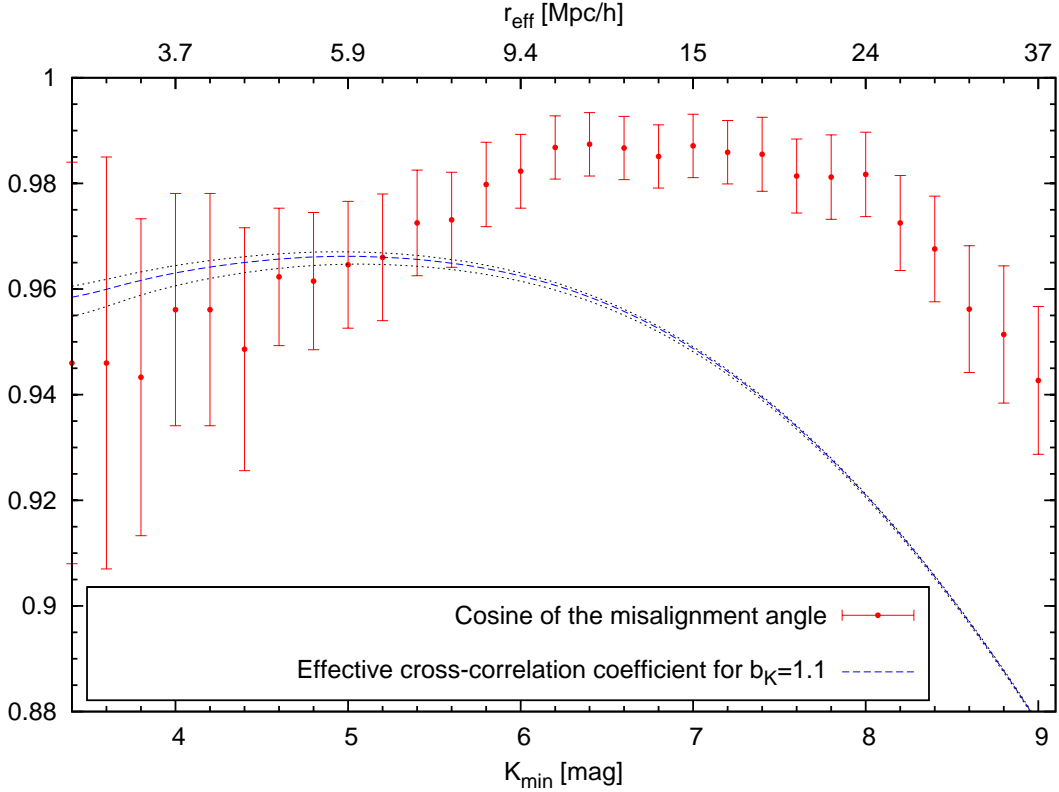}
\caption[Effective cross-correlation coefficient and the cosine of the misalignment angle]{\label{Fig:rhoprim_mi}\small Comparison of the effective cross-correlation coefficient between LG gravity and velocity, $\tro$ (blue dashed line for $b_K=1.1$ and two dotted lines for $b_K=0.9$ and $b_K=1.3${, top and bottom respectively}) with the cosine of the observed misalignment angle between LG velocity and the 2MASS clustering dipole (red points with errorbars), both as a function of increased minimum magnitude of the sample, and for a constant maximum one, $\Kmax=13.6$~mag.{The top axis gives effective cut-off distances related to $\Kmin$.}}
\end{figure}
\ind{The cross-correlation coefficient for $\Kmin=6.3$~mag, where we have the minimum of the misalignment angle, $\theta_\rmmin=8.9\degr$, is $\rho|_{\theta_\rmmin}=0.964$. It is interesting to check how likely it would be in \LCDM\ to observe such a misalignment angle, or smaller, for this value of the cross-correlation coefficient. For that purpose, we follow \CCBCC\ and use the conditional distribution of the m.a.\ with the LG velocity as a constraint \citepc{JVW90,LKH90}. In the small-angle approximation, the conditional m.a.\ is Rayleigh-distributed:
\beq\label{eq:te_approx}
p(\theta|\bmv) \simeq \frac{\theta}{\theta^2_\ast}
\exp\left(- \frac{\theta^2}{2 \theta^2_\ast}\right) , 
\eeq
with the characteristic measure of misalignment
\beq \label{eq:te_ast}
\theta_\ast = \frac{\sqrt{1 - \varrho^2}}{\varrho\, y} \,,
\eeq
where $y\equiv v_{_{\rm LG}}/\tsig_\bmv$. Integrating the distribution (\ref{eq:te_approx}) from $\theta=0\degr$ to $\theta_\rmmin=8.9\degr$ with fixed $\rho=0.964$, we obtain that observing such a misalignment angle, or smaller, at this magnitude cut has more than 56\% probability. For $\Kmin=6.3$~mag, the observed m.a.\ is almost the same as its expectation value:
\beq\label{eq:exp_te|v}
\lan\theta|\bmv\ran = \sqrt{\frac{\pi}{2}} \theta_\ast
\eeq
(\CCBCC), which in this case is equal to $8.74\degr$.}\\
\ind{Coming back to the expression for the log-likelihood, Eq.\ (\ref{eq:loglike}), in order to calculate $\beta$, we will need to apply the \textit{effective} cross-correlation coefficient, $\tro$. It includes gravity errors (Eq.\ \ref{eq:rho_prim}), but depends also on the r.m.s.\ of gravity (hence its observational window), as well as on the biasing parameter, (although very weakly)}. These facts are illustrated in Figure~\ref{Fig:rhoprim_mi}. The blue curve shows $\tro$ for $b_K=1.1$. The lower and upper dotted ones are respectively for $b_K=0.9$ and $b_K=1.3$. These values span the $1\sigma$ interval of $b_K$ as determined by \citec{Maller05}. As can be seen, the value of biasing has {a very small} influence both on the position and on the value of $\tro_\rmmax$. {What is more}, there is a wide range of $\Kmin$ for which the effective cross-correlation coefficient is virtually constant (for a given $b_K$) {and maximal}. For example, if we take  $b_K=1.1$, we have {$\tro_\rmmax=0.967$ for $\Kmin\in\langle4.8,5.2\rangle$}.  Situation is similar for other reasonable values of $b_K$. {The effective cross-correlation coefficient is slightly lowered with respect to the theoretical one, although its maximum is not shifted towards bigger $\Kmin$.}  In the plot we also show the cosine of the measured misalignment angle, $\mu_\mrm=\cos\theta_\mrm$, as this is the quantity that enters into the expression for the likelihood, Eq.\ (\ref{eq:loglike}). {We can see that as was in the case of the theoretical cross-correlation coefficient, the $\Kmin$, for which the \textit{effective} coefficient attains its maximum, is also below $6.3$~mag, where $\mu_\mrm$ is the biggest. We cannot thus unambiguously select a single window that would be optimal, based on this comparison alone.} However, already at this stage we can {support the claims} of \CCBCC\ that $\Kmin=8$~mag is too big a magnitude at which one should limit the sample from below to optimize the window. Such a cutoff was tried by \citec{Maller03}, who claimed that removing the 375 brightest galaxies from their sample with $K<8$~mag resulted in a substantial decrease in the misalignment, to about $5.2\degr$. This was then tested by \citec{Erdogdu06a} from 2MRS data with no confirmation. We cannot confirm this result from our data either. If we remove the sources brighter than $K=8$~mag (453 galaxies in our case\footnote{This number is slightly different from the 375 given by \citec{Maller03}, but note that it includes LGA galaxies and also the `clones' in the ZoA.}), the misalignment angle is equal to $\sim\!11\degr$, which is indeed less than the $\sim\!19\degr$ obtained from the whole sample, but still significantly larger than $5.2\degr$. What is even more important for the likelihood analysis, both the `pure' cross-correlation coefficient and the effective one are for $\Kmin=8$~mag {several per cent below} their maximum values. {These two effects will translate into the instable behavior of $\beta$ estimated for such minimum-magnitude cuts. We present it in the following Section.}

\section{Precise measurement of the $\bm\beta$ parameter}
\label{Sec:Opti.beta}
The analysis presented in the previous Section did not provide a single value of $\Kmin$ for which the window (\ref{eq:W_r_opti}) {may be unambiguously claimed} optimal. We can however propose $\Kmin\simeq6$~mag as a `reasonable' value {here} (the misalignment angle is close to its minimum and the effective cross-correlation coefficient is almost maximal). Such a minimum-magnitude cut {would mean} removing the 35 brightest galaxies from the sample {and is equivalent to an effective distance of $9.4\Mpch$}. {For the interested reader, the properties of these galaxies, such as their near-infrared magnitudes, morphological types, distance estimates and radial velocities, can be found in Appendix \ref{App:Brightest}. As can be seen in Table \ref{Tab:brightest}, most of them are located at a distance smaller than $\sim\!11\Mpc$ ($\simeq8\Mpch$), so indeed removing them from the sample is roughly equivalent to including in the analysis only the sufficiently distant ones, thus mitigating large non-linear influences from those nearby.}

{We can also calculate the $\beta$ parameter for different cutoffs of the window. We should expect $\beta$ to stabilize for an interval of minimum magnitudes, in a similar way what was observed for both $\tilde\varrho$ and $\mu_\mrm$}. Let us then come back to the expression for the logarithmic likelihood, Eq.\ (\ref{eq:loglike}). We calculate its partial derivative with respect to $\beta$ and equate it to zero, to find the maximum of $\ln\calL$. This gives the following quadratic equation (\textsc{\textcolor{kolor_cyt}{CC04}}; \CCBCC):
\begin{equation}
\label{eq:quadr beta}
3(1-\tro^2)\,\beta^2 + \frac{\tro\,\mu_\mrm\, g_\mrm v_\mrm}{b \,\tsig_\bmg\, \sig_\bmv}\, \beta - \frac{v_{\rm m}^2}{b^{2\,} \sig_\bmv^2}
= 0 \,.
\end{equation}
The relevant solution for $\beta$ has the form:
\beq
\label{eq:beta for b}
\bmle=\frac{v_\mrm}{6 (1-\tro^2)\, b\, \sig_\bmv\, \tsig_\bmg} \left(\sqrt{\tro^2 \mu_\mrm^2\, g_\mrm^2 + 12(1-\tro^2)\tsig_\bmg^2}-\tro\, \mu_\mrm\, g_\mrm \right)\;,
\eeq
{where the effective cross-correlation coefficient, $\tro$, is given by Eq.\ (\ref{eq:rho_prim}) and the 1D variance of gravity $\tsig_\bmg$ is calculated via Eq.\ (\ref{eq:sigma_g}).}

\begin{figure}[!t]
\centering\includegraphics[angle=270,width=\textwidth]{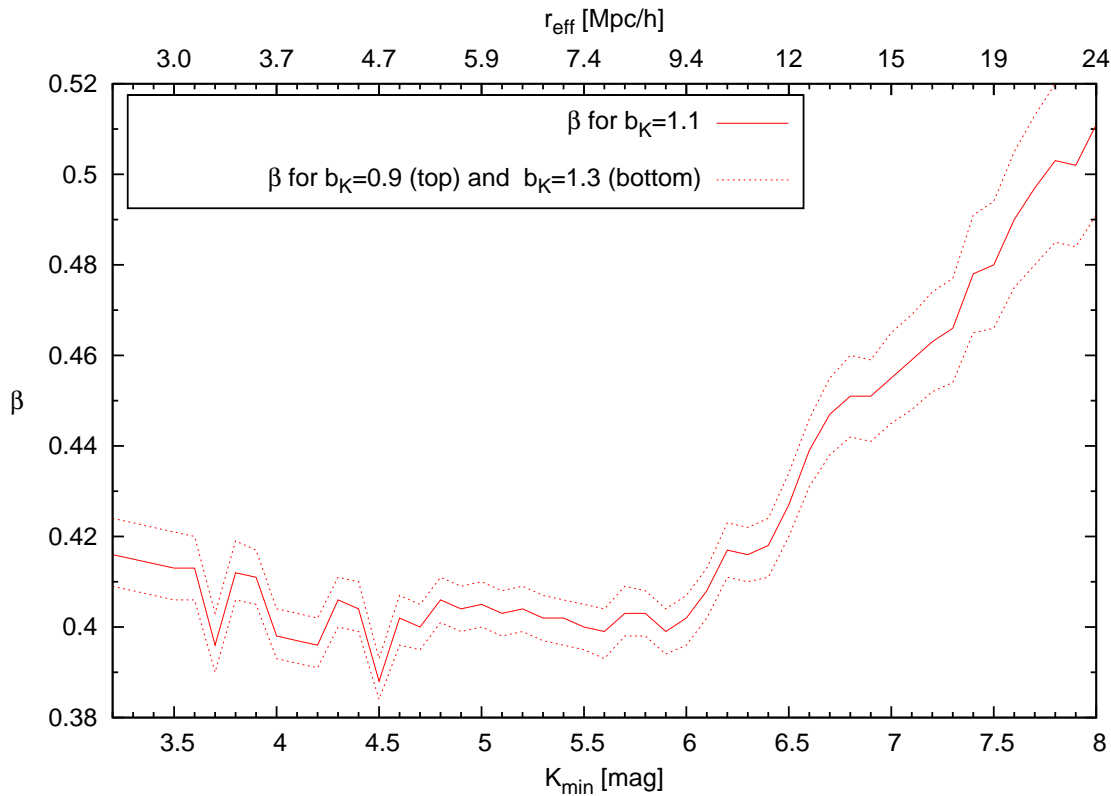}
\caption[Maximum-likelihood estimate of the $\beta$ parameter]{\label{Fig:MLE beta}\small Maximum-likelihood estimate of the $\beta$ parameter for three values of $b_K=0.9,1.1,1.3$, from top to bottom, as a function of increased minimum magnitude of the 2MASS sample, and for a constant maximum one, $\Kmax=13.6$~mag. {Note only slight variations of $\beta$ for $4.6\leq\Kmin\leq6.0$.}}
\end{figure}
 
The important thing to note is that the dependence of $\bmle$ on $b$ is more complicated (and much weaker) than in the case of the `simple' $\beta\propto b^{-1}$ relation, because both $\tro$ and $\tsig_\bmg$ are functions of $b$. This is readily visible in Figure~\ref{Fig:MLE beta}, which shows how $\bmle$ changes as a function of increased minimum magnitude of the sample, for three values of the $K$-band biasing parameter ($b_K=0.9,1.1,1.3$, from top to bottom). We can see that for $b_K$ within $1\sigma$ confidence interval from \citealtc{Maller05}, {i.e.\ changing by over 40\%,} $\bmle$ varies by no more than $\sim\!6\%$ (for every given $\Kmin$). What is more, for a fixed $b_K$, there is a wide range of $\Kmin$ for which $\bmle$ is almost constant, as expected. For instance, if $b_K=1.1$, then $0.399\leq\bmle\leq0.408$ for $\Kmin\in\langle4.6,6.1\rangle$. More generally, for parameter space $b_K\in\langle0.9,1.3\rangle$ and $\Kmin\in\langle4.6,6.1\rangle$, we have $0.393\leq\bmle\leq0.413$. In order to account for possible systematic errors, such as those dealt with in Chapters~\ref{Ch:Nonlin} and \ref{Ch:Data}, we take conservative $1\sigma$ confidence intervals for our measurement and propose the following estimate of the $\beta$ parameter based on this analysis:
\beq\label{eq:beta.opti}
\beta=0.40\pm0.02\;.
\eeq
This value is slightly larger than, but consistent within the errors, with the one obtained from the analysis of the growth of the 2MASS dipole, presented in the previous Chapter. It is also in agreement with already mentioned results of the other authors who equally used 2MASS data: $\beta=0.40\pm0.09$ \citepc{Erdogdu06a} and $\beta=0.35\pm0.10$ \citepc{NBD11.GR}{, but has much better precision. We also note that repeating our analysis for `2MRS-like' data, i.e.\ setting the constant maximum magnitude of the sample at $\Kmax=11.75$~mag, gives $\beta=0.43\pm0.04$ -- a value consistent with those mentioned here, but with twice as big errors as our measurement from the whole sample. This increase in errors is caused by weaker correlation of the LG velocity and gravity if the maximum-magnitude cutoff is diminished to mimic the shallower extent of the 2MRS dataset, which results in less stable behavior of $\bmle$.}

{If we now use our value of $\beta$ to estimate the parameter $\Omm=(\beta\, b_K)^{1/0.55}$, we get from (\ref{eq:beta.opti}) 
\beq
\Omm=b_K^{1.82}\times(0.19\pm0.02)\;.
\eeq
Finally, we can also constrain $\Omm$ alone, using $b_K=1.1\pm0.2$ \citepc{Maller05}, to obtain
\beq
\Omm=0.225\pm0.077\;,
\eeq
where, as in Chapter~\ref{Ch:Growth}, such a big $1\sigma$ error comes mostly from the uncertainty in the biasing parameter. This value of the density parameter is in good agreement with the estimates mentioned in Section \ref{Sec:Gr.SummConcl}.}

\section{Summary and conclusions}\label{Sec:Opti.Summary}
{This Chapter presented a `modern' approach to the measurement of the $\beta$ parameter from the comparison of the LG velocity and acceleration. We used the same Two Micron All Sky Survey data as in Chapter~\ref{Ch:Growth} and took advantage of the fact that the growth of the 2MASS dipole is consistent with the \LCDM\ model. This allowed us to focus only on the optimization of our measurement, for which we used the maximum-likelihood (MLE) approach. Application of this method for this problem is not a new idea; it has been used by various authors and in different variants for two decades now \citepc{Strauss92,Schmoldt99,CiCh04}. However, only recently a really \textit{precise} measurement of the $\beta$ parameter has become possible, thanks to new datasets, such as the 2MASS XSC.}

{In our search for the optimal window, we followed the considerations of \citec{CCBCC} and reduced the galaxy sample by removing the sources with the largest observed fluxes in the $K$ band. We confirmed their predictions that the misalignment angle $\theta$ between the LG velocity and the 2MASS dipole should decrease to a minimum value for a certain cutoff magnitude $\Kmin$. From our analysis we found that $\theta_\rmmin=9\degr\pm2\degr$ for $\Kmin=6.3$~mag. However, we have not managed to select a \textit{single} optimal window for the MLE of $\beta$. In particular, we have shown that the minimum of the misalignment angle is not observed for the same magnitude cut as is the maximum of the \textit{effective} cross-correlation coefficient between the two vectors -- contrarily to the theoretical expectations of \CCBCC, which were however strictly valid only in the absence of observational errors. The latter (estimated here via bootstrap resampling) are not very large
, but still influence (slightly reduce) the cross-correlation coefficient.}\\
\ind{Although we do not select a single value of $\Kmin$ at which the sample should be limited to obtain an optimal measurement, we can still determine the $\beta$ parameter with excellent precision. Basing on the formula first derived by \textsc{\textcolor{kolor_cyt}{CC04}} for the likelihood function of cosmological parameters given $\vLG$ and $\gLG$, we presented the equation for the maximum-likelihood estimate of $\beta$ and argued that it depends very weakly on the biasing parameter $b$. By changing the observational window, we have observed very stable behavior of $\bmle$ for a wide range of $\Kmin$ and $b_K$. Our measurement is $\beta=0.40\pm0.02$, which is consistent with the result presented in the previous Chapter and twice as precise, and leads to the constraint on the density parameter $\Omm=0.23\pm0.08$.}\\
\ind{Such small errors in the $\beta$ parameter measured via the maximum-likelihood method are a great advancement when compared to older estimates, that used the data obtained by the IRAS satellite. Despite a rigorous statistical approach, neither \citec{Strauss92}, \citec{Schmoldt99} nor even \textsc{\textcolor{kolor_cyt}{CC04}} (who made the most precise of these measurements) were able to reduce relative errorbars in $\beta$ to less then $\sim\!30$\%. The main reason was much sparser sampling of the IRAS dataset when compared to 2MASS XSC, as well as the fact that the galaxies shining bright in the far infrared (wavelengths of IRAS) are much more biased tracers of stellar and total mass than those from the catalog we used. In fact, the choice of the near-infrared bands for the survey that later became 2MASS had been partly motivated by the fact that the distribution of its galaxies was expected to follow the one of the underlying density field, as apparently is the case \citepc{Westover,Davis11}. Adding to that the relatively dense sampling and unprecedented depth as for an all-sky survey, we have a catalog that is a great tool for precision cosmology, as we have shown in this and the preceding Chapter.}

\chapter{\textsf{\textbf{Conclusions and future prospects}}}\label{Ch:Conclusions}
{In this Thesis, we presented two approaches to measure the cosmological parameter $\beta\equiv\Omm^{0.55}/b$, where $\Omm$ is the mean matter density of the Universe and $b$ is linear biasing. Both our methods relied on the comparison between the clustering dipole of the galaxies from the 2MASS Extended Source Catalog and the peculiar velocity of the Local Group. Before describing the two approaches and the results, we started in Chapter~\ref{Ch:Nonlin} by studying a particular non-linear effect, related to the fact that the nearby Local Void is partially hidden behind the Zone of Avoidance. We showed that even random filling of this region and elongation of the void along the line of sight should not bring crucial systematics into the determination of the amplitude and direction of the dipole. Next, in Chapter~\ref{Ch:Data} we presented the 2MASS data that we used and explained in detail how we handled them for the purpose of the analyses that followed. Of particular importance was the statistical criterion to remove Milky Way sources from the sample, which -- if kept -- could seriously bias the measurement. }

{Next two Chapters dealt with the application of the data. In Chapter~\ref{Ch:Growth} we studied how the 2MASS clustering dipole grows when the sample depth is increased. By comparing the observations to the theoretical expectations of the \LCDM\ cosmological model, we showed that once the proper window of the survey is included, the observed growth is consistent with the theoretical, conditional one (using the fact that the velocity of the Local Group is known independently). This comparison requires rescaling the datapoints by the $\beta$ parameter, which provided us with a straightforward method to estimate it. The result was $\beta=0.38\pm0.04$, where these 1$\sigma$ errors included such systematics as those dealt with in Chapters~\ref{Ch:Nonlin} and \ref{Ch:Data}.}\\
\ind{Finally, in Chapter~\ref{Ch:MLE} we refined our measurement of $\beta$. We used the maximum-likelihood machinery developed for this purpose first in the context of IRAS data by \citec{Strauss92,Schmoldt99} and \citec{CiCh04} and then adjusted for the 2MASS case by \citec{CCBCC}. We showed that removing from the sample the $\sim\!50$ galaxies brighter than $\Kmin=6.3$~mag allows to reduce the misalignment angle between the LG gravity and velocity to $\sim\!9\degr$. However, the cross-correlation coefficient between the two vectors, even when corrected for observational errors, has its maximum for a significantly smaller $\Kmin$, which precluded us from determining a \textit{single} optimal window for the maximum-likelihood estimate. Still, for a wide range of minimum magnitude cutoffs we noted very stable behavior of $\beta$ calculated within this framework, which allowed us to evaluate this parameter with double the precision of the method presented in Chapter~\ref{Ch:Growth}. The final result is $\beta=0.40\pm0.02$, which has much smaller errorbars than not only the older maximum-likelihood measurements that used the IRAS data, but also when compared to more recent estimates for which 2MASS data were used, with various methods.}\\
\ind{Statistical precision (small errorbars) of a measurement is not necessarily tantamount to its \textit{accuracy}, i.e.\ the ability to reproduce the real value of the estimated quantity. Unknown systematic errors can bias the estimate without the researcher(s) being aware of that fact. It seems however that our results are both precise and accurate, at least to a certain extent, as they are consistent with separate and independent analyses of a subset of the 2MASS catalog, namely the 2MASS Redshift Survey. Both \citec{Erdogdu06a} and \citec{NBD11.GR} found values of $\beta$ that agree within the errors with those presented here. Additionally, our results are roughly compatible with the findings of \citec{PH05} and \citec{Davis11}, where apart from the 2MASS data, also peculiar velocity catalogs where used to compare the density and velocity fields.}

{There is of course still room for improvement. First of all, it might be worthwhile to confront our findings with computer simulations. Generating thousands of mock `2MASS-like' catalogs and repeating the same measurements on them as we did here could help to validate our estimates of the errorbars, especially that simulated data would not suffer from such biases as the Zone of Avoidance or noise induced by Galactic sources that were not removed by our filter. Simulations could also be useful to verify how much our results may be biased due to the presence of nearby non-linear structures hidden behind the ZoA, other than the Local Void (such as the Norma Cluster, part of the Great Attractor, e.g.\ \citealtc{Norma}).}

{Progress is also possible once new data become available. Firstly, an all-sky \textit{redshift} catalog with a depth and coverage comparable to those of the 2MASS XSC would allow to mitigate the effects related to flux-limiting the dataset, by using the selection function which one cannot calculate not having galaxy redshifts in hand. Unfortunately, there is currently no observational project aiming at creating such a catalog in the near future, i.e.\ within next several years. There may be however a way to overcome this problem, by estimating so-called \textit{photometric} redshifts for all those galaxies in the 2MASS XSC that do not have spectroscopic ones measured. The method bases on correlating redshifts with other properties of galaxies, such as colors in different bands, morphologies etc. The literature dealing with relevant methods is vast and is beyond the scope of this work. Anyway, a future photometric-redshift all-sky catalog based on 2MASS would be a great tool not only for such analyses as those presented here, but also for studies of galaxy bulk flows or the cosmography of the `local' Universe, which still holds some secrets despite immense progress in extragalactic science  we are now witnessing. We hope that more and more of these secrets will be revealed in the coming years.}

\appendix
{\chapter{\textsf{\textbf{Data of the brightest 2MASS galaxies}}}
\label{App:Brightest}
The following Table \ref{Tab:brightest} presents the data of the 50 brightest galaxies in our sample, beyond the Local Group. The list includes 2MASS XSC galaxies with Galactic latitude $|b|>5\degr$ and the 3 brightest LGA objects, which are located in the ZoA. The parameters given are the following: (1) name of the galaxy; (2) Galactic longitude $l$ and latitude $b$; (3) radial velocity in the Local Group frame and (4) redshift-independent distance estimate, both taken from the NASA/IPAC Extragalactic Database (NED); (5) morphological type, as given by SIMBAD; (6)  extinction-corrected $20\,\mathrm{mag\slash{}sq.''}$ isophotal fiducial elliptical aperture magnitudes ($J$, $H$ and $K_s$ bands), without the offset of $\Delta=-0.2$~mag that should be applied to compensate for lost flux (see Section \ref{Sec:Data.prep}). Note that the distance estimate given here in column (4) is the `mean' taken from NED, which is an unweighted average of several independent measurements that may vary significantly depending on the method. We thus do not quote the errors, which formally may be equal to several dozen per cent. For details see relevant NED entries.}
\newpage
\begin{center}
\begin{small}
\begin{longtable}{l c c c c c c c c}
\caption{ {First fifty 2MASS galaxies with the smallest $K_S$-band apparent magnitudes. Galaxies marked with an asterisk were added from the Large Galaxy Atlas.}} \label{Tab:brightest}  \\

\multicolumn{1}{l}{(1)}	&	\multicolumn{2}{c}{(2)}	&	\multicolumn{1}{c}{(3)}	&	\multicolumn{1}{c}{(4)}	&	\multicolumn{1}{c}{(5)}	&	\multicolumn{3}{c}{(6)}	\\

\multicolumn{1}{l}{name}	&	\multicolumn{1}{c}{$l$ [$\degr$]}	&	\multicolumn{1}{c}{$b$ [$\degr$]}	&	\multicolumn{1}{c}{$v_{_\mathrm{LGf}}$ [km/s]}	&	\multicolumn{1}{c}{$d$ [Mpc]}	&	\multicolumn{1}{c}{type}	&	\multicolumn{1}{c}{J [mag]}	&	\multicolumn{1}{c}{H [mag]}	&	\multicolumn{1}{c}{K [mag]}	 \\ \hline

\endfirsthead

\multicolumn{9}{c}%
{\small{\tablename\ \thetable{} -- continued from previous page}} \\

\multicolumn{1}{l}{(1)}	&	\multicolumn{2}{c}{(2)}	&	\multicolumn{1}{c}{(3)}	&	\multicolumn{1}{c}{(4)}	&	\multicolumn{1}{c}{(5)}	&	\multicolumn{3}{c}{(6)}	\\

 \multicolumn{1}{l}{name}	&	\multicolumn{1}{c}{$l$ [$\degr$]}	&	\multicolumn{1}{c}{$b$ [$\degr$]}	&	\multicolumn{1}{c}{$v_{_\mathrm{LGf}}$ [km/s]}	&	\multicolumn{1}{c}{$d$ [Mpc]}	&	\multicolumn{1}{c}{type}	&	\multicolumn{1}{c}{J [mag]}	&	\multicolumn{1}{c}{H [mag]}	&	\multicolumn{1}{c}{K [mag]}	 \\ \hline
\endhead

\hline \multicolumn{9}{r}{\small{...continued on next page}} \\
\endfoot

\hline \hline
\endlastfoot

NGC~253	&	97.367	&	-87.965	&	276$\pm$3	&	3.2	&	Sc	&	4.857	&	4.132	&	3.815 \\ 
M~81	&	142.092	&	40.9	&	108$\pm$9	&	3.7	&	Sb	&	4.782	&	4.132	&	3.897 \\ 
Centaurus~A	&	309.517	&	19.418	&	301$\pm$16	&	3.7	&	E	&	4.927	&	4.246	&	3.947 \\ 
Maffei~1*	&	135.862	&	-0.551	&	298$\pm$15	&	3.5	&	E	&	4.791	&	4.235	&	4.128 \\ 
NGC~4945	&	305.272	&	13.34	&	299$\pm$16	&	4.0	&	Scr	&	5.503	&	4.792	&	4.468 \\ 
Maffei~2*	&	136.498	&	-0.326	&	212$\pm$15	&	3.2	&	Sbc	&	5.262	&	4.721	&	4.606 \\ 
M~82	&	141.409	&	40.567	&	348$\pm$10	&	3.9	&	I	&	5.738	&	5.004	&	4.633 \\ 
IC~342	&	138.173	&	10.58	&	245$\pm$13	&	3.1	&	Scd	&	5.479	&	4.96	&	4.674 \\ 
M~83	&	314.584	&	31.973	&	301$\pm$13	&	4.7	&	SBc	&	5.593	&	4.952	&	4.721 \\ 
Circinus*	&	311.326	&	-3.808	&	189$\pm$15	&	4.2	&	Sb	&	5.193	&	4.701	&	4.790 \\ 
M~104	&	298.461	&	51.149	&	828$\pm$13	&	10.5	&	Sa	&	5.896	&	5.229	&	4.99 \\ 
M~94	&	123.362	&	76.008	&	352$\pm$3	&	5.0	&	Sb	&	6.068	&	5.408	&	5.163 \\ 
M~64	&	315.683	&	84.423	&	364$\pm$5	&	5.4	&	Sb	&	6.299	&	5.623	&	5.381 \\ 
NGC~6946	&	95.719	&	11.673	&	344$\pm$18	&	6.5	&	Scd	&	6.138	&	5.926	&	5.42 \\ 
M~49	&	286.923	&	70.196	&	873$\pm$10	&	16.1	&	E	&	6.369	&	5.732	&	5.498 \\ 
M~51	&	104.851	&	68.561	&	692$\pm$6	&	8.0	&	Sc	&	6.486	&	5.797	&	5.588 \\ 
M~106	&	138.319	&	68.842	&	507$\pm$5	&	7.5	&	Sbc	&	6.498	&	5.832	&	5.592 \\ 
NGC~1316	&	240.163	&	-56.69	&	1633$\pm$1	&	19.0	&	Sa	&	6.546	&	5.947	&	5.68 \\ 
M~63	&	105.997	&	74.288	&	546$\pm$4	&	9.0	&	Sbc	&	6.681	&	5.947	&	5.722 \\ 
M~77	&	172.104	&	-51.934	&	1191$\pm$4	&	12.3	&	S	&	6.984	&	6.267	&	5.8 \\ 
M~60	&	295.874	&	74.318	&	1017$\pm$8	&	16.5	&	S0	&	6.739	&	6.065	&	5.815 \\ 
NGC~1269	&	247.524	&	-57.042	&	703$\pm$8	&	8.6	&	SBa	&	6.73	&	6.094	&	5.847 \\ 
NGC~3521	&	255.532	&	52.829	&	602$\pm$12	&	12.2	&	Sbc	&	6.779	&	6.095	&	5.847 \\ 
M~87	&	283.778	&	74.491	&	1203$\pm$9	&	16.7	&	E	&	6.806	&	6.145	&	5.896 \\ 
NGC~3115	&	247.783	&	36.781	&	422$\pm$15	&	10.1	&	S0	&	6.808	&	6.148	&	5.92 \\ 
M~66	&	241.962	&	64.419	&	590$\pm$9	&	10.1	&	Sb	&	6.881	&	6.196	&	5.939 \\ 
M~101	&	102.037	&	59.771	&	379$\pm$9	&	7.2	&	Sc	&	6.893	&	6.204	&	5.94 \\ 
NGC~891	&	140.382	&	-17.415	&	738$\pm$13	&	10.2	&	Sb	&	7.277	&	6.397	&	5.97 \\ 
IC~356	&	138.463	&	13.11	&	1105$\pm$13	&	18.9	&	Sbc	&	6.891	&	6.196	&	5.99 \\ 
NGC~7331	&	93.722	&	-20.724	&	1119$\pm$18	&	14.2	&	Sb	&	7.075	&	6.334	&	6.08 \\ 
NGC~2903	&	208.712	&	44.54	&	443$\pm$7	&	9.4	&	Sb/Sc	&	7.002	&	6.376	&	6.089 \\ 
M~65	&	241.332	&	64.222	&	671$\pm$9	&	12.6	&	Sa	&	7.018	&	6.343	&	6.104 \\ 
NGC~4565	&	230.761	&	86.438	&	1194$\pm$6	&	13.3	&	Sb	&	7.224	&	6.408	&	6.124 \\ 
NGC~2841	&	166.942	&	44.151	&	685$\pm$4	&	17.8	&	Sb	&	7.083	&	6.381	&	6.152 \\ 
NGC~6744	&	332.224	&	-26.147	&	706$\pm$8	&	9.5	&	S	&	6.97	&	6.434	&	6.181 \\ 
NGC~3628	&	240.852	&	64.781	&	709$\pm$8	&	12.2	&	Sbc	&	7.351	&	6.528	&	6.211 \\ 
M~85	&	267.713	&	79.237	&	649$\pm$5	&	17.4	&	S0	&	7.144	&	6.496	&	6.249 \\ 
NGC~2974	&	239.512	&	35.014	&	1691$\pm$19	&	25.3	&	E	&	7.208	&	6.847	&	6.249 \\ 
M~86	&	279.084	&	74.637	&	-348$\pm$8	&	16.3	&	S0	&	7.164	&	6.489	&	6.272 \\ 
NGC~1023	&	145.023	&	-19.089	&	828$\pm$12	&	11.8	&	SB0	&	7.206	&	6.547	&	6.31 \\ 
NGC~2784	&	251.971	&	16.354	&	402$\pm$39	&	8.5	&	S0	&	7.205	&	6.539	&	6.317 \\ 
M~88	&	282.33	&	76.508	&	2187$\pm$6	&	19.5	&	Sbc	&	7.256	&	6.604	&	6.32 \\ 
M~84	&	278.205	&	74.478	&	955$\pm$9	&	17.4	&	E	&	7.224	&	6.564	&	6.332 \\ 
NGC~1553	&	265.631	&	-43.691	&	869$\pm$17	&	15.6	&	S0	&	7.245	&	6.57	&	6.345 \\ 
M~105	&	233.49	&	57.633	&	762$\pm$9	&	10.5	&	E	&	7.246	&	6.573	&	6.353 \\ 
NGC~2683	&	190.455	&	38.761	&	365$\pm$3	&	10.2	&	Sb	&	7.349	&	6.645	&	6.375 \\ 
NGC~4725	&	295.085	&	88.357	&	1176$\pm$4	&	13.6	&	SBb/Sb	&	7.341	&	6.55	&	6.38 \\ 
NGC~5195	&	104.881	&	68.488	&	558$\pm$11	&	(=M51)	&	SB0	&	7.329	&	6.62	&	6.389 \\ 
M~96	&	234.435	&	57.011	&	744$\pm$10	&	10.9	&	Sab	&	7.323	&	6.634	&	6.394 \\ 
NGC~1097	&	226.915	&	-64.68	&	1198$\pm$5	&	16.8	&	SBbc	&	7.372	&	6.702	&	6.429 \\ \hline
\end{longtable}
\end{small}
\end{center}


\chapter{\textsf{\textbf{Effective distance for a given flux}}}
\label{App:r_eff}
In this Appendix we show how we calculate the \textit{effective distance of galaxies with a given flux} $S$, knowing their luminosity function (LF) in the given band, $\Phi(L)$, but not knowing their redshifts. Let us start by deriving the \emph{effective depth} of a flux-limited sample.

The number of galaxies $\de\mathcal{N}$ in a volume element $\de V=4 \,\pi\,r^2\, \de r$ of a spherical sample limited by minimum flux $S_\rmmin$ (equivalent in the $K$ band to some limiting magnitude $\Kmax$) is given by
\begin{equation}\label{eq:mathcalN}
\de\mathcal{N}=4 \,\pi\,r^2\, \de r\intlim_{L_\rmmin}^{+\infty}\Phi(L)\de L\;,
\end{equation}
where $L_\rmmin=4 \,\pi\,r^2\,S_\rmmin$ is the minimum luminosity of galaxies in the sample at a given distance $r$. The \textit{mean depth} of the sample is defined as
\begin{equation}
\langle R \rangle=\frac{\intlim_0^{+\infty}{r\, \frac{\de\mathcal{N}}{\de r}}\,\de r}{\intlim_0^{+\infty}{ \frac{\de\mathcal{N}}{\de r}}\,\de r}\;,
\end{equation}
which, for the Schechter LF \citepc{Sche76} with a faint-end slope $\alpha$ and characteristic luminosity $L_*$, gives
\begin{equation}\label{eq:mean.depth}
\langle R \rangle=\frac{\intlim_0^{+\infty}r^3\, \Gamma(1+\alpha,4 \,\pi\,r^2\, S_\rmmin\slash L_*)\,\de r}{\intlim_0^{+\infty}r^2\, \Gamma(1+\alpha,4 \,\pi\,r^2\, S_\rmmin\slash L_*)\,\de r}\;,
\end{equation}
where $\Gamma(a,x)$ is the upper incomplete Gamma function.

Using the distribution (\ref{eq:mathcalN}), we can also easily derive the \emph{median depth} of a flux-limited sample, which we shall denote as $\bar{R}$. We obtain it by solving the integral equation:
\begin{equation}\label{eq:median.depth}
\intlim_0^{\bar{R}}\frac{\de\mathcal{N}}{\de r}\,\de r=\intlim_{\bar{R}}^{+\infty} \frac{\de\mathcal{N}}{\de r}\,\de r\;.
\end{equation}

The following Table \ref{Tab:depth} presents the mean and median depth of the 2MRS ($\Kmax=11.75$~mag) and 2MASS XSC ($\Kmax=13.6$~mag) for the LF as given by \citec{6dF_Fi}. The difference between the mean and the median is in that case of the order of a few percent.

\begin{table}[!ht]
\begin{center}
\begin{tabular}{|l | c | c|}\hline
survey & mean depth $\langle R \rangle$ & median depth $\bar{R}$ \\ \hline
2MRS & $104\Mpch$  & $96\Mpch$ \\ \hline
2MASS XSC & $243\Mpch$ & $225\Mpch$ \\ \hline
\end{tabular}
\end{center}
\caption{\label{Tab:depth} \small Mean and median effective depth of flux-limited surveys: the 2MASS Redshift Survey (2MRS, $\Kmax=11.75$ mag) and the 2MASS Extended Source Catalog (XSC, $\Kmax=13.6$ mag), calculated from Eqs.\ (\ref{eq:mean.depth}) and (\ref{eq:median.depth}), respectively, for the $K$-band luminosity function in the Schechter form with $\alpha=-1.16$ and $M_*=-23.83+5\log h$ \citepc{6dF_Fi}.}
\end{table}

The effective depth of a flux-limited survey is however not a good measure of the effective \textit{distance} of a thin shell in `flux space' (in which the third coordinate is the flux, by analogy with the redshift space) that we seek. What we need is the mean value of distances of all galaxies \textit{with given flux} $S$. The mean that we calculate will be thus a conditional one. This derivation is qualitatively the same as for the mean redshift  of galaxies with a given flux, presented on pages 120--121 of \citec{Pe93}.

We start by deriving the joint probability distribution of galaxy distances $r$ and fluxes $S$. It is easily obtained by differentiating Eq.\ (\ref{eq:mathcalN}) with respect to luminosity $L$. Hence, the differential number $\delta^2 N$  of galaxies with a LF $\Phi(L)$ in a volume element $\delta V$ is given by
\begin{equation}
\delta^2\! N=\Phi(L)\,\delta L\,\delta V
\end{equation}
with $\delta V=4\pi r^2\,\delta r$. Now, passing from luminosity to flux, $L=4\pi r^2 S$, for fixed $r$ we have $\delta L=4\pi r^2\, \delta S$. This gives the joint probability as
\begin{equation}
p(r,S)=\frac{\partial^2\! N}{\partial r\, \partial S}=16\,\pi^2\,r^4\,\Phi(4\pi r^2 S).
\end{equation}
The conditional probability for $r$ given $S$ is
\begin{equation}
p(r|S)=\frac{p(r,S)}{p(S)}=\frac{16\,\pi^2\,r^4\,\Phi(4\pi r^2 S)}{\int_0^{+\infty}{16\,\pi^2\,r^4\,\Phi(4\pi r^2 S)}\,\de r}\;,
\end{equation}
hence the \emph{conditional mean} for $r$ given $S$ will be
\begin{equation}\label{eq:r_mean}
\langle r \rangle_S=\frac{\intlim_0^{+\infty}{r^5\,\Phi(4\pi r^2 S)\,\de r}}{\intlim_0^{+\infty}{r^4\,\Phi(4\pi r^2 S)}\,\de r}\;.
\end{equation}
For the Schechter form of the luminosity function, let us define a characteristic distance $r_S$, which is the distance to a galaxy with a given flux $S$ and the characteristic luminosity $L_*$ of the LF:
\begin{equation}
r_S=\sqrt{\frac{L_*}{4\pi S}}\;.
\end{equation}
Then for the Schechter LF, Eq.\ (\ref{eq:r_mean}) simplifies to
\begin{equation}
\langle r \rangle_S=
r_S\,\frac{\Gamma(\alpha+3)}{\Gamma(\alpha+5\slash 2)}
\end{equation}
with $\Gamma(a)$ being the Gamma function.

Similarly, the \emph{conditional median} for $r$ given $S$, denoted here as $\bar{r}_S$, will be obtained from the implicit equation
\begin{equation}
{\intlim_0^{\bar{r}_S}{r^4\,\Phi(4\pi r^2 S)}\,\de r}={\intlim_{\bar{r}_S}^{+\infty}{r^4\,\Phi(4\pi r^2 S)}\,\de r}\;,
\end{equation}
which for the Schechter LF is equivalent to solving
\begin{equation}
\gamma\left(\alpha+\frac{5}{2},\frac{\bar{r}_S^2}{r_S^2}\right)=\Gamma\left(\alpha+\frac{5}{2},\frac{\bar{r}_S^2}{r_S^2}\right)
\end{equation}
with respect to $\bar{r}_S$, where $\gamma(a,x)$ and $\Gamma(a,x)$ are respectively the lower and upper incomplete Gamma functions.

In the particular case of the $K$ band, taking $\alpha=-1.16$ \citepc{6dF_Fi}, we obtain
\begin{equation}
\langle r \rangle_S=1.056\, r_S
\end{equation}
and
\begin{equation}
\bar{r}_S=1.013\, r_S\;.
\end{equation}
Note that especially the median conditional distance gives a value very close to the characteristic distance $r_S$, which could be the `first-guess' effective distance (Tully 2008, private communication).

For the purpose of the present work, we have decided to use the conditional median as our measure of the effective distance of galaxies with a given flux. This particular choice is somewhat arbitrary and influences the scaling of top axes in Figures \ref{Fig:growth.r}, \ref{Fig:misal}, \ref{Fig:predicted} and \ref{Fig:g_m and v_c}. However, since the difference between the conditional mean and median is very small (about 5\%), this choice has a negligible impact on general conclusions and results of Chapter~\ref{Ch:Growth}.

Relating the flux to the $K$ magnitude via Eq.\ (\ref{eq:S_i}), $S=S_0\,10^{-0.4 K}$, and owing to the luminosity-magnitude relation $L_*=4 \pi (10\, \mathrm{pc})^2 S_0\,10^{-0.4 M_*}$, we obtain
\begin{equation}
r_\mathrm{eff}\equiv \bar{r}_S=1.013\times 10^{0.2(K-M_*)-5}\,\mathrm{Mpc}\;,
\end{equation}
so finally, for $M_*=-23.83+5\log h$ \citepc{6dF_Fi},
\begin{equation}\label{eq:r.eff.app}
r_\mathrm{eff}=0.591\times10^{0.2K}\Mpch\;.
\end{equation}
For example, galaxies at the limit of the 2MASS catalog (with $K=13.6$ mag) are assigned an effective distance of $r_\mathrm{eff}=310\Mpch$.

{\chapter{\textsf{\textbf{Error estimation via bootstrap resampling}}}
\label{App:Bootstrap}
Direct estimation of the errors of the measured dipole, as well as of the misalignment angle (Chapters \ref{Ch:Growth} and \ref{Ch:MLE}), is very difficult if feasible at all. Many factors contribute to these errors, such as photometric and astrometric uncertainties in the catalog (which are however tiny), biases related to the data preparation procedure as described in Chapter~\ref{Ch:Data}, etc. For that reason we have chosen to use the widely adopted \textit{bootstrap resampling} procedure to estimate the confidence intervals of our measurements. In general, \textit{bootstrapping} is a statistical method for constraining the distribution of an estimator by sampling with replacement from the original sample, most often with the purpose of deriving robust estimates of standard errors and confidence intervals of population parameters \citepc{ET94}. In our case, it was the 2MASS catalog that was resampled. Based on the original catalog, we created `mock' datasets by drawing with replacement from the original one as many sources as actually exist. Such a mock catalog thus created contained the same number of galaxies as the real 2MASS XSC, however some of the original galaxies were missing, while some were repeated more than once, in line with the bootstrap ideology. For our purposes we created one thousand such catalogs (independent of each other), we checked however that the procedure converged already for 100 mocks. For each of these 1000 catalogs, we applied the similar procedure as for the real data. Namely, for the purpose of the analysis presented in Chapter~\ref{Ch:Growth}, we calculated the 2MASS clustering dipole and the misalignment angle by increasing the maximum magnitude of the sample. The values obtained for each of the mocks up to a given $\Kmax$ were averaged and used to calculate the $1\sigma$ variance. These variances were applied as error estimates for the data in Figure~\ref{Fig:g_m and v_c} and in Subsection \ref{Subsec:Compare}. Similarly, for the purpose of Chapter~\ref{Ch:MLE}, we did the bootstrap for every subsample limited by a given $\Kmin$ and $\Kmax$, which allowed us to estimate the errors in the measured gravity and (the cosine of) the misalignment angle (cf.\ Figures \ref{Fig:teta_rho} and \ref{Fig:rhoprim_mi}).}

\addcontentsline{toc}{chapter}{{\textsf{\textbf{List of figures}}}}
\listoffigures

\addcontentsline{toc}{chapter}{\textsf{\textbf{Bibliography}}}
\begin{small}
\bibliographystyle{mojstyl}
\bibliography{tezaMB}
\end{small}


\end{document}